\documentclass[longauth]{aa}
\usepackage{placeins}

\usepackage{graphicx}
\usepackage[flushleft]{threeparttable}
\usepackage{comment}
\usepackage{txfonts}
\usepackage[citecolor=blue, linkcolor=blue]{hyperref}
\hypersetup{colorlinks=True}
\newcommand{\Msun}{M$_{\odot}$}
\newcommand{\Mstar}{M$_{\star}$}
\newcommand{\xiion}{$\xi_{\rm ion}$}
\newcommand{\halpha}{H${\alpha}$}
\newcommand{\hbeta}{H${\beta}$}
\newcommand{\lya}{Ly${\alpha}$}
\newcommand{\flya}{fesc$_{\rm{Ly}\alpha}$}
\newcommand{\flyc}{fesc$_{\rm{LyC}}$}
\newcommand{\oiii}{[O~III]}
\newcommand{\oii}{[O~II]}
\newcommand{\neiii}{[Ne~III]}
\newcommand{\oiiihb}{[O~III]+\hbeta}

\begin{document} 

\title{Extreme equivalent width-selected low-mass starbursts at $z=4-9$: insights into their role in cosmic reionization}
 
\author{Llerena, M.\inst{1}\fnmsep\thanks{\email{mario.llerenaona@inaf.it}}
\and Pentericci, L.\inst{1}
\and Amor\'in, R. \inst{2}
\and Ferrara, A. \inst{3}
\and Dickinson, M. \inst{4}
\and Arevalo-Gonzalez, F. \inst{1,5}
\and Calabr{\`o}, A.\inst{1}
\and Napolitano, L. \inst{1,5}
\and Mascia, S. \inst{6}
\and Arrabal Haro, P.\inst{7} 
\and Begley, R.\inst{8}
\and Cleri, N. J. \inst{9,10,11}
\and Davis, K. \inst{12,13}\fnmsep\thanks{{NSF Graduate Research Fellow}}
\and Hu, W. \inst{14,15}
\and Kartaltepe, J. S. \inst{16} 
\and Koekemoer, A. M.\inst{17}
\and Lucas, R. A. \inst{17}
\and McGrath, E. \inst{18}
\and McLeod, D. J.\inst{19}
\and Papovich, C.\inst{14,15}
\and Stanton, T. M.\inst{19}
\and Taylor, A. J. \inst{20,21}
\and Tripodi, R. \inst{1}
\and Wang, X.\inst{22,23,24}
\and Yung, L.~Y.~A. \inst{17}
          }

   \institute{INAF - Osservatorio Astronomico di Roma, Via di Frascati 33, 00078, Monte Porzio Catone, Italy
   \and Instituto de Astrofísica de Andalucía (CSIC), Apartado 3004, 18080 Granada, Spain
   \and Scuola Normale Superiore, Piazza dei Cavalieri 7, 50126 Pisa, Italy
   \and NSF NOIRLab, 950 N. Cherry Ave., Tucson, AZ 85719, USA
   \and Dipartimento di Fisica, Università di Roma Sapienza, Città Universitaria di Roma - Sapienza, Piazzale Aldo Moro, 2, 00185, Roma, Italy
   \and Institute of Science and Technology Austria (ISTA), Am Campus 1, A-3400 Klosterneuburg, Austria 
   \and Astrophysics Science Division, NASA Goddard Space Flight Center,
8800 Greenbelt Rd, Greenbelt, MD 20771, USA
   \and{Armagh Observatory and Planetarium, College Hill, Armagh, BT61 9DG, N. Ireland, UK}
   \and {Department of Astronomy and Astrophysics, The Pennsylvania State University, University Park, PA 16802, USA}
\and {Institute for Computational and Data Sciences, The Pennsylvania State University, University Park, PA 16802, USA}
\and {Institute for Gravitation and the Cosmos, The Pennsylvania State University, University Park, PA 16802, USA}
\and Department of Physics, 196 Auditorium Road, Unit 3046, University of Connecticut, Storrs, CT 06269, USA
\and Los Alamos National Laboratory, Los Alamos, NM 87545, USA
\and{Department of Physics and Astronomy, Texas A\&M University, College Station, TX, 77843-4242 USA}
\and{George P.\ and Cynthia Woods Mitchell Institute for Fundamental Physics and Astronomy, Texas A\&M University, College Station, TX, 77843-4242 USA}
 \and Laboratory for Multiwavelength Astrophysics, School of Physics and Astronomy, Rochester Institute of Technology, 84 Lomb Memorial Drive, Rochester, NY 14623, USA
   \and Space Telescope Science Institute, 3700 San Martin Drive, Baltimore, MD 21218, USA
   \and Department of Physics and Astronomy, Colby College, Waterville, ME 04901, USA
  \and{Institute for Astronomy, University of Edinburgh, Royal Observatory, Edinburgh, EH9 3HJ, UK}
  \and Department of Astronomy, The University of Texas at Austin, Austin, TX, USA
  \and Cosmic Frontier Center, The University of Texas at Austin, Austin, TX, USA
\and {School of Astronomy and Space Science, University of Chinese Academy of Sciences (UCAS), Beijing 100049, China}
\and {National Astronomical Observatories, Chinese Academy of Sciences, Beijing 100101, China}
\and {Institute for Frontiers in Astronomy and Astrophysics, Beijing Normal University, Beijing 102206, China}
    } 

   \date{Received ; accepted }

  \abstract
   {We investigate the properties of extreme emission line galaxies (EELGs) at $z=4-9$ and their role in reionization. Compact, low-mass galaxies with intense optical emission lines are linked to elevated specific star formation rates (sSFRs) and recent bursts of star formation. Feedback in these systems may enable the leakage of ionizing radiation into the intergalactic medium. Using JWST/NIRSpec spectroscopy from the CAPERS, CEERS, and RUBIES surveys, we compile 160 NIRCam-selected EELGs in the EGS field. These galaxies show extreme rest-frame equivalent widths (EWs), with a median EW(\oiiihb)=1616\r{A} and EW(\halpha)=763\r{A}. They are low-mass (median log(\Mstar/\Msun)=8.26) with high sSFRs (median 43 Gyr$^{-1}$), above the $z\sim6$ main sequence. UV slopes are diverse, with a {median} $\beta=-2.0$, and only 7\% have extremely blue continua ($\beta<-2.6$). Emission-line diagnostics suggest stellar populations as the primary ionizing source, although an AGN fraction of 14\% {cannot be entirely ruled out}. These galaxies are efficient ionizing photon producers, with {median} log(\xiion [Hz erg$^{-1}$])={25.37}, exceeding typical values at similar redshifts. Escape fractions, however, are heterogeneous: {16}\% of EELGs {at $z<7$} show escape fractions $>${5}\% for both \lya~and LyC photons, while 82\% lack detectable \lya~emission. The median inferred LyC escape fraction is modest ({5}\%) but enhanced in {compact} super-Eddington systems with sSFR >25 Gyr$^{-1}$. {These results indicate that EELGs contribute approximately $16-40$\% of the total ionizing emissivity required to sustain hydrogen reionization.} {EELGs} are extremely compact, with a median effective radius of 0.49 kpc, and exhibit a recent star-formation burst. Our analysis indicates that sSFR and star-formation rate surface density are the primary drivers of their extreme emission line strengths. 
   }

   \keywords{Galaxies: starburst --
                Galaxies: high-redshift --
                Galaxies: evolution --
                Galaxies: formation --
                Galaxies: ISM
               }
\titlerunning{Extreme equivalent width-selected low-mass starbursts at $z=4-9$}
\authorrunning{Llerena, M. et al.}

\maketitle
%

\section{Introduction} \label{sec:intro}

Extreme emission line galaxies (EELGs) are characterized by intense emission lines that stand out prominently against their stellar continuum, resulting in large rest-frame equivalent widths (EWs). Some examples of these lines include rest-frame optical [O~III]$\lambda5007$\r{A} (hereafter \oiii) and \halpha. The strong nebular emission lines are typically powered by ionizing photons emitted from massive, short-lived O and B stars (or active galactic nuclei, AGN), while the surrounding rest-frame optical continuum arises mainly from longer-lived, lower-mass stars \citep[e.g.,][]{Eldridge2022}. Therefore, EW provides insight into the ratio of very young to older stellar populations, serving as an indicator of variations in star formation history (SFH) over time. Identifying EELGs within a galaxy population can reveal galaxies experiencing a burst in star formation \citep[e.g.,][]{Endsley2023}, often characterized by high specific star formation rates (sSFR) and compact morphologies \citep[e.g.,][]{Tang2022}. 

Based on the recent Attenuation-Free Model \citep[AFM,][]{Ferrara2024}, elevated sSFR~>~25~Gyr$^{-1}$ might be linked to galaxies that become super-Eddington {(SE)} and trigger a radiation-driven outflow that clears the dust, making the galaxy UV brighter and bluer. The presence of such outflows is crucial for understanding the escape of ionizing Lyman continuum (LyC, $\lambda_{\text{rest}} < 912$~\r{A}) photons, as they can create ionized, dust-free channels that enable the leakage of LyC photons \citep[e.g.,][]{Hogarth2020,Amorin2024,Carr2025,Komarova2025}. Outflows can significantly increase the visibility of the \lya~line even at $z>5.5-6$ when the intergalactic medium (IGM) is not fully ionized \citep[e.g.,][]{Fan2006,Mason2020}. In samples of $z>7$ galaxies, where in principle \lya~should be strongly attenuated
by the neutral IGM, high EW(\lya) are associated with high sSFRs and ionized bubbles large enough to produce frequency shifting that reduces the effects of resonant scattering for \lya~\citep[e.g.,][]{Endsley2022alma,Bunker2023jades,Tang2023,Witstok2025}. Evidence of outflows has been observed in samples of \oiii~emitters in a wide range of redshifts \citep[e.g.,][]{Llerena2023,Saldana-Lopez2025,Cooper2025,Leon-Contreras2025,Zamora2025} with a fraction of $\sim20$\% of galaxies at $z<6$ showing signatures of outflows \citep[e.g.,][]{Calabro2024sigmasfr,Carniani2024}. Furthermore, strong [O III]$\lambda\lambda4959,5007$\r{A}+\hbeta~(hereafter \oiiihb) emission and high O32=log(\oiii/[O II]$\lambda\lambda$3727,3729\r{A}) have been suggested as a necessary requirement for high escape fractions of LyC photons (\flyc) given that these conditions are observed in a high proportion of confirmed LyC
leaking galaxies \citep[e.g.,][]{Vanzella2016,Izotov2021}. Additionally, galaxies exhibiting high EW(\oiii) tend to show enhanced ionizing photon production efficiencies (\xiion) at different redshifts \citep[e.g.,][]{Chevallard2018,Tang2019,Llerena2024xi}. However, very high \flyc\ or very low metallicities could affect the strength of \oiii~emission, so the connection is still not fully understood \citep[e.g.,][]{Endsley2022,Laseter2025}.

Due to their extreme emission properties and intense star formation activity, EELGs serve as valuable laboratories for studying the physical conditions and mechanisms that enable the escape of both LyC and \lya~photons, offering key insights into their role in the Epoch of Reionization (EoR). EELGs are observed at different redshifts, from the local universe \citep[e.g.,][]{Cardamone2009,Amorin2010,Izotov2011}, intermediate redshifts $z\sim 1-3$ \citep[e.g.,][]{vanderWel2011,Amorin2015, Maseda2018,Tang2019,Du2020,Llerena2023}, and during the EoR \citep[e.g.,][]{Smit2015,DeBarros2019,Boyett2024, Llerena2024EELG, Davis2023,Begley2025,Daikuhara2025}. These high-EW systems can be identified through either direct spectroscopy or indirect photometric flux excesses, where strong line emission significantly enhances the flux in determined filters.  JWST \citep{Gardner2006, Gardner2023} has enabled highly sensitive near-infrared spectroscopy with NIRSpec \citep{Jakobsen2022}, covering wavelengths up to 5.3~$\mu$m. This makes it possible to directly measure rest-frame optical emission lines at high-$z$, such as \halpha~up to $z \lesssim 7$ and \oiii~up to $z \lesssim 9.5$. These capabilities allow for robust identification of EELGs and a detailed characterization of their physical properties as spectroscopic samples continue to grow.

In \citet{Llerena2024EELG}, a sample of EELG candidates at $z = 4-9$ in the Extended Groth Strip \citep[EGS, 14h19m00s +52$^{\circ}$48'00'', ][]{Davis2007} field was identified based on their JWST/NIRCam \citep{Rieke2023} photometry, specifically using the F277W, F356W, F444W, and F410M filters. The selection was restricted to galaxies with rest-frame EW(\oiiihb) $>$ 680\r{A}, motivated by the average spectra of rare, metal-poor starbursts in the local universe \citep{SA2012, EPM2021}. Based on their photometry, these candidates exhibit low stellar masses, high sSFRs, and compact morphologies, which are characteristics associated with young, rapidly evolving galaxies. In this paper, we analyze the JWST/NIRSpec spectra of a subsample of these candidates to confirm the presence of such extreme emission lines. With this spectroscopic confirmation, we investigate their physical properties, including their ionization conditions and star formation activity, and explore the mechanisms driving the elevated EWs. We further investigate the potential link between strong optical line emission and the production and escape of LyC photons, thereby assessing the role of EELGs in cosmic reionization. Since our analysis is restricted to the EGS field, we note that cosmic variance must be taken into account, as this region has been suggested to be already nearly fully ionized by $z \sim 7$ \citep[e.g.,][]{Napolitano2024lya,Chen2025,Napolitano2025capers}

This paper is organized as follows: in Sec. \ref{sec:sample}, we describe the {spectroscopic} sample selection. In Sec. \ref{sec:analysis}, we outline the procedures for measuring emission lines {and the validation of the photometric selection. In Sec. \ref{sec:physical_properties}, we present the physical properties of the sample based on spectral energy distribution (SED) fitting. We also derive star formation rates from Balmer lines, the rest-frame UV continuum slopes ($\beta$), absolute UV magnitudes, the dust reddening, and the gas-phase metallicity.} In Sec. \ref{sec:nature}, we analyze the source of ionization of the EELGs based on emission line diagnostics. In Sec. \ref{xiion}, we present the results of the estimated \xiion~in this sample. {In Sec. \ref{sec:escapelyc} and Sec. \ref{sec:escapelya}, we present the results regarding the escape of LyC and \lya~photons, respectively}.  In Sec. \ref{sec:drivers}, we discuss the potential drivers of the elevated EWs, which include high sSFRs, compactness, and burstiness. Finally, we present our summary and conclusions in Sec. \ref{sec:conclusions}.

Throughout this paper, we adopt a $\Lambda$-dominated flat universe with $\Omega_\Lambda = 0.7$, $\Omega_M = 0.3,$ and H$_0= 70$ km s$^{-1}$ Mpc$^{-1}$. All magnitudes are quoted in the AB system \citep{Oke1983}. EWs are quoted in the rest-frame and are positive for emission lines. We adopt log(O/H)$_{\odot}=8.69$ \citep{Asplund2009}. We assumed the \cite{Kroupa2001} initial mass function (IMF) for the SED fitting. We report Spearman’s correlation coefficients ($\rho$) with their p-values ($p$), to assess the strength and significance of correlations.

\section{Data and sample selection}\label{sec:sample}
In this study, we aim to evaluate the effectiveness of the selection method proposed in \cite{Llerena2024EELG} for identifying EELGs at $z=4-9$ based on their photometry and then to study their properties to understand the origin of the extreme emission lines. To this end, we construct a subsample of galaxies from the parent sample with available JWST/NIRSpec observations. This includes spectra from the following JWST surveys: the CANDELS-Area Prism Epoch of Reionization Survey (CAPERS, GO-6368; PI Mark Dickinson), the Cosmic Evolution Early Release Science survey \citep[CEERS,][]{Finkelstein2023,Finkelstein2023b,Finkelstein2025}, and the Red Unknowns: Bright Infrared Extragalactic Survey \citep[RUBIES,][]{deGraaff2025}.

\subsection{{Selection of the spectroscopic sample}}
\subsubsection{CAPERS sample}
CAPERS is an ongoing JWST Cycle 3 Treasury program of
moderately deep NIRSpec multiobject spectroscopy in three fields from CANDELS \citep{Grogin2011,Koekemoer2011}, which includes the EGS field. CAPERS is obtaining low-resolution NIRSpec/PRISM spectra for thousands
of distant galaxies, including objects at $z>10$ \citep[e.g.,][]{Kokorev2025}. We cross-matched the list of EELG candidates in \cite{Llerena2024EELG} with the available NIRSpec/PRISM spectra from the CAPERS survey in the EGS field, and we considered the spectra from the data release of the collaboration for the analysis. The data reduction is described in \cite{Taylor2025capers}. We identified 62 candidates in the survey, but excluded 7 from the sample because their bright emission lines, based on the photometric redshift, coincided with a detector gap. We discarded an additional galaxy because its reduced spectrum was contaminated by a secondary source in the slit.

\subsubsection{CEERS and RUBIES samples}
We also cross-matched the list of EELG candidates in \cite{Llerena2024EELG} with the available spectra in the Public NIRSpec datasets (v3) published on the DAWN JWST Archive \citep[DJA\footnote{\url{https://dawn-cph.github.io/dja/spectroscopy/nirspec/}},][]{Heintz2024dja, deGraaff2025} and we used the NIRSpec data reduced by the Cosmic Dawn Center. In particular, we selected galaxies from the CEERS and RUBIES surveys with NIRSpec/PRISM and NIRSpec/G395M (medium resolution, MR) spectra. Regarding the CEERS subsample, we found 32 galaxies with PRISM spectra in the DJA database. We excluded 3 galaxies because their bright emission lines coincide with a detector gap, as determined by the photometric redshift. We also found 10 galaxies with MR spectra. Regarding the RUBIES subsample, we identified 48 galaxies with PRISM spectra in the DJA database, excluding 4 because of the detector gaps, as in the CEERS subsample. In addition, we found 76 galaxies with MR spectra and excluded 5 of them for the same reason.

In summary, our final sample consists of 208 spectra: 127 EELG candidates observed with the PRISM configuration and 81 with the MR configuration. A subsample of 42 galaxies was observed in both PRISM and MR within CEERS and RUBIES, for which we analyze both spectra. Additionally, two galaxies are included in both CAPERS and RUBIES, and four more are observed in both CEERS and RUBIES. Given the small overlap among surveys, we retain all spectra in our analysis. Therefore, the final sample comprises 160 unique galaxies. The coordinates of the galaxies, the corresponding survey and disperser, are listed in Table \ref{lt_ewo3_err}. Throughout this paper, individual galaxies are identified using their IDs from the CEERS photometric catalog.

\subsection{Spectroscopic redshifts}\label{sec:spec_z}

\begin{figure}[t!]
    \centering
    \includegraphics[width=\linewidth]{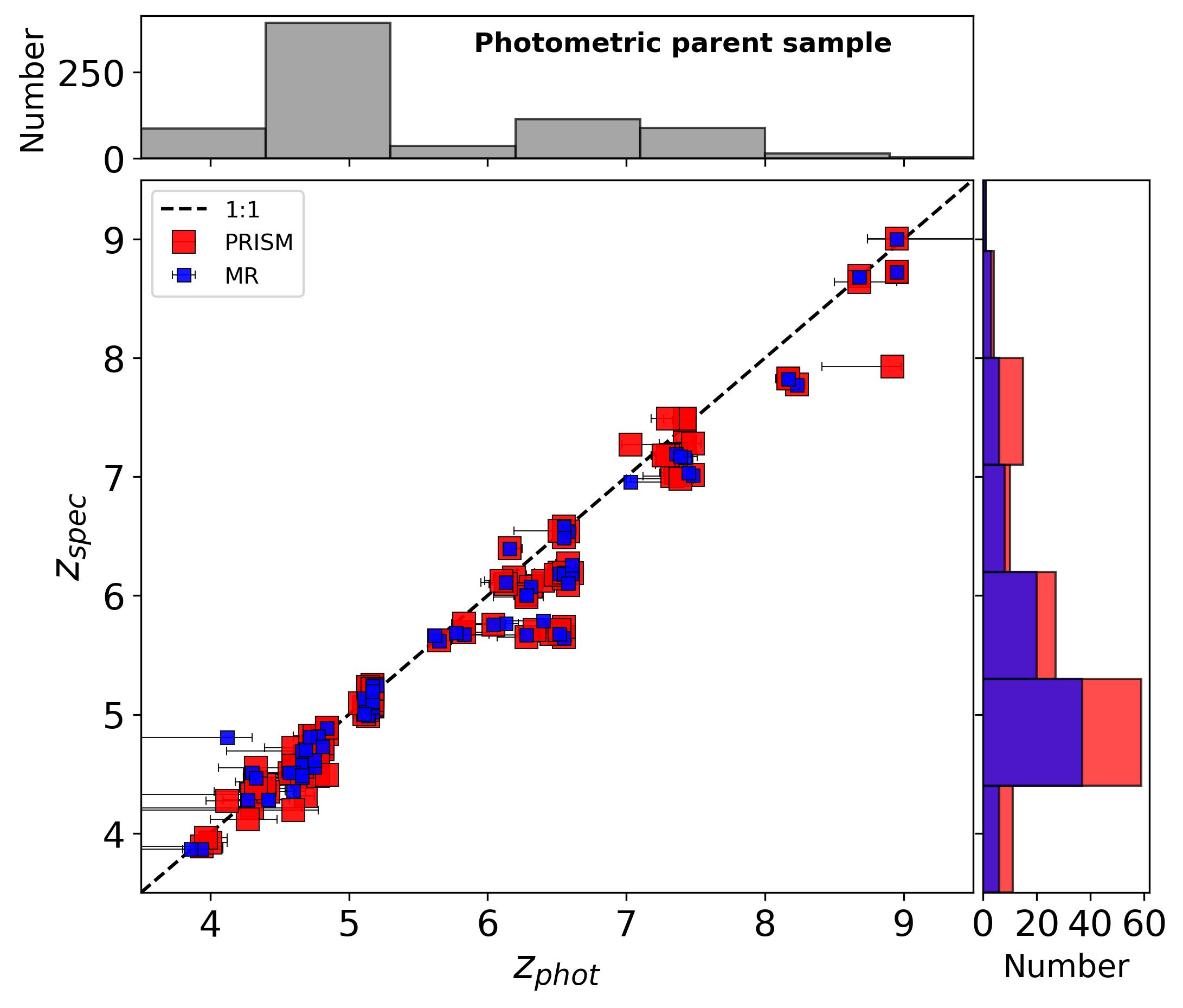}
    \caption{{\textit{Central panel}}: Comparison between $z_{phot}$ and $z_{spec}$ {in the sample of selected EELGs.} The dashed black line is the 1:1 relation. \textit{Right panel}: Distribution of $z_{spec}$, shown as superimposed histograms. {In both panels,} in red (blue), the galaxies with PRISM (MR) spectra. {\textit{Top panel}: Distribution of $z_{phot}$ in the photometric parent sample \citep{Llerena2024EELG}.} }
    \label{fig:zphot_zspec}
\end{figure}

As a first step, we assessed the accuracy of the photometric redshifts ($z_{phot}$). To this aim, we used LiMe\footnote{\url{https://lime-stable.readthedocs.io/en/latest/}} \citep{Fernandez2024Lime}, a library that provides a set of tools to fit lines in astronomical spectra, to estimate the spectroscopic redshifts ($z_{spec}$) in both the PRISM and MR spectra. We assumed a single Gaussian for each line {and a local linear continuum}, with \oiiihb~modeled together as three blended Gaussians {(see two showcases in Fig. \ref{fig:z_fitting} in the Appendix \ref{sec:App_o3})}, while \halpha~was modeled with a Gaussian profile, merged (blended) with [N II]$\lambda\lambda$6548,6583\r{A} for PRISM (MR) spectra. We adopted $z_{phot}$ as an initial estimate of the redshift. We determined the observed central wavelength of the Gaussian fits of \oiii~to calculate $z_{spec}$. When \oiii~is not covered by the spectral range, the observed central peak of \halpha~is used to determine $z_{spec}$. The distribution of $z_{spec}$ of the sample is displayed in {the right panel} in Fig. \ref{fig:zphot_zspec} and reported in Table \ref{lt_ewo3_err}. The sample covers redshifts from $z=3.86$ to $z=9.00$ with a median $z=5.22$ ($\sigma=1.17$). {We note that the peak of the $z_{spec}$ distribution approximately coincides with the peak of the $z_{phot}$ distribution of the photometrically selected parent sample (top panel of Fig.~\ref{fig:zphot_zspec}).}

In {the central panel in} Fig. \ref{fig:zphot_zspec} we show a comparison between $z_{phot}$ and $z_{spec}$, including the 1$\sigma$ uncertainty of the $z_{phot}$ based on their probability distribution functions \citep{Finkelstein2023}. We find a good agreement ($\rho=0.97$, $p\sim0$) between both quantities with a mean $\Delta z=z_{phot}-z_{spec}=0.13$ ($\sigma=0.23$){, considering only one spectrum per galaxy in cases where a galaxy has both PRISM and MR spectra.} The galaxies at $z\sim4$ with large $z_{phot}$ uncertainties often have low-$z$ solutions with high probability, which are ruled out by the spectroscopic data. We highlight that no catastrophic outliers with $\Delta z>3$ are found. We note that larger $\Delta z$ are observed at $z\sim6-7$. In this range, \oiiihb~falls within the F356W filter, while \halpha~lies in the broader F444W filter. Although the F410M filter sits between them, it primarily samples the continuum rather than overlapping a strong spectral feature, and thus provides limited improvement in redshift precision. The combination of these two broad filters appears to result in less accurate $z_{phot}$ estimates compared to the $z\sim4-6$ range, where \halpha~may be captured by the narrower F410M filter, or where strong emission lines fall into the F277W and F356W filters, both of which are narrower than F444W. 

\section{Data analysis}\label{sec:analysis}

Before proceeding with the spectral analysis, and given the different data reduction applied to the sample, we evaluate potential slit losses by comparing NIRCam photometry with synthetic photometry derived from the spectra. This consistency check ensures that the extracted spectra are properly calibrated and that any significant discrepancies, potentially arising from slit losses or extraction artifacts, are identified and corrected. We used the same photometry described in Sec. \ref{sec:sed_fitting}. We find that, on average, the correction factor, defined as the ratio between observed and synthetic photometry, does not show a significant wavelength dependence, which is consistent with previous works \citep[e.g.,][]{Roberts-Borsani2024, Napolitano2024}, and yields a median value of 1.21. Therefore, to account for slit losses, we scale each spectrum by a constant correction factor, which is unique to each galaxy and calculated as the uncertainty-weighted median of the correction factors across all observed filters for that galaxy.

\subsection{Emission line measurements}

{To determine $z_{spec}$ in Sec. \ref{sec:spec_z}, we fit only the bright \oiiihb~and \halpha~emission lines. In order to measure fluxes of additional emission lines, we then performed a second set of measurements with LiMe, fixing $z_{spec}$ to the value derived in Sec.~\ref{sec:spec_z}.} In Fig. \ref{fig:stack} and \ref{fig:stack_mr}, we display the median spectra of galaxies observed with PRISM and MR, respectively, highlighting key spectral features. We modeled [O II]$\lambda\lambda$3727,3729 (hereafter \oii) and [Ne III]$\lambda$3868\r{A} (hereafter \neiii) in all galaxies using a single Gaussian profile for each line and a local linear continuum, estimated from 100 \r{A} (50 \r{A})-wide windows on either side of each line for the PRISM (MR) data. For lines with a signal-to-noise ratio (S/N) below 3, we adopt a 3$\sigma$ upper limit. 

To estimate the rest-frame EWs {of bright emission lines}, we use the continuum measured directly from the spectra when it is detected with S/N > 2 (48\% of the sample for \oiiihb~and 32\% of the sample for \halpha~with PRISM), and propagate both the line flux and continuum uncertainties to derive the uncertainty of EWs. When the continuum is undetected, we adopt the continuum estimated from the SED model (see Sec. \ref{sec:sed_fitting}), and derive its uncertainty by propagating the uncertainties of the photometry from the two closest filters in wavelength to each line. On average, the ratio between the continuum measured from the spectra and from the SED model is $0.9-1.1$. The EWs and uncertainties {of \hbeta, \oiii~and \halpha}~are reported in Table \ref{lt_ewo3_err}.

Given the complexity of \lya, we performed a separate set of measurements using the PRISM spectra, which provide suitable wavelength coverage. Owing to the low spectral resolution of the PRISM, \lya~is modeled within a 2-pixel window centered on the observed emission peak. As a first step, we compute the line flux via direct integration over this window. The continuum is estimated by a linear fit to the red side of \lya, from 1900~\r{A} to 3 pixels redward of the emission peak. We adopt a 3$\sigma$ upper limit for \lya~emission if the S/N < 3 or if the measured flux is negative. In our sample, we find 21 galaxies with \lya~in emission (above 3$\sigma$), i.e., Ly$\alpha$ emitters (LAE). We note that, in contrast to ground-based observations with larger slit widths, and given the resonant nature of \lya~emission where \lya~photons can scatter to spatial scales exceeding the stellar continuum {\citep[e.g.,][]{Ning2024},} JWST slit losses of 35\% have been reported for \lya~samples at $z=6-7$ \citep[][and references therein]{Napolitano2025capers}. We adopt this average value to correct for possible \lya~slit losses.

\subsection{{Confirmation of  extreme EWs}}

\begin{figure}[t!]
    \centering
    \includegraphics[width=\linewidth]{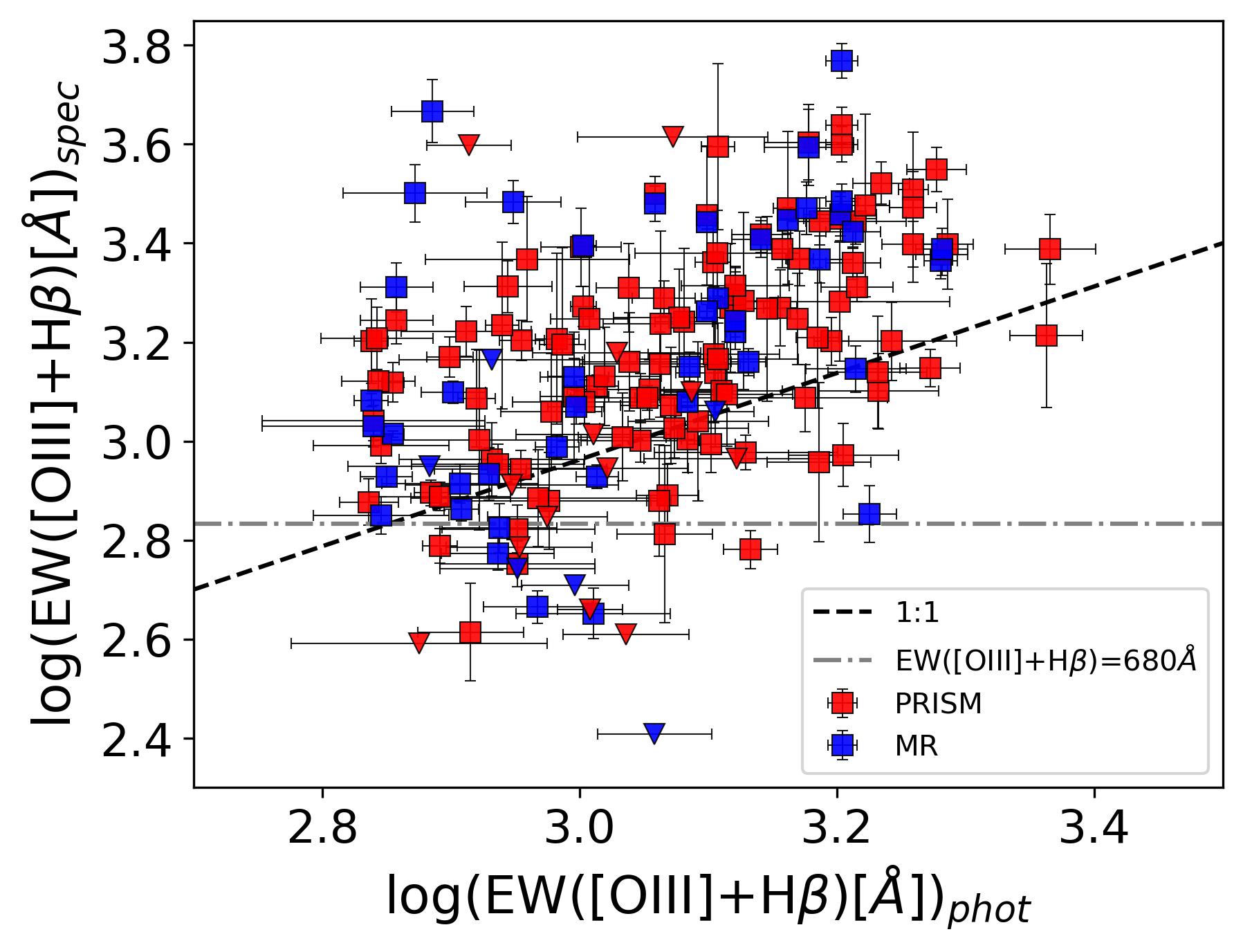}\\
    \includegraphics[width=\linewidth]{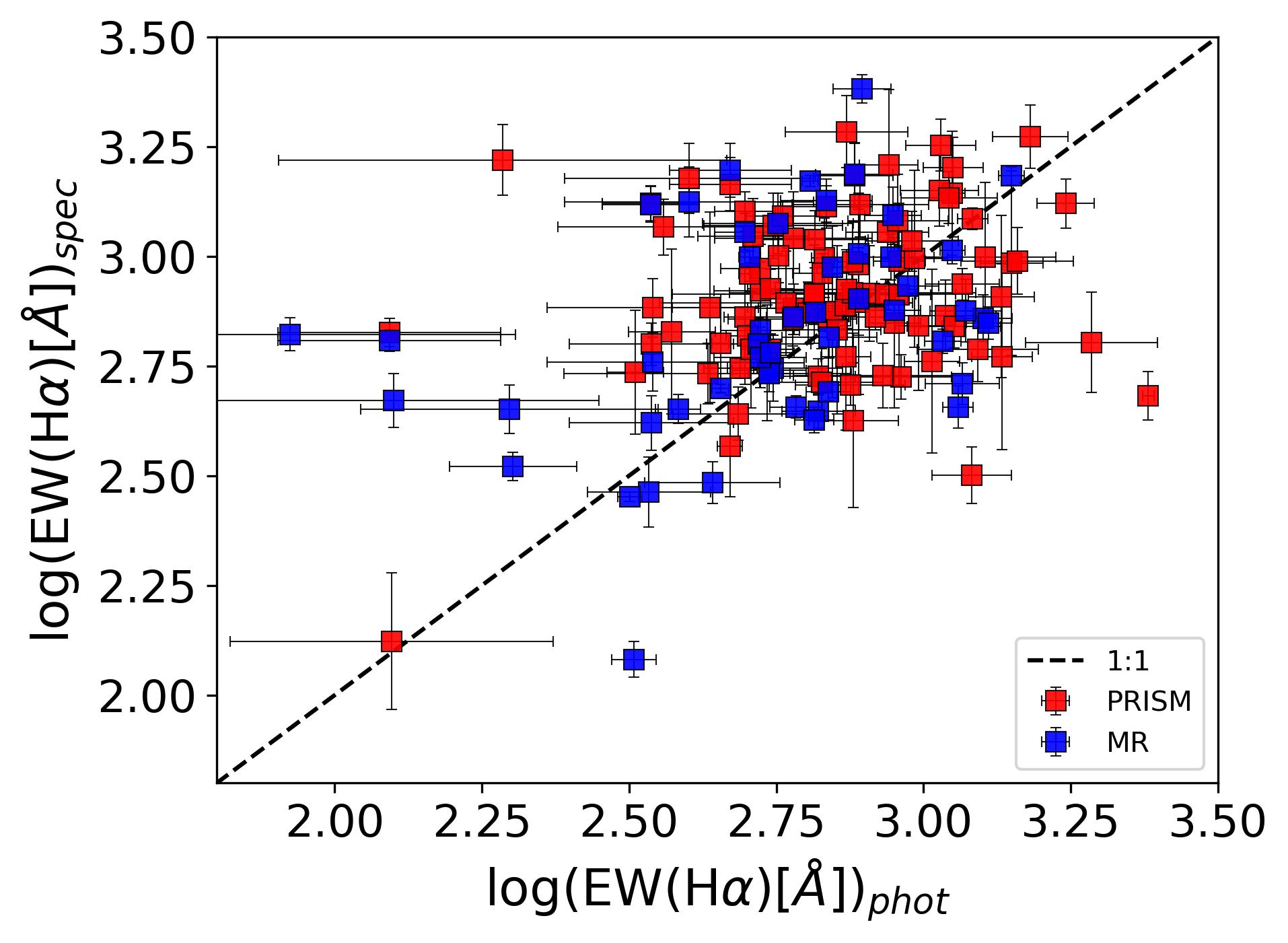}
    \caption{Comparison between EWs estimated from the photometry and from spectroscopy {in the sample of EELGs with PRISM (red squares) or MR (blue squares) spectra}. \textit{Top panel}: EW(\oiiihb). The triangle symbols are upper limits due to low S/N of \hbeta. {The gray dotted-dashed line is the lower limit for the photometric selection.} \textit{Bottom panel:} EW(\halpha). The dashed black line is the 1:1 relation.}
    \label{fig:ew_comp}
\end{figure}
{In this subsection, we focus on validating the selection method used to identify EELG candidates based on photometric criteria.} Fig. \ref{fig:ew_comp} shows a comparison between the EWs measured from the spectra and those estimated from the photometry in \cite{Llerena2024EELG}. In the top panel, we display the EW(\oiiihb), and we show that there is a moderate correlation ($\rho=0.50$, $p\sim0$) between the estimation and the spectroscopic confirmation with a median underestimation of 0.11~dex ($\sigma=0.22$) in the photometric values. This indicates that most of the candidates selected in the sample are EELGs with high EWs. We note that in 154 spectra (74\%) both \oiii~and \hbeta~are detected, with 93\% of them showing EW(\oiiihb)$>$680~\r{A}, which corresponds to the selection threshold adopted in \cite{Llerena2024EELG}. In 21 spectra (10\%), \oiii~is detected while only an upper limit is available for \hbeta; among these, 62\% present an upper limit in EW(\oiiihb) higher than the selection threshold. In 8 spectra (4\%), only \oiii~is detected while \hbeta~is not covered by the spectral range, and 75\% of these already exhibit lower limits for EW(\oiiihb) above the selection threshold. Finally, in 25 spectra (12\%), neither \oiii~nor \hbeta~are covered by the analyzed spectra. 

In summary, excluding the spectra where neither \oiii~nor \hbeta~are covered, the success rate of spectra with EW(\oiiihb) above the selection threshold (or consistent upper/lower limits)  reaches 89\%. For subsequent analysis, we removed galaxies that, based on either measured or limit values, do not fulfill the selection criterion of EW(\oiiihb)$>680$ \r{A}, since these sources were not targeted in photometric selection and would otherwise introduce incompleteness into the sample.

In this analyzed sample, we found a median log(EW(\oiiihb)[\r{A}])=3.20 (1616~\r{A}) and a 16–84th percentile range from 2.99 to 3.44 (997 to 2785~\r{A}). Such high EWs have been observed in galaxies in the EoR in photometric-selected samples \citep[e.g.,][]{DeBarros2019,Daikuhara2025} and with slitless spectroscopy \citep[e.g.,][]{Matthee2023}. Regarding only EW(\oiii), we found a median log(EW(\oiii)[\r{A}])=3.04 (1095~\r{A}) and a 16–84th percentile range from 2.80 to 3.28 (635 to 1946~\r{A}). In Fig. \ref{fig:ew_balmer_o3}, we show that EW(\oiii) and EW(\hbeta) are strongly correlated, suggesting a common source of ionization. 

On the other hand, in the bottom panel of Fig. \ref{fig:ew_comp}, we instead display the comparison with EW(\halpha), and we find a weaker correlation ($\rho=0.21$, $p\sim8\times10^{-3}$), but confirm the high EWs estimated from the photometry with a median underestimation of 0.08~dex ($\sigma=0.44$). We found a median log(EW(\halpha)[\r{A}])=2.88 (763~\r{A}) and a 16–84th percentile range from 2.71 to 3.11 (518 to 1307~\r{A}). {The ranges of EWs in the analyzed sample are summarized in Table \ref{tab:sample-median}.} In {Fig. \ref{fig:fluxes_prism_mr}}, we compared the fluxes and EWs of bright emission lines obtained with each disperser in the subsample with both PRISM and MR spectra. We show that they correlate strongly, with higher PRISM fluxes by $\sim 0.06$~dex on average. 

{\tiny
\begin{table}[t!]
    \centering
    \caption{{Median properties of the analyzed sample of EELGs. For each parameter, we report the median value and the corresponding 16th and 84th percentiles.}}
    \begin{tabular}{l|cc}
    \hline
         & Median & 16-84th interval \\\hline
         EW(\oiiihb)[\r{A}]&1616&(997, 2785)\\
        EW(\oiii)[\r{A}]&1095&(635, 1946)\\
        EW(\halpha)[\r{A}]&763&(518, 1307)\\
        log(\Mstar/\Msun) & 8.26&(7.84, 8.67)\\
        log(SFR$_{\rm Balmer}$[\Msun~yr$^{-1}$])& 0.90& (0.47, 1.42)\\
        $\beta$&-2.00& (-2.33, -1.66)\\
        M$_{\rm UV}$&-19.17& (-20.12, -18.46)\\
        Stellar E(B-V)&0.09& (0.04, 0.15)\\
        E(B-V)$_{\rm neb}$&0.18& (0.06, 0.30)\\
        log(O/H)+12&7.74& (7.45, 7.90)\\
        \hline\hline
    \end{tabular}
    \label{tab:sample-median}
\end{table}}

\section{Characterisation of  the sample of EELGs}\label{sec:physical_properties}
{In this section, we investigate the physical properties of the EELGs in our sample. Our analysis incorporates quantities derived from SED fitting, as well as measurements from the spectra, including star formation rates, UV continuum slopes, dust attenuation, and gas-phase metallicities. A summary of the median properties of the analyzed sample of EELGs is reported in Table \ref{tab:sample-median}.}

\subsection{{Spectral Energy Distribution fitting}}\label{sec:sed_fitting}

To estimate the physical properties of the galaxies in our sample, namely stellar mass, star-formation rate (SFR), and dust attenuation E(B–V), we update the SED fitting performed in \cite{Llerena2024EELG}. First, we used BAGPIPES version 1.2.0 \citep{Carnall2018} with the \cite{Bruzual_2003} stellar population models and photometry from the CEERS Photometric Catalogs v0.51.2 \citep{Finkelstein2025}. The photometry was performed with SExtractor \citep[v2.25.0; ][]{Bertin1996} with F277W and F356W as the detection bands. The fiducial fluxes were measured in small Kron apertures corrected by large-scale flux, following the methodology in \cite{Finkelstein2023}. The catalog includes photometric measurements in seven NIRCam filters: F115W, F150W, F200W, F277W, F356W, F410M, and F444W. In addition, it incorporates archival HST imaging of the EGS field from programs such as CANDELS \citep{Koekemoer2011, Grogin2011}, covering the filters F606W, F814W, F105W, F125W, F140W, and F160W.

We fixed the redshift to $z_{spec}$. We adopt a non-parametric SFH from \cite{Leja2019},  using eight time bins where SFR is fit with a constant value in each
bin. The first four bins are set to lookback times of 0–3,
3–10, 10–30, and 30–100 Myr, while the last four
bins are logarithmically spaced between 100 Myr and
$t_{max} = t_{universe} (z=z_{spec}) - t_{universe} (z=20)$. We allow the metallicity to freely vary up to 0.5~Z$_\odot$. The upper limit in metallicity is based on the stellar mass-metallicity relation observed at $z=3.5$ for a stellar mass of 10$^{11}$\Msun \,\citep{Llerena2022,Stanton2024}, which is in agreement with the range of stellar masses of the sample \citep{Llerena2024EELG}. For the dust component, we consider the \cite{Calzetti2000} attenuation curve and let A$_V$ vary between $0-2$ mag. We also include a nebular component in the model, and we let the ionization parameter $\log U$ freely vary between $-4$ and $0$.

Given that incorporating JWST/MIRI \citep{Wright2015} detections into SED modeling of galaxies may lower their stellar mass by 0.25~dex at $z=4-6$ and 0.38~dex at $z=6-9$ \citep{Papovich2023}, we checked the MIRI imaging provided by CEERS which includes eight pointings, four of which provide deep imaging with the bluer bands (F560W and F770W) and four of which provide contiguous wavelength coverage in F1000W, F1280W, F1500W, and F1800W, where two of these also include coverage in F770W and F2100W \citep{Yang2023}. Only six galaxies in our sample (at $z\sim4.5-8.7$) are in the MIRI footprint with F560W and F770W imaging, and they have detections in both bands. Although MIRI bands are not included in the SED fitting, we find good agreement, within the uncertainties, between the synthetic photometry from the SED model and the observed F560W and F770W photometry in this subsample, except for one galaxy (ID: 47521) whose observed photometry is larger by a factor of 2.5 compared to the model. This suggests that the SED modeling reliably reproduces the reddest part of the SED, not covered by NIRCam, across the sample of EELGs. 

\subsection{{Star-formation rates from Balmer lines}}
\begin{figure}[t!]
    \centering
    \includegraphics[width=\linewidth]{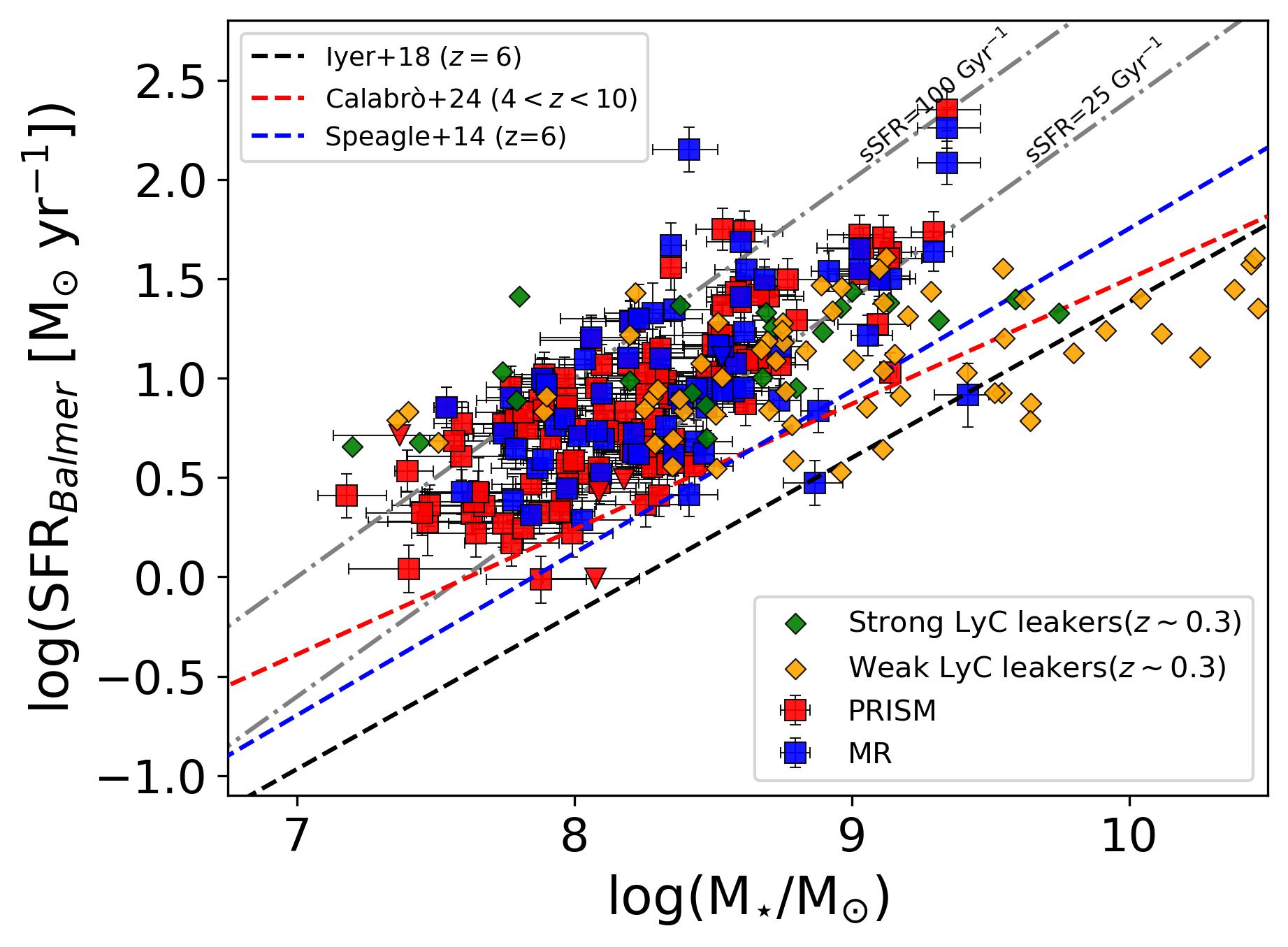}
    \caption{Distribution of the sample in the \Mstar-SFR plane. {The red and blue squares represent the EELGs with PRISM and MR spectra, respectively.} The triangle symbols are upper limits due to low \hbeta~S/N. The dashed black and blue lines are the main-sequence of SF galaxies at $z=6$ from \cite{Iyer2018} and \cite{Speagle_2014}, respectively, while the dashed red line represents the main-sequence at $4<z<10$ from \cite{Calabro2024sigmasfr}. The {green (orange) diamonds} are the sample of strong (weak) LyC leakers at $z\sim0.3$ \citep{Flury2022I,Jaskot2024a,Jaskot2024b}. The dashed-dotted lines represent constant sSFR lines.}
    \label{fig:MS}
\end{figure}
We corrected the line fluxes for dust reddening, adopting the \cite{Calzetti2000} attenuation law, which is consistent with the SED fitting. We considered the stellar E(B-V) value from the SED fitting. A comparison with the nebular E(B-V) obtained by Balmer lines is presented in Sec. \ref{sec:dust red}. In particular, we converted the dust-corrected \halpha~luminosity L(\halpha) to SFR assuming the calibration from \cite{Shapley2023} as
$\text{SFR(Balmer)}~\rm{[M_{\odot}~yr^{-1}]}=\rm{L(H\alpha)}~\rm{[erg~s^{-1}]}\times 10^{-41.37}$. Some systematics of these calibrations include the IMF and extinction \citep[e.g.,][]{Kennicutt2012}. This is the most suited calibration according to subsolar metallicities expected for our galaxies at $z > 4$. We rescaled these values by a factor of 0.94 \citep{Madau2014}, accounting for the IMF mismatch in the calibration \citep{Chabrier2003} and in the SED fitting \citep{Kroupa2001}. When \halpha~is not covered, we used \hbeta~luminosity assuming \halpha/\hbeta = 2.79 under case B approximation for T$_e$ = 15 000 K and n$_e$ = 100 cm$^{-3}$ \citep{EPM2017}. In Appendix \ref{sec:sfr10Myr}, we checked that the estimated SFRs trace the average SFR in the last 10~Myr. We use these Balmer-based SFRs to estimate the sSFRs which are reported in Table \ref{lt_ewo3_err}.

The distribution of SFR vs \Mstar~for the sample is shown in Fig. \ref{fig:MS}. As a reference, we show the main sequence of star-forming (SF) galaxies from \cite{Calabro2024sigmasfr} derived from a spectroscopic sample of galaxies in a wide range of redshifts $z=4-10$. Our sample is scattered above the main sequence \citep[e.g.,][]{Speagle_2014,Iyer2018,Cole2023}, and the galaxies cover $\sim2$~dex in stellar mass and SFR. The sample is composed of low-mass galaxies from log(\Mstar/\Msun)=7.18 up to log(\Mstar/\Msun)=9.42, with a median of log(\Mstar/\Msun)=8.26. This offset above the main-sequence is also observed in samples of low-$z$ EELGs \citep[e.g.,][]{Calabro2017}. For comparison, we also plot  galaxies at $z=0.2-0.4$ with LyC detections from the Low-redshift Lyman Continuum Survey \citep[LzLCS;][]{Flury2022I} and archival programs \citep{Jaskot2024a,Jaskot2024b}. Throughout the paper and in subsequent plots, this reference sample is divided into strong (\flyc\,$\geq{5}$\%) and weak (\flyc\,$<{5}$\%) {LyC leakers. We adopt the 5\% threshold to indicate cosmologically significant LyC leakers in order to remain consistent with recent literature \citep{Flury2022I,Flury2022,Jaskot2025review}}. From Fig. \ref{fig:MS}, we note that for a similar range in stellar mass, the galaxies in our sample show SFRs consistent with those of the LzLCS galaxy sample, even though the latter extends to galaxies with higher stellar masses{, which are mostly weak LyC leakers.}

\subsection{{UV $\beta$ slope and M$_{\rm UV}$ magnitude}}\label{sec:beta-muv}

The rest-frame UV continuum slope can be represented by a power-law $f_\lambda \propto \lambda^\beta$, as modeled by \cite{Calzetti1994}. For the calculation of $\beta$, we considered galaxies with PRISM spectra and the fitting windows in the original definition of \cite{Calzetti1994} that exclude contamination by UV lines. We considered the rest-frame range $1309~\AA < \lambda < 2580~\AA$. We excluded shorter wavelengths due to the impact of \lya~damping wing, resulting in redder $\beta$ \citep[see][]{Dottorini2024}. We fit the windows with a linear relation $\log f_\lambda = \beta \log \lambda + q$. The fit is performed using \textsc{emcee} \citep{Foreman-Mackey2013} with 50 walkers and 5000 steps. We adopt a flat prior on $\beta$ between -4 and 0. The best-fit value is taken as the posterior median, and the uncertainty is defined by the 16th–84th percentile interval. 

To account for possible residual slit losses that are not corrected by the assumption of a no wavelength-dependent correction factor, we estimated the $\beta$ slope using only the photometry and checked consistency with the value estimated directly from spectra. To this aim, we considered NIRCam filters with pivot rest-frame wavelength between 1309 and 3645\r{A} (Balmer break). Depending on the redshift, it implies fitting between 2 and 4 filters. In Fig. \ref{fig:delta_beta} we show the difference between the $\beta$ calculated from the spectrum and from the photometry. We find a median difference $\Delta\beta=|\,\beta_{spec}-\beta_{photo}|=0.36$ ($\sigma=0.32$). This significant discrepancy arises from the limited number of filters employed in the photometric estimation. In the following analysis regarding $\beta$ slope, we consider the estimation based on the spectra only in the galaxies where $\Delta\beta<0.68$ (dashed lines in Fig. \ref{fig:delta_beta}), which is 75\% of the sample, excluding galaxies where $\beta$ is more different than 1$\sigma$ the median $\Delta\beta$. For this subsample, we obtained {a median value of -2.00, with a 16-84th percentile interval between -2.33 and -1.66.} The mean uncertainty in the $\beta$ slopes is 0.25. The estimated spectroscopic $\beta$ slopes are reported in Table \ref{lt_ewo3_err}.

Additionally, we estimated the absolute UV magnitudes (M$_{\rm UV}$) of the galaxies from the best-fit of the UV continuum of the spectra. At each step of the fitting, we computed the flux density at 1500~\r{A} and we determined the median of the posterior distribution, while the uncertainty is given by the 16th-84th percentiles. {We obtained a median $M_{\rm UV}=-19.17$ in the sample with a 16-84th percentile interval between $-20.12$ to $-18.46$.}

\subsection{{How blue are EELGs?}}\label{sec:blue}

\begin{figure}[t!]
    \centering
    \includegraphics[width=\linewidth]{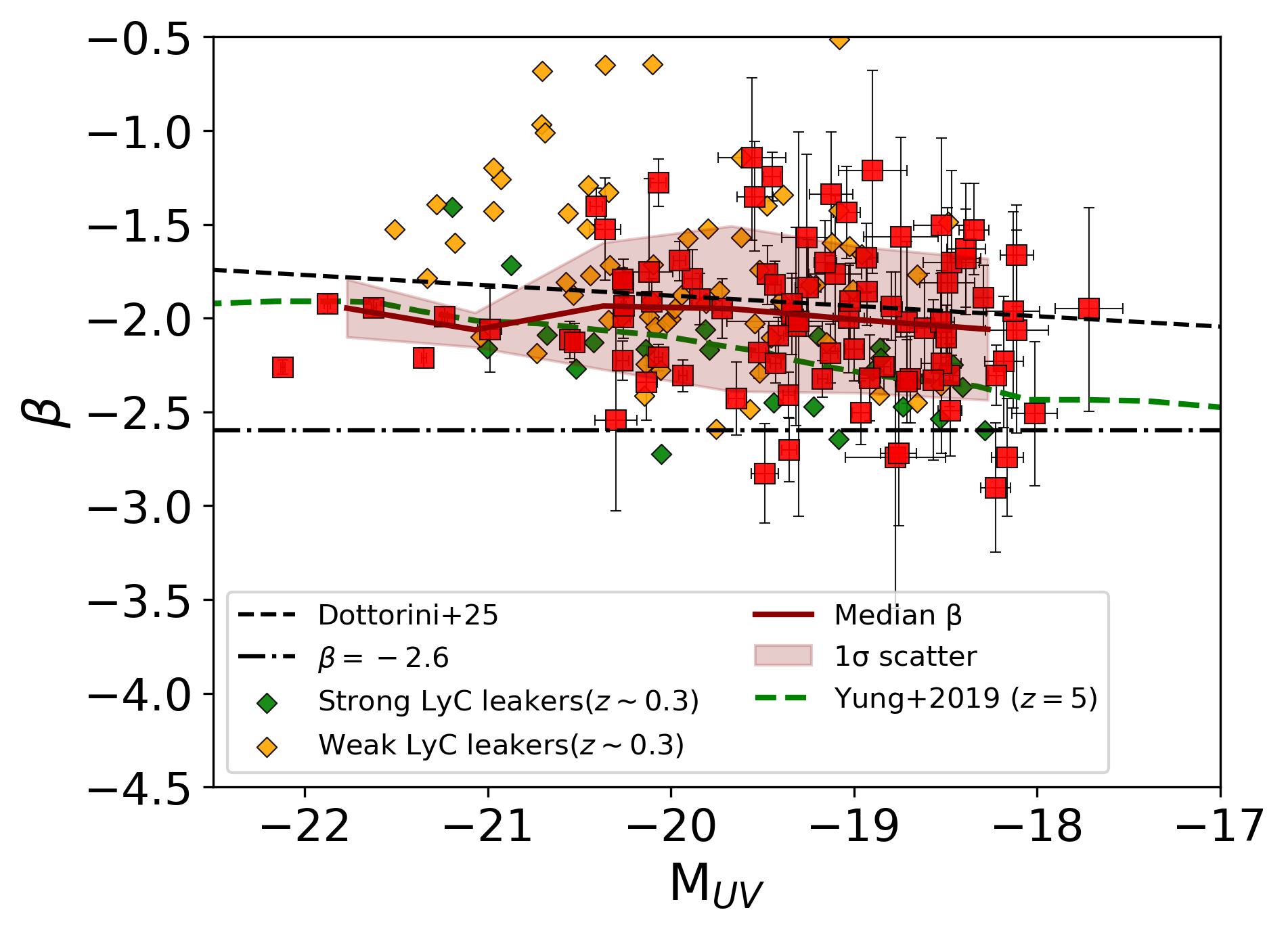}
    \caption{Relation between M$_{\rm UV}$ and $\beta$ {for the sample of EELGs with PRISM spectra (red squares).} The black dashed line is the relation from \cite{Dottorini2024}. The dashed green line represents the predictions from SAMs at $ z=5$ \citep{Yung2019}. The red line traces the median values of $\beta$ as a function of M$_{\rm UV}$, computed in bins of M$_{\rm UV}$. The shaded region corresponds to the 1$\sigma$ observed scatter in the sample. {Green and orange} symbols are as in Fig. \ref{fig:MS}.}
    \label{fig:muv_beta}
\end{figure}

In Fig. \ref{fig:muv_beta}, we show the relation between M$_{\rm UV}$ and $\beta$ slope for our sample. The median distribution of our sample (red line in Fig. \ref{fig:muv_beta}) tends to follow the decreasing relation reported in \cite{Dottorini2024}, which includes a spectroscopic sample of galaxies at similar redshifts. This decreasing trend is also predicted by semi-analytic
models (SAMs) at $z=4-10$ \citep{Yung2019}, in agreement with our results. For galaxies with M$_{\rm UV}\lesssim-21.06$, the galaxies tend to be bluer than the general population with values lower than the best-fit in \cite{Dottorini2024} for a given M$_{\rm UV}$. However, in this M$_{\rm UV}$ range, the comparison sample contains relatively few galaxies, limiting the statistical significance of this trend. For galaxies with M$_{\rm UV}\gtrsim-21.06$, we find a wide range of $\beta$ slopes, bluer and redder than the general population of galaxies at similar redshifts for a given M$_{\rm UV}$. Compared to the LzLCS sample, the values of $\beta$ are consistent in the range of M$_{\rm UV}$ that both samples cover, except for galaxies M$_{\rm UV}\lesssim-21.06$ where we do not find extremely red galaxies in our sample, and they are likely to be weak LyC leakers at $z\sim0.3$. {We note, however, that a subset of EELGs in our sample aligns more closely with the population of strong LyC leakers, which are typically bluer than their weaker counterparts.}

We find that most of the galaxies have $\beta>-2.6$, which indicates they are not super-blue and they are above the theoretical lower limit produced by dust-free pure
stellar and nebular continuum emission \citep[e.g.,][]{Cullen2024}. Only $7$\% of the sample exhibits $\beta<-2.6$; however, none of these measurements are below this value at a level greater than $1\sigma$.

\subsection{{Dust reddening}}\label{sec:dust red}

\begin{figure}[t!]
    \centering
    \includegraphics[width=\linewidth]{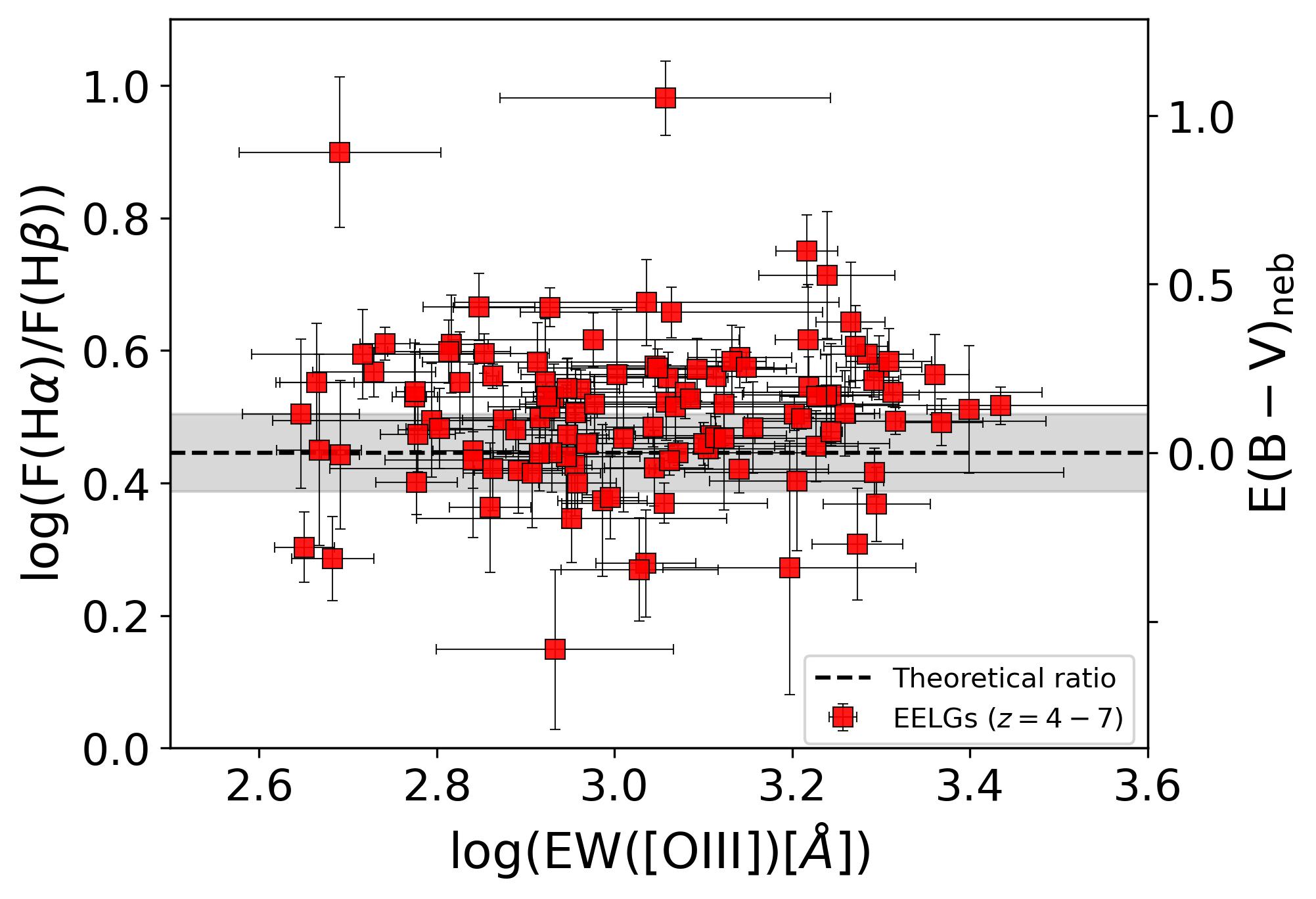}    \caption{Relation between \halpha/\hbeta~{ratio} and EW(\oiii) {for the sample of EELGs (red squares)}. In the right y-axis, we show the corresponding E$(\mathrm{B-V})_\mathrm{neb}$. The gray shaded region corresponds to the mean error $\sim0.06$ dex in the \halpha/\hbeta~ratio.}
    \label{fig:ebv-ew-mass}
\end{figure}
The results obtained from the SED fitting indicate a wide diversity in global dust attenuation among the galaxies in our sample. The derived stellar E(B–V) values span a broad range {with a median value of 0.09, but reaching values from} $0.009$ to 0.46, highlighting significant variations in the amount of dust present within these systems. As an alternative approach to estimating dust attenuation, we use the Balmer emission lines in the subset of galaxies where both \halpha~and \hbeta~are detected. By comparing the observed Balmer decrement (\halpha/\hbeta) to the theoretical ratio expected for case B recombination, we derive independent estimates of dust reddening for this subsample.

To estimate the nebular E$(\mathrm{B-V})_\mathrm{neb}$, we considered the theoretical ratio \halpha/\hbeta=2.79 under case B approximation for T$_e$ = 15 000 K and n$_e$ = 100 cm$^{-3}$ \citep{EPM2017}, and the \cite{Calzetti2000} extinction law. In Fig. \ref{fig:ebv-ew-mass} we show the Balmer ratios as a function of EW(\oiii). As can be seen, most of the galaxies in the sample show \halpha/\hbeta~ratios higher than the theoretical value, indicating they are affected by dust reddening. In these cases, we find a {median} \halpha/\hbeta={3.44} ($\sigma=0.96$), which implies a {median} E$(\mathrm{B-V})_\mathrm{neb}=${0.18} ($\sigma=0.17$). This value is slightly higher than the {median} stellar E$(\mathrm{B-V})=0.09$ ($\sigma=0.07$) obtained from the SED fitting for the same subsample, but the difference is not statistically significant at the 1$\sigma$ level. Previous studies have reported that the offset between nebular and stellar E(B–V) decreases with increasing sSFR \citep{Koyama2019}. Interestingly, there is a subsample of galaxies that shows \halpha/\hbeta~ratios lower than the theoretical value, taking into account the typical uncertainty of the \halpha/\hbeta~ratios in the sample. They represent $11$\% of the sample where \halpha~and \hbeta~were detected. These cases can be explained by a scenario in which case B of recombination is not satisfied, and ionizing photons can escape into the IGM. Nevertheless, wavelength-dependent slit losses cannot be excluded for these galaxies.

Overall, we do not find a clear trend of dust attenuation with EW(\oiii), and for a given EW(\oiii), the \halpha/\hbeta~ratios and the equivalent E$(\mathrm{B-V})_\mathrm{neb}$ are diverse. We will discuss this in Sec. \ref{sec:ssfr-size} in the context of the AFM model.

\subsection{{Gas-phase metallicity}}

We estimate the gas-phase metallicity using the R23=($\rm{[OIII]}\lambda\lambda$4959,5007+\oii)/\hbeta~calibration presented in \cite{Sanders2024} for galaxies at $z = 2.1-8.7$. Given that this relation is bivalued, we also used the O32 calibration in \cite{Sanders2024}. We consider the metallicity estimation to be the value from the R23 calibration that is closer to the value from the O32 calibration. We found a {median} log(O/H)+12={7.74}, equivalent to {0.11}~Z$_\odot$, but the values range from 0.03 to 0.8~Z$_\odot$. The same results are obtained in this approach if we use the Ne3O2=log(\neiii/\oii) calibration instead of O32.

In the sample of galaxies with MR spectra, we find 8 galaxies where [O III]$\lambda$4363\r{A} is detected with S/N~>~3. The \oiii/$\rm{[O~III]}\lambda$4363\r{A} ratios \citep{EPM2017} indicate T$_e>1.46\times10^4$K in this subsample. For one galaxy (ID: 81061) we also detect \oii~and \hbeta, and can estimate the gas-phase metallicity with the direct method following \cite{EPM2017}. We found a gas-phase metallicity $0.11-0.19$~Z$_{\odot}$, in agreement with the {median} value obtained with the calibrations.

\section{Are EELGs primarily AGN or SF galaxies?}\label{sec:nature}
Understanding whether the ionizing radiation in galaxies arises from massive stars or AGN is fundamental for interpreting the physical conditions within EELGs. This distinction is particularly relevant in the context of the early universe, where EELGs are considered potential analogs of the galaxies that contributed to cosmic reionization. In this section, we investigate the nature of the ionizing sources in our sample to determine whether stellar populations alone can account for the observed emission or if additional sources, such as AGN, are required.

\subsection{Emission line diagnostic diagrams}\label{sec:ohno-mex}

\subsubsection{{OHNO diagram}}

{The OHNO diagram \citep{Backhaus2022} has been proposed as an alternative to the classical BPT diagram \citep{BPT1981} for identifying narrow-line AGN (NLAGN) in high-redshift sources. Owing to the wavelength coverage of our spectra, this diagnostic can be applied to the EELG sample.} For the subsample of galaxies where \neiii~is detected or with upper limits, we analyze the OHNO diagram. We detect \neiii~and \oii~in only 39 galaxies across the entire sample. For 9 additional galaxies, the detection of \oii~allows us to place an upper limit on the \neiii/\oii~ratio. We used this subsample, representing 25\% of the total sample, for the OHNO analysis. For an additional 45 galaxies, both \neiii~and \oii~fall within the spectral coverage, yet only upper limits are obtained for both lines. We analyze the stacked PRISM spectra of these galaxies following the methodology described in Section~\ref{sec:stack} to construct the composite spectrum. 

\begin{figure}[t!]
    \centering
    \includegraphics[width=\linewidth]{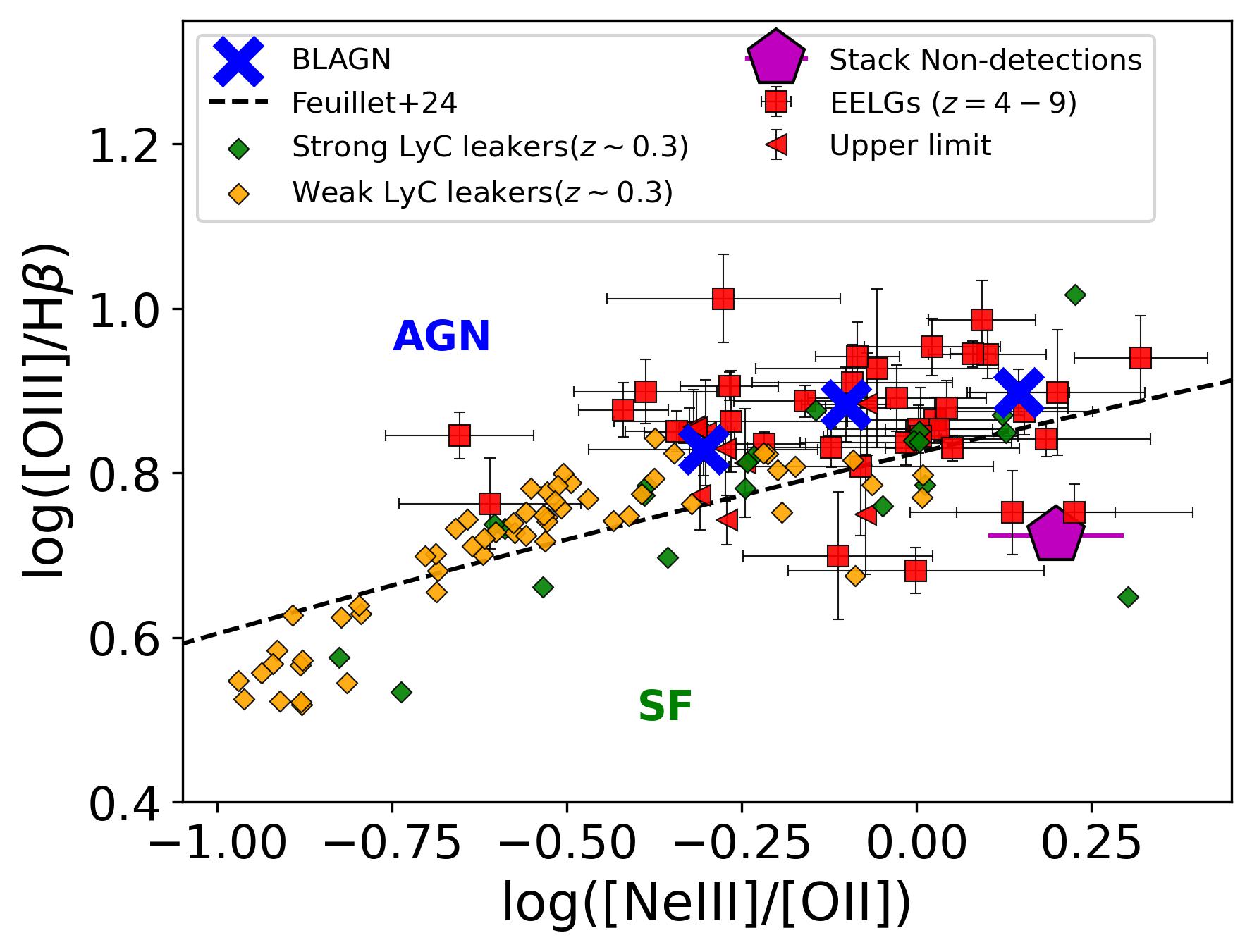}
    \caption{OHNO diagram. {The red symbols represent the sample of EELGs.} The black dashed line indicates the demarcation between SF galaxies and AGN (see labels), according to  \cite{Feuillet2024}. The {magenta} pentagon is the stacked spectrum of EELGs with undetected \neiii~and~\oii. The blue X-symbols are EELGs classified as BLAGN. {Green and orange} symbols are as in Fig. \ref{fig:MS}.}
    \label{fig:ohno}
\end{figure}

In Fig. \ref{fig:ohno}, we show the distribution of our sample in the OHNO diagram. Based on the demarcation line from \cite{Feuillet2024} calibrated for galaxies $z \leq 1.06$, only 2 galaxies are in the SF region at 3$\sigma$ level, which represents 2.2\% of the sample where \neiii~and \oii~fall within the spectral coverage. 13 (18) galaxies are in the AGN region at the 3$\sigma$ (2$\sigma$) level, which represents 14\% (19.4\%) of the same sample. 33 (28) galaxies overlap with the demarcation at the 3$\sigma$ (2$\sigma$) level, suggesting that the dominant ionization source cannot be clearly identified, which represents 35.5\% (30.1\%) of the same sample. However, we note that this region of the diagram close to the demarcation line is also populated by the sample of low-$z$ LyC leakers with elevated \oiii/\hbeta~ratios. In addition, photoionization model predictions for SF and AGN often overlap in this region \citep{Calabro2023,Calabro2024}, with the ionization parameter emerging as the main driver independently of the ionizing source \citep{Cleri2025}. 

We also note that the mean log(\oiii/\hbeta)$=0.85$ for the subsample included in the OHNO diagram is higher than the mean log(\oiii/\hbeta)$=0.72$ measured for the remaining 48.3\% of the sample where neither \neiii~nor \oii~is detected with S/N$>3$. The stacked spectrum of this latter subsample falls within the SF region {(see magenta symbol in Fig. \ref{fig:ohno})}. This suggests that the galaxies in the OHNO diagram ({red squares}) represent the most extreme in terms of \oiii/\hbeta~ratios {compared to the rest of the sample, which could alternatively be explained by harder stellar ionizing spectra and differences in gas-phase metallicity \citep[e.g.,][]{Kewley2002,Bian2018}, without requiring a dominant AGN contribution.}

{Additionally, revised versions of the OHNO diagram have demonstrated that it struggles to clearly separate AGN and SF models \citep[e.g.,][]{Arevalo-Gonzalez2025}. Recent studies have shown that the \neiii/\oii~ratio is a useful probe of the ionization state, but it does not effectively discriminate between excitation mechanisms, owing to the similar shapes of the ionizing SEDs at the relevant ionization potentials \citep[e.g.,][]{Zhu2023,Flury2025diag}.
With these caveats in mind, the estimated AGN fraction of $\sim$14\% in our sample of EELGs should therefore be considered as a limit, as it can neither be firmly confirmed nor completely ruled out. Independent diagnostics are required to robustly assess the presence of AGN.}

\subsubsection{{Mass-Excitation diagram}}

\begin{figure}[t!]
    \centering
    \includegraphics[width=\linewidth]{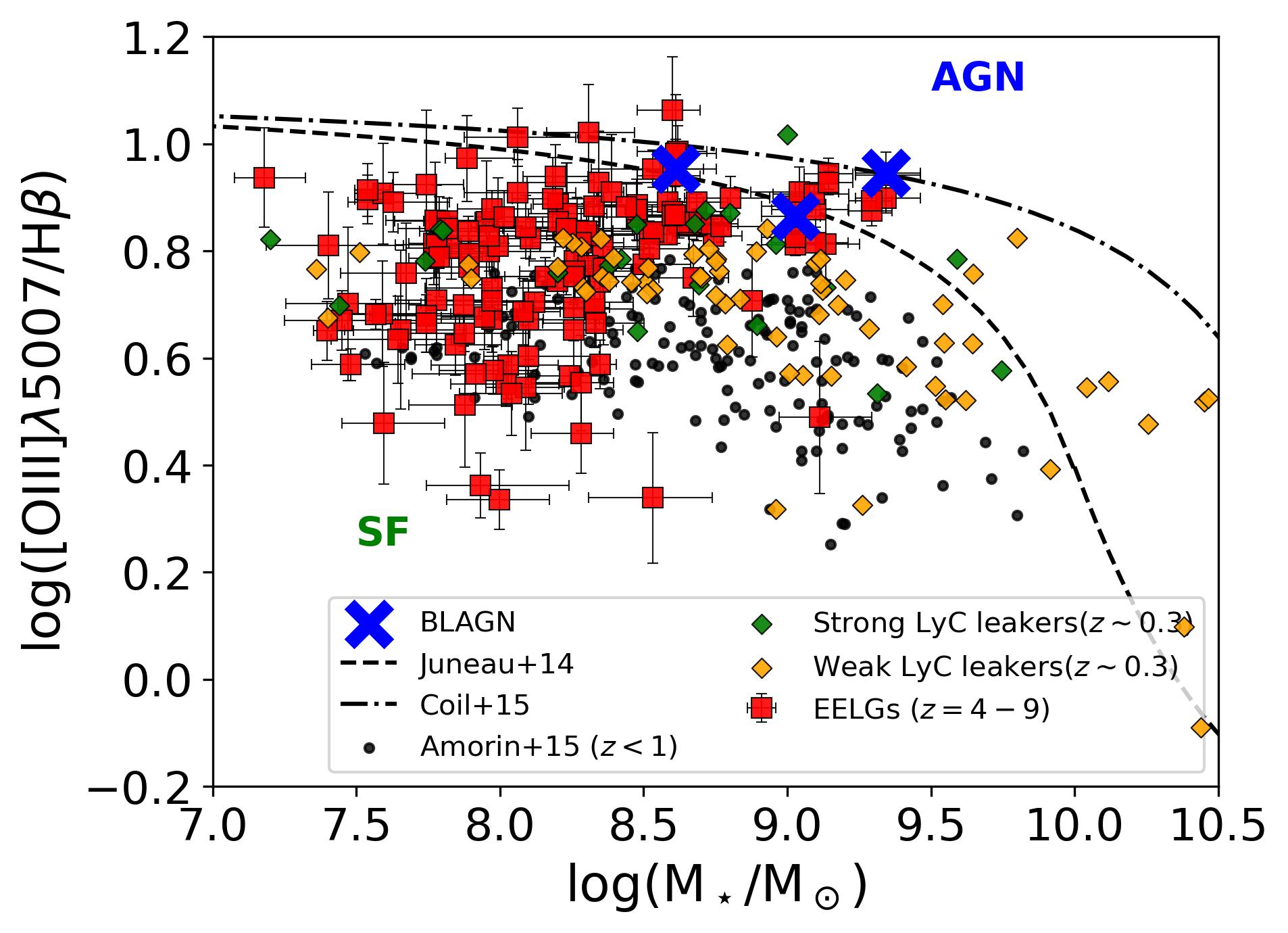}
    \caption{MEx diagram. {The red symbols represent the sample of EELGs.} The black dashed line indicates the demarcation between SF galaxies and AGN (see labels), according to \cite{Juneau2014}. The black dashed-dotted line is the demarcation at $z\sim 2$ \citep{Coil2015}. The blue X-symbols are EELGs classified as BLAGN. The black circles are EELGs up to $z\sim1$ \citep{Amorin2015}. {Green and orange} symbols are as in Fig. \ref{fig:MS}.}
    \label{fig:mex}
\end{figure}

To further investigate this conclusion, we employ an alternative diagnostic. We use the Mass-Excitation \citep[MEx, ][]{Juneau2011} diagram to study the ionization source in the sample of confirmed EELGs. Although the MEx diagnostic is generally less favored compared to other line-ratio methods—since it has not been thoroughly tested at high-$z$ and is affected by systematics in stellar mass estimates \citep[e.g.,][]{Coil2015,Cleri2023clear}—we retain it in our analysis because it can be applied to a large fraction of galaxies in our sample. 

In Fig. \ref{fig:mex}, we present the location of our EELG sample on the MEx diagram, where they predominantly occupy the SF region, largely avoiding the AGN regime. We find that their position is consistent with the location of metal-poor starbursts with high log(\oiii/\hbeta)$\gtrsim 0.33-1.06$ (mean value of 0.78). This region is populated by EELGs at $z\sim 1$ \citep{Amorin2015} that occupy the same locus with slightly lower \oiii/\hbeta~ratios for galaxies with \Mstar$\gtrsim10^{9}$\Msun. Similarly, our sample shows similar \oiii/\hbeta~ratios as the ones of LyC leakers at low-$z$ at the same stellar mass range. However, we note that our sample typically spans lower stellar masses compared to the reference samples. Based on the MEx diagram, we conclude that the galaxies in our sample are predominantly powered by massive stars rather than by AGN. {This conclusion is broadly consistent with the OHNO diagram, where the stacked spectrum, representing approximately half of the sample with covered but undetected \neiii~and \oii~lines, is consistent with ionization driven by star formation.} Additionally, no galaxies in the sample show an X-ray counterpart within 2arcsec in the 800-ks Chandra imaging in EGS \citep{Nandra2015}. We note that, within the redshift range of our sample, the luminosity limit corresponds to L$_{\rm 2-10,keV}\gtrsim10^{43}$ erg s$^{-1}$, implying that fainter AGN may still be present but remain undetected. 

Overall, the excitation properties of low-mass EELGs are predominantly consistent with ionization driven by young starbursts. {However, in approximately 14\% of the galaxies a NLAGN may also contribute to the ionization. This fraction should be regarded as a limit, as it cannot be robustly confirmed by our analysis and may alternatively be explained by harder stellar ionizing spectra producing similarly high line ratios.}

\subsection{Identification of broad-line AGN and stellar outflows}\label{sec:blagn}

We compared our sample with the literature-based classifications of broad-line AGN (BLAGN) in the same field \citep{Harikane2023,Roberts-Borsani2024,Brooks2024,Taylor2025,Hviding2025,Kocevski2025}. We found that only five galaxies (IDs: 25074, 71325, 81061, 70867, 3391) in our sample are classified as BLAGN. To further evaluate whether there are BLAGN in the sample, we modeled the \halpha~and \oiiihb~lines assuming two Gaussians using LiMe. For the fit, we assume that the second (broad) component has a larger velocity width than the first (narrow) component, while allowing both components to have independent systemic velocities. Given the spectral resolution of the instrument, \halpha~is unresolved for FWHM\,$<940-2600$\,km s$^{-1}$ (PRISM) and $<240-371$\,km s$^{-1}$ (MR), depending on redshift. For \oiii, the corresponding FWHM ranges are $<1045-4360$\,km s$^{-1}$ (PRISM) and $<236-410$\,km s$^{-1}$ (MR). We assumed these limits for the broad component. To evaluate the necessity of the broad component, we use the Bayesian Information Criterion (BIC), which incorporates both the goodness-of-fit ($\chi^2$) and the number of free parameters. To accept the broad emission line in the model, we require a $\Delta$BIC > 6 \citep{Fabozzi2014} and with S/N\,$>3$. 

In the subsample with PRISM spectra, we found that in six cases the broad component in \halpha~is accepted, but after a visual inspection, we found that only in two cases (IDs: 25074, 3391) is there evidence of a BLAGN in which both components have similar peak velocities ($\Delta\sim100-360$km s$^{-1}$). Regarding the MR spectra, we found that in nine cases \halpha~is better modeled with a broad component, and similarly, after a visual inspection, in only four cases (IDs: 71325, 75857, 15050, 25074) is there evidence of BLAGN ($\Delta\sim4-70$\,km s$^{-1}$).
Importantly, our methodology enables us to recover the galaxies previously classified as BLAGN. We note that source 81061 is not recovered, as it is classified by \citet{Harikane2023} as a faint AGN with S/N\,$<3$ in the broad component. Source 70867 is also not recovered because \hbeta~is not covered within the spectral range analyzed in this paper. 

The EELGs identified as BLAGN in our sample are located in the AGN region of the OHNO diagram ({blue X-symbols in }Fig.~\ref{fig:ohno}) and in the high-mass regime of the MEx diagram ({blue X-symbols in }Fig.~\ref{fig:mex}), consistent with an additional ionization contribution from an AGN in these galaxies. {We note that only three of these sources are shown in the diagnostic diagrams, as the remaining EELGs classified as BLAGN lack \hbeta~coverage in the spectra used for their classification.}

Moreover, in four galaxies with accepted broad component (two observed with PRISM (ID: 45212, 47379) and two with MR (IDs: 75511, 28312)) the broad component is velocity–shifted ($\Delta>360$\,km s$^{-1}$ for PRISM and $\Delta>90$\,km s$^{-1}$ for MR) relative to the narrow component, which may indicate an outflow rather than a BLAGN. In all the above cases, \oiii~modeling reveals the absence of a broad component. 

We identify a subsample of five galaxies (IDs: 66855, 75478, 78647, 47504, 59920) with MR spectra where \halpha~is not covered (partially in ID: 75478) and where a broad \oiii~component is accepted, which may indicate the presence of outflows. However, no unambiguous outflow signatures are detected in the EELG sample. Outflows may not be ongoing at the time of observation but could have previously cleared dust, as suggested by the AFM. Alternatively, the lack of detection of outflows could be due to the limited spectral resolution of the PRISM and MR data; confirming their presence would require follow-up with higher-resolution spectroscopy and improved sensitivity.

Based on this Gaussian modeling, the lines are well described by a single Gaussian, indicating that while BLAGN may be present, they do not dominate the Balmer line emission. Given the small fraction of BLAGN in our sample, we do not remove them from the analysis. Their minimal contribution ($<4$\%) is unlikely to significantly impact the statistical properties or overall trends of the EELG population. We conclude that massive stars are the dominant source of ionization in EELGs without significant fractions of BLAGN, although about 14\% of the galaxies are still consistent with a {NLAGN} contribution based on diagnostic diagrams.

\section{The contribution of EELGs to cosmic reionization}\label{sec:xi_fesc}

Several studies indicate that SF galaxies may be the primary drivers of cosmic reionization \citep[e.g.,][]{Finkelstein2019, Yung2020b, Robertson2022}. However, the efficiency with which these galaxies produced and leaked LyC photons into the IGM is still poorly constrained. Due to their low metallicities, compact morphologies, and intense radiation fields, EELGs emerged as promising key players in cosmic reionization \citep[e.g.,][]{Begley2025}, given that they could have large \xiion~and potentially elevated \flyc, both of which are crucial parameters in determining their contribution to reionization. In this section, we study these two parameters in EELGs to understand their role during the EoR.

\subsection{Ionizing photon production efficiency }\label{xiion}

\begin{figure}[t!]
    \centering
    \includegraphics[width=\linewidth]{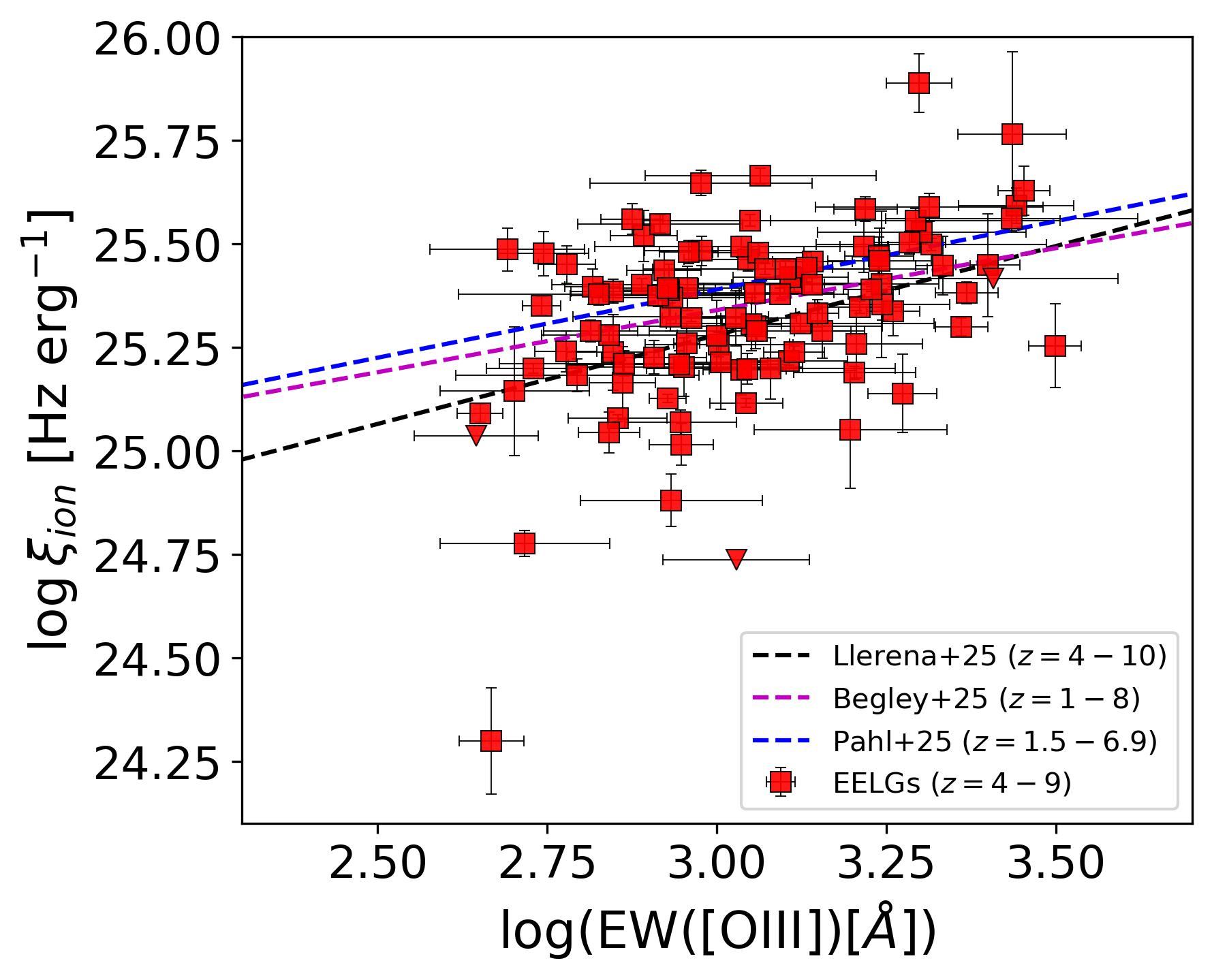}
    \caption{Relation between EW(\oiii) and \xiion. The black, magenta, and blue dashed lines are the relations from \cite{Llerena2024xi}, \cite{Begley2025xi}, and \cite{Pahl2025xi}, respectively.}
    \label{fig:ewo3_xi}
\end{figure}

Ionizing photon production efficiency is a key parameter that quantifies the number of hydrogen-ionizing photons produced per unit of UV luminosity. SAMs predict that, due to a strong metallicity dependency, UV-faint, low-mass galaxies have \xiion~values that are up to a factor of 2 higher than those of more massive galaxies, with the efficiency further enhanced at higher sSFRs \citep{Yung2020a}. We estimated \xiion~following \cite{Llerena2024xi}, with \xiion$=\dfrac{\rm{N(H^0)}}{\rm{L}_{UV}}$, where $\rm{N(H^0)}$ is the ionizing photon rate in units s$^{-1}$ and L$_{UV}$ is the UV luminosity density in units of erg s$^{-1}$ Hz$^{-1}$ at rest-frame 1500\r{A}. We estimate $\rm{N(H^0)}$ using the dust-corrected \halpha~luminosity as L$(\rm{H\alpha})[\rm{erg\, s^{-1}}]=1.36\times 10^{-12} N(H^0) [\rm{s^{-1}}]$, which was derived from \cite{Leitherer1995} assuming no ionizing photons escape from the galaxy and case B recombination. For \flyc\,$>0$, \xiion~scales by a factor 1/(1-\flyc).

In Fig. \ref{fig:ewo3_xi}, we show the relation between EW(\oiii) and 
\xiion. This relation arises because large EWs are typically linked to young, metal-poor stellar populations, which are expected to produce copious amounts of hydrogen-ionizing photons \citep[e.g.,][]{Chevallard2018}. We found a {median} log(\xiion [Hz erg$^{-1}$])~=~{25.37, with a 16-84th percentile interval between 25.19 and 25.50 (Table \ref{tab:correlations}). The median \xiion} for the sample is higher than the {median} log(\xiion [Hz erg$^{-1}$])~=~25.22 found for the general population of star-forming galaxies at similar redshift \citep{Llerena2024xi} and also predicted from SAMs at $z=6-8$ in the same stellar mass range of the sample with \cite{Bruzual_2003} models \citep{Yung2020a}. Our sample follows the relations found in a wide range of redshifts \cite[e.g.,][]{Llerena2024xi,Begley2025xi,Pahl2025xi} with an increasing \xiion~with increasing EW(\oiii). We note that even in the most extreme EW(\oiii), the \xiion~values do not reach the theoretical limits of log(\xiion [Hz erg$^{-1}$])~=~26 expected for young metal-poor stellar populations \citep{Raiter2010,Maseda2020} and from simulations of PopIII populations \citep{Lecroq2025}. Therefore, EELGs are efficient producers of ionizing photons, although they are not the most efficient ones, as they do not reach the maximum theoretical values of \xiion.

\subsection{The escape of LyC photons}\label{sec:escapelyc}

Given that at the redshift of our EELGs the \flyc\ cannot be measured directly due to the IGM opacity, we {inferred} it using models based on multivariate indirect diagnostics \citep{Jaskot2024a,Jaskot2024b}. {Results of the LzLCS survey indicate that 
\flyc~correlates with several galaxy properties, including dust attenuation, 
\flya, and the O32 ratio \citep[e.g.,][]{Flury2022,Chisholm2022,Saldana-Lopez2022lzlcs,Xu2023}. However, these correlations exhibit significant scatter, suggesting that no single observable can reliably predict 
\flyc~on its own and that a more accurate description requires a combination of multiple galaxy properties.}

{To address these limitations, \citet{Jaskot2024a,Jaskot2024b} applied survival analysis techniques, in particular Cox models. These semi-parametric models estimate the probability of detecting 
\flyc~as a function of multiple galaxy properties, providing a more robust framework for predicting \flyc~compared to traditional methods like linear regression. The performance of the Cox models is quantified using the concordance index (C), a goodness-of-fit metric that ranges from 0 (perfect disagreement) to 1 (perfect concordance). These Cox models were subsequently recalibrated by \citet{Mascia2025} using a subset of LzLCS galaxies \citep{Flury2022I} selected to resemble the physical properties of high-redshift sources, enabling more reliable predictions of 
\flyc~during the EoR. In this work, we adopt the revised ELG–O32 and revised ELG–EW Cox models from \citet{Mascia2025}, which achieve concordance indices of C=0.83 and C=0.79, respectively.}

{First,} we consider the revised ELG-O32 Model which estimates \flyc~ from  M$_{\rm UV}$, stellar mass, E(B-V), and O32 ratio. In the cases where O32 is not determined in our sources due to low S/N of \oii~or the wavelength coverage, we consider the revised ELG-EW Model, which depends on the same indirect parameters but uses EW(\oiiihb) instead of O32. In our sample, O32 values range from 0.44 to 1.35, with an average of 0.89, exceeding the O32 = 0.69 threshold commonly observed in strong leakers at low-$z$ \citep{Flury2022}. Notably, among the 46 galaxies with detected \oii~emission, 34 ($74\%$) exhibit O32 ratios above 0.69. We employed a Monte Carlo approach, perturbing the model parameters according to their uncertainties to generate 1000 random realizations. From the resulting distributions of \flyc, we report the median as the best estimate (reported in Table \ref{lt_ewo3_err}) and the 16th–84th percentiles as their uncertainties.

\begin{figure}[t!]
    \centering
    \includegraphics[width=\linewidth]{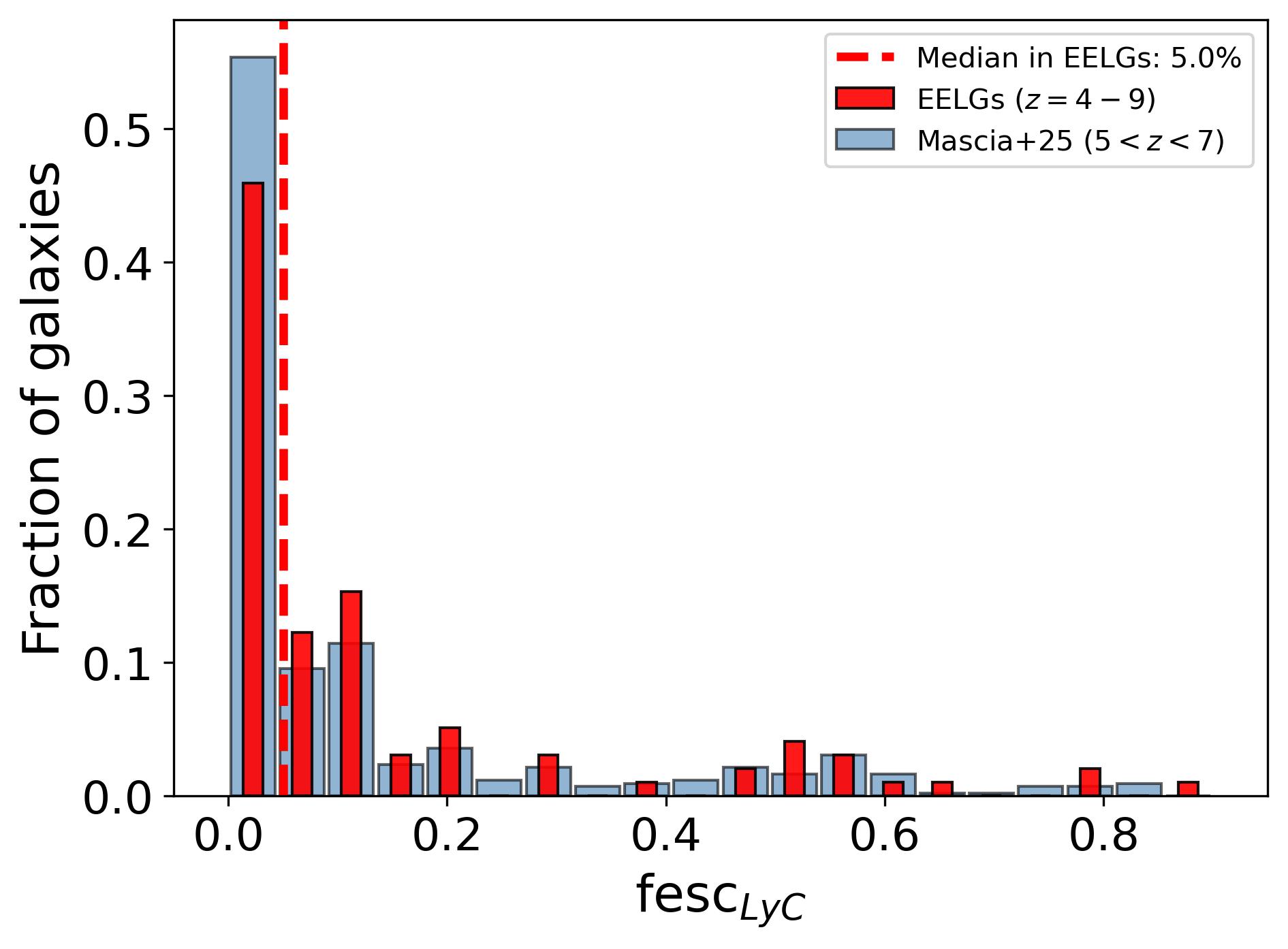}
    \caption{Distribution of \flyc~{in the sample of EELGs.} The red dashed line is the median of the distribution. The blue histogram is the \flyc~distribution in galaxies at $5<z<7$ from \cite{Mascia2025}.}
    \label{fig:dist_fesc}
\end{figure}

In Fig. \ref{fig:dist_fesc}, we show the distribution of \flyc~in the {EELG} sample. We find a median (mean) value of \flyc\,$={5}\%$ (${14.5}$\%){ with a 16-84th percentile interval from 0.8 to 29\% (Table \ref{tab:correlations})}. In {51}\% of the sample, we find \flyc\,$>5\%$ {which indicates that half of EELGs are strong LyC leakers but they are not a dominant population since half of the sample are weak LyC leakers}. The median value is consistent with the \flyc~predicted by the AFM at $z\sim5-6$ \citep{Ferrara2025} and {slightly higher than the} median value of $4$\% from inferred \flyc~ in a sample of 436 galaxies at $z=5-7$ with a wide range of EW(\oiiihb) and using the same methodology \citep{Mascia2025}. This suggests that, although {half} EELGs exhibit elevated \flyc, {half} are unlikely to be strong LyC leakers. Moreover, the \flyc\ distribution in EELGs closely follows that of the general galaxy population at similar redshifts \citep{Mascia2025}, not necessarily selected as EELGs, indicating that EELGs are not particularly more efficient LyC {leakers} than other galaxies at comparable epochs.
\subsection{The escape of \lya~photons}\label{sec:escapelya}

\begin{figure}[t!]
    \centering
    \includegraphics[width=\linewidth]{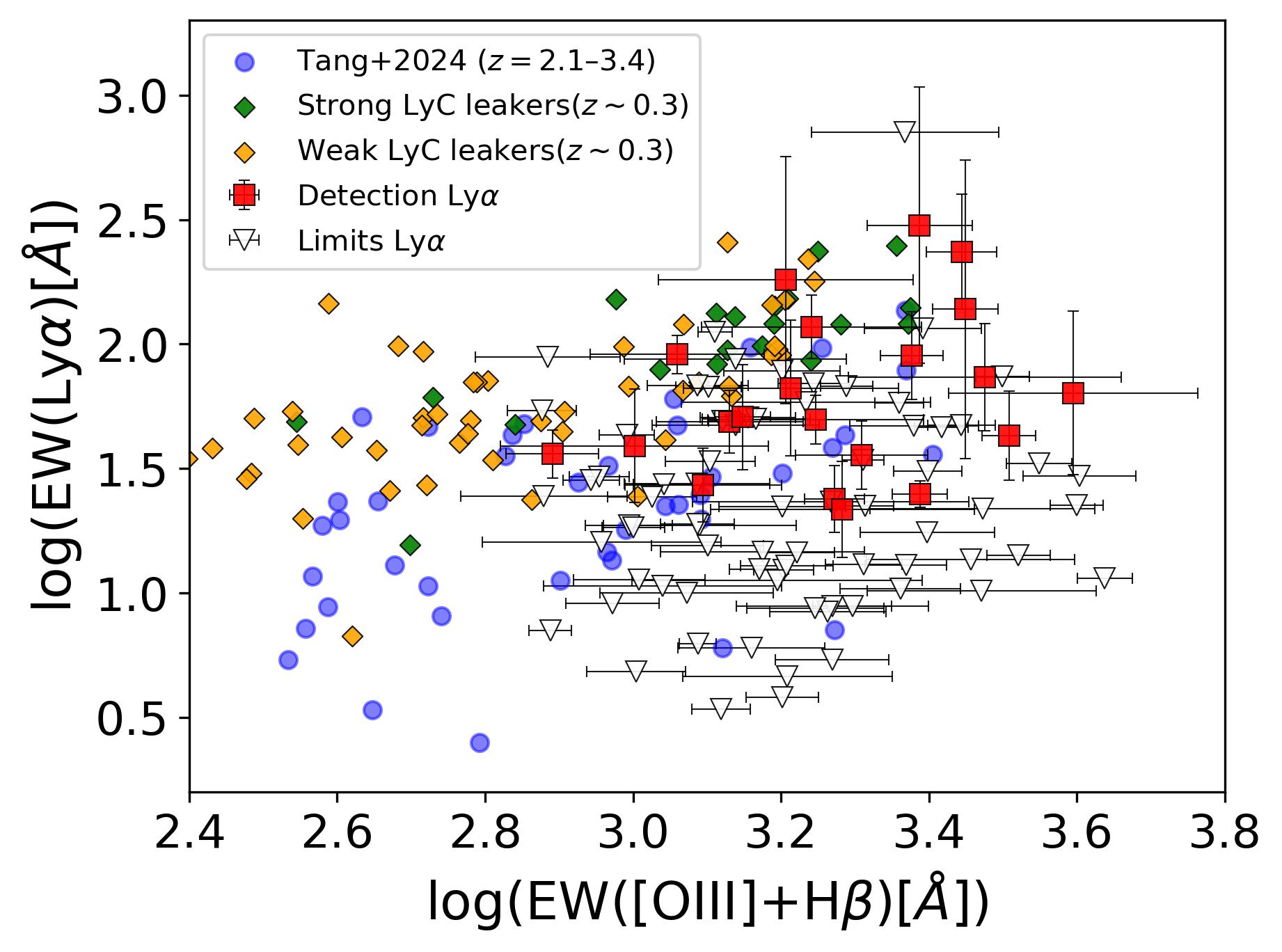}
    \caption{Relation between the EW([OIII]+\hbeta) and EW(Ly$\alpha$). The red {squares} are the sample of EELGs with detected \lya, while the {white} triangles are upper limits. The blue circles are the sample from \cite{Tang2024}. {Green and orange} symbols are as in Fig. \ref{fig:MS}.}
    \label{fig:ewo3_ewlya}
\end{figure}

\lya~emission is considered one of the key proxy for LyC escape. At low and intermediate-$z$, several \lya~properties (like its EW or the  separation between blue and red peaks) correlate with LyC leakage, albeit with significant scatter \citep{Marchi2017,Izotov2020,Flury2022}. {In this section, we investigate the \lya~properties in the sample of EELGs to assess whether they are consistent with the modest inferred \flyc~for these galaxies. First, we focus on the EW(\lya).} In Fig. \ref{fig:ewo3_ewlya} we show the relation between EW(\oiiihb) and EW(\lya). For this figure, we considered the subsample of LAEs ({red squares}) and the non-LAEs subsample with upper limits ({white} triangles). In the subsample of LAEs, EW(\lya) ranges from 21 to 300~\r{A}, with a median value of 50~\r{A}. A subsample of EELGs show high EW(\lya)\,$\gtrsim100$\r{A} (24\% of the subsample of LAEs).

We compared our results with previous works at $z=2.1-3.4$ \citep[][]{Tang2024}, and we find that our sample of LAEs follows the trend of increasing EW(\lya) with increasing EW(\oiiihb), showing a moderate correlation ($\rho=0.62$, $p=7\times10^{-8}$) when including the sample at intermediate-$z$. Our sample covers the region of more extreme EW(\oiiihb) and consequently higher EW(\lya) compared to LAEs at intermediate-$z$ \citep{Tang2024}. This increasing trend is also observed in low-$z$ LyC leakers. However, we note that \lya~emission is not detected in all EELGs of the sample, with only 18\% classified as LAEs. {This indicates that a high EW(\oiiihb) does not necessarily imply a high EW(\lya), which is consistent with a diversity in the properties of EELGs that facilitate the escape of \lya~photons. This supports a mixed population of weak and strong LyC leakers, since the escape fraction of \lya~ photons (\flya) has been shown to correlate with \flyc~as both processes are facilitated by low neutral hydrogen column densities and porous ISM geometries \citep[e.g.,][]{Verhamme2015,Begley2024vandels}.}

To quantify {\flya~in our sample}, we use the flux ratio between \lya~and dust-corrected \halpha. To this aim, we considered the theoretical ratio Ly$\alpha$/\halpha=8.7 \citep{Hayes2015}. We restrict our analysis to galaxies with 
$z<7$, where the EGS field has recently been reported to be largely ionized \citep[e.g.,][]{Napolitano2025capers}, to minimize the impact of neutral IGM on \lya~emission. In our LAE sample, \flya~spans 2\% to 55\%, with a median of 14.7\% {with a 16-84th percentile interval from 6\% to 30\% (Table \ref{tab:correlations})}, consistent with the median \flya~of 18\% in the low-$z$ LyC leakers. Considering also the upper limits {in the non-LAE subsample}, which in most cases are below {5}\%, we conclude that the average escape of \lya~in EELG is modest. {This is consistent with results from samples of LAEs at $z\approx6$, where it has been found that, although these systems can reach high \xiion~values, their \flya~is typically $\lesssim10\%$, with bluer and UV-fainter LAEs exhibiting higher \flya \citep{Ning2023}}.

Overall, although some EELGs appear to exhibit high \flya~and \flyc, not all are likely to be strong leakers. Only {16}\% of the sample {at $z<7$} shows \flya~and \flyc\,>{5}\%. This is supported by the absence of \lya~emission in several cases and the fact that the median inferred \flyc~is not particularly elevated. 
Coupled with the fact that the EELGs show moderately elevated \xiion~values, {this highlights the need for a proper estimation of their fractional contribution to the cosmic reionization.}

\subsection{{Fractional contribution to the ionizing budget}}

{In this section, we assess the median fractional contribution of the population represented by our sample to the total ionizing emissivity ($\dot{n}^{\rm SF}_{\rm ion}$) of SF galaxies, defined as the comoving density of LyC photons produced per unit time that are available to ionize hydrogen in the IGM and expressed as
$
\dot{n}^{\rm SF}_{\rm ion} = \rho^{\rm SF}_{\rm UV}~\xi^{\rm SF}_{\rm ion}~{\rm fesc}^{\rm SF}_{\rm LyC}$,
where $\rho^{\rm SF}_{\rm UV}$ is the comoving total UV luminosity density of SF galaxies. On the other hand, the total ionizing emissivity attributable to EELGs can be expressed as
$
\dot{n}^{\rm EELG}_{\rm ion} = \rho^{\rm EELG}_{\rm UV}~\xi^{\rm EELG}_{\rm ion}~{\rm fesc}^{\rm EELG}_{\rm LyC}.
$}

{A robust determination of the fractional contribution $\dot{n}^{\rm EELG}_{\rm ion}/\dot{n}^{\rm SF}_{\rm ion}$ requires a full characterization of $\rho^{\rm EELG}_{\rm UV}$, which is beyond the scope of this work. Instead, we adopt a set of simplifying assumptions to estimate a representative median fractional contribution. Recent studies have found that galaxies with EW(\oiiihb)~$>1500$~\r{A} represent a fraction of $\sim$28\% of the galaxy population at $z\sim7-8$ \citep{Endsley2022}. 
However, \citet{Endsley2023} showed that bright ($-22<\rm{M}_{\rm UV}<-19.5$), intermediate ($-19.5<\rm{M}_{\rm UV}<-18$) and faint ($-18<\rm{M}_{\rm UV}<-16.5$) EELGs at $z\sim6-9$ exhibit distinct EELG fractions ($f_{\rm EELG}$)  decreasing systematically toward fainter UV magnitudes. We estimate these fractions in our photometric parent sample and find a similar trend, with galaxies with EW(\oiiihb)$>680$~\r{A} representing $f_{\rm EELG}=42\%, 27\%, 25\%$ of bright, intermediate, and faint UV galaxies at $z\sim6$, respectively. Very-faint ($-16.5<\rm{M}_{\rm UV}<-13$) galaxies are also crucial for determining $\dot{n}^{\rm SF}_{\rm ion}$, since they are expected to contribute $\sim$60-65\% of the total budget \citep[e.g.,][]{Mascia2024ceers}. However, our sample does not probe this very faint regime where we must rely on extrapolation. Under the assumption of a linear decline in the EELG fraction, we infer a median $f_{\rm EELG}=13\%$ in this range of UV magnitudes.}

{We compute $\dot{n}^{\rm SF}_{\rm ion}$ in the same four bins of UV luminosity. We adopt the UV luminosity function at $z\sim6$ from \cite{Bouwens2021}. For $\xi_{\rm ion}^{\rm SF}$, we use the median log(\xiion [Hz erg$^{-1}$]) values in each UV bin, based on the \xiion-M$_{\rm UV}$ relation presented in \cite{Llerena2024xi}. For the escape fraction, we adopt the median \flyc=4\% at $z\sim5-7$ from \cite{Mascia2025}, inferred using a methodology consistent with that adopted in this work. This yields log($\dot{n}^{\rm SF}_{\rm ion} [\rm{s}^{-1} \rm{Mpc}^{-3}])=50.1^{+0.13}_{-0.19}$, consistent with the emissivity required to maintain reionization at $z=6$ \citep{Madau1999}. Approximately 72\% of the SF budget is produced by very-faint sources, partially driven by the elevated extrapolated median log(\xiion [Hz erg$^{-1}$])=25.95, which is consistent with measurements in small samples of very-faint sources in lensed fields at $z\sim6$ \citep[e.g.,][]{Atek2024,Asada2026}, as well as in individual galaxies at similar redshifts with M$_{\rm UV}\gtrsim-12$, which show elevated \xiion~with both high \citep[e.g.,][]{Vanzella2024} and low \citep[e.g.,][]{Vanzella2023} EW(\oiii).}

{To compute $\dot{n}^{\rm EELG}_{\rm ion}$, we assume $\rho^{\rm EELG}_{\rm UV}=f_{\rm EELG}~\rho^{\rm SF}_{\rm UV}$. For $\xi_{\rm ion}^{\rm EELG}$, we found a median offset of 0.14 dex in our sample with respect to the \xiion-M$_{\rm UV}$ relation presented in \cite{Llerena2024xi}. Assuming these median offsets for the median $\xi_{\rm ion}^{\rm SF}$, and the median \flyc=5\% in our sample, we estimate that EELGs contribute $27^{+13}_{-11}\%$ of $\dot{n}^{\rm SF}_{\rm ion}$ when summing all contributions from bright to very-faint sources. Therefore, EELGs with large EW(\oiiihb) play a non-negligible role in the total ionizing photon budget of SF galaxies.}

{In our analysis, the value of $\dot{n}^{\rm EELG}_{\rm ion}/\dot{n}^{\rm SF}_{\rm ion}$ is driven by the fact that at very-faint UV luminosities, where the bulk of the budget is produced, the fractional contribution of EELGs is small ($\sim18\%$). Indeed, assuming $f_{\rm EELG}=0$ in the very-faint regime yields 
$\dot{n}^{\rm EELG}_{\rm ion}/\dot{n}^{\rm SF}_{\rm ion}=13\%$, which can be interpreted as the minimum median fractional contribution of EELGs. We note that our sample at faint and very-faint UV luminosities is highly incomplete, comprising only four galaxies and none, respectively, highlighting the need for a more comprehensive characterization of \xiion~and \flyc~in this regime for both SF galaxies and EELGs given that the median properties of very-faint sources are largely uncertain.  
Moreover, we note that the fraction f$_{\rm EELG}$ of galaxies with EW(\oiii)$>750$~\r{A} has been shown to increase with redshift \citep{Boyett2024}, which may further enhance the fractional contribution of EELGs at earlier epochs. In addition, as discussed in Section~\ref{sec:fesc_in_SE}, the LyC escape fraction may be elevated in specific subpopulations of EELGs characterized by compact star-forming regions, potentially increasing their contribution to cosmic reionization.}

\section{The drivers of the high EWs}\label{sec:drivers}
In this section, we explore the potential physical properties responsible for the elevated EW(\oiii) observed in our sample. We examine various properties that could contribute to this enhancement. Furthermore, we place our findings within the broader context of galaxies with SE star-formation. {A summary of the results of this section are reported in Table \ref{tab:correlations}, including the correlation parameters of the relations with EW(\oiii).}

{\tiny
\begin{table}
\centering
\setlength{\tabcolsep}{3pt}
\caption{{Median values and 16-84th percentile intervals for the different parameters analyzed in Sec. \ref{sec:xi_fesc} and Sec. \ref{sec:drivers}. We also summarize the correlation coefficientes $\rho$ for each parameter with EW(\oiii) shown in Sec. \ref{sec:drivers} and the partial correlation coefficientes $\rho_{\mathrm{partial}}$ shown in Sec. \ref{sec:partial_corr}. We note that these correlations are valid only in the range of high EWs and $z$ covered by our sample ($2.6\lesssim$log(EW(\oiii))$\lesssim3.6$).}}
\label{tab:correlations}
\begin{tabular}{l|cccc}
\hline
 &Median& Interval&$\rho$&
$\rho_{\mathrm{partial}}$ \\
\hline
log(\xiion [Hz erg$^{-1}$])&25.37&(25.19, 25.50)&--&--\\
\flyc&5\%&(0.8\%, 29\%)&--&--\\
\flyc$^{*}$&5.1\%&(1.3\%, 43.5\%)&--&--\\
\flyc$^{\dagger}$ &11.8\%&(1.2\%, 40.5\%)&--&--\\
\flyc $^{\ddagger}$ &1.3\%&(0.44\%, 3.4\%)&--&--\\
\flya $^{\mathsection}$&14.6\%&(6\%, 30\%)&--&--\\\hline
sSFR [Gyr$^{-1}$] & 43 &(24, 103)&0.64&0.42\\
$r_{\rm opt}$ [kpc]& 0.49&(0.19,  0.86)&-0.37&-0.09 \\
$\Sigma_{\rm SFR}$ [\Msun~yr$^{-1}$ kpc$^{-2}$] & 10.7&(1.9, 56) &0.50&0.41  \\
log(M$_{\rm gas}$/\Msun)&9.28&(8.95, 9.71)&--&--\\
f$_{\rm gas}$&87\%&(80\%, 93\%)&--&--\\
log(SFR$_{\rm 3Myr}$/SFR$_{\rm 100Myr}$)&1.16&(0.84, 1.59)&0.46&0   \\
\hline\hline
\end{tabular}
\begin{flushleft}
{\footnotesize
{Notes:}\\~$^{*}$ For EELGs with sSFR$>$25Gyr$^{-1}$.\\ 
$^{\dagger}$ For EELGs with sSFR$>$25Gyr$^{-1}$ and r$_{\rm opt}<200$ pc. \\$^{\ddagger}$ For EELGs with sSFR$<$25Gyr$^{-1}$. \\$^{\mathsection}$ For the sample of LAEs at $z<7$ in our sample.
}
\end{flushleft}
\end{table}}

\subsection{Elevated and compact star-formation}\label{sec:ssfr-size}

\begin{figure*}[t!]
    \centering
    \includegraphics[width=0.5\linewidth]{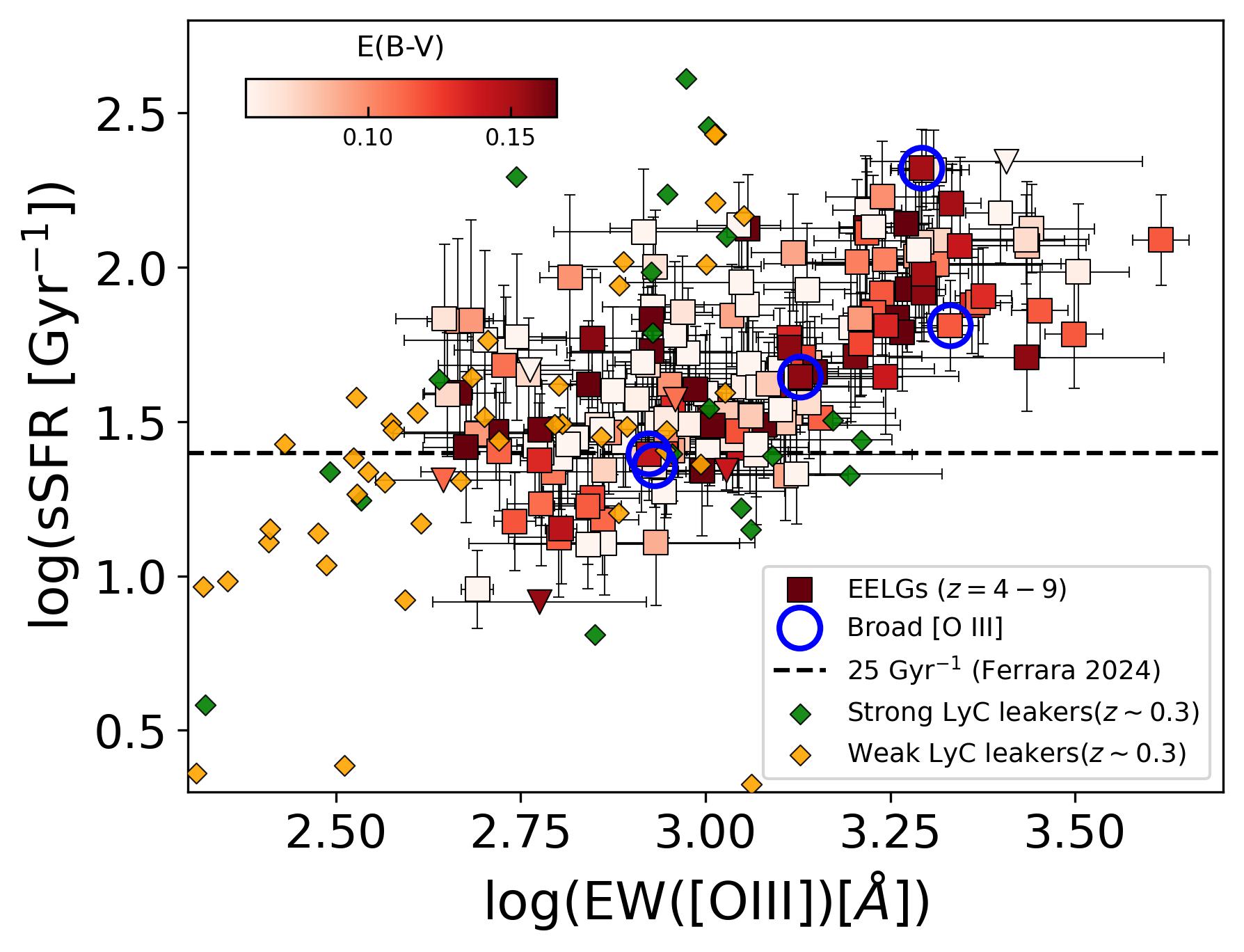}\,\includegraphics[width=0.5\linewidth]{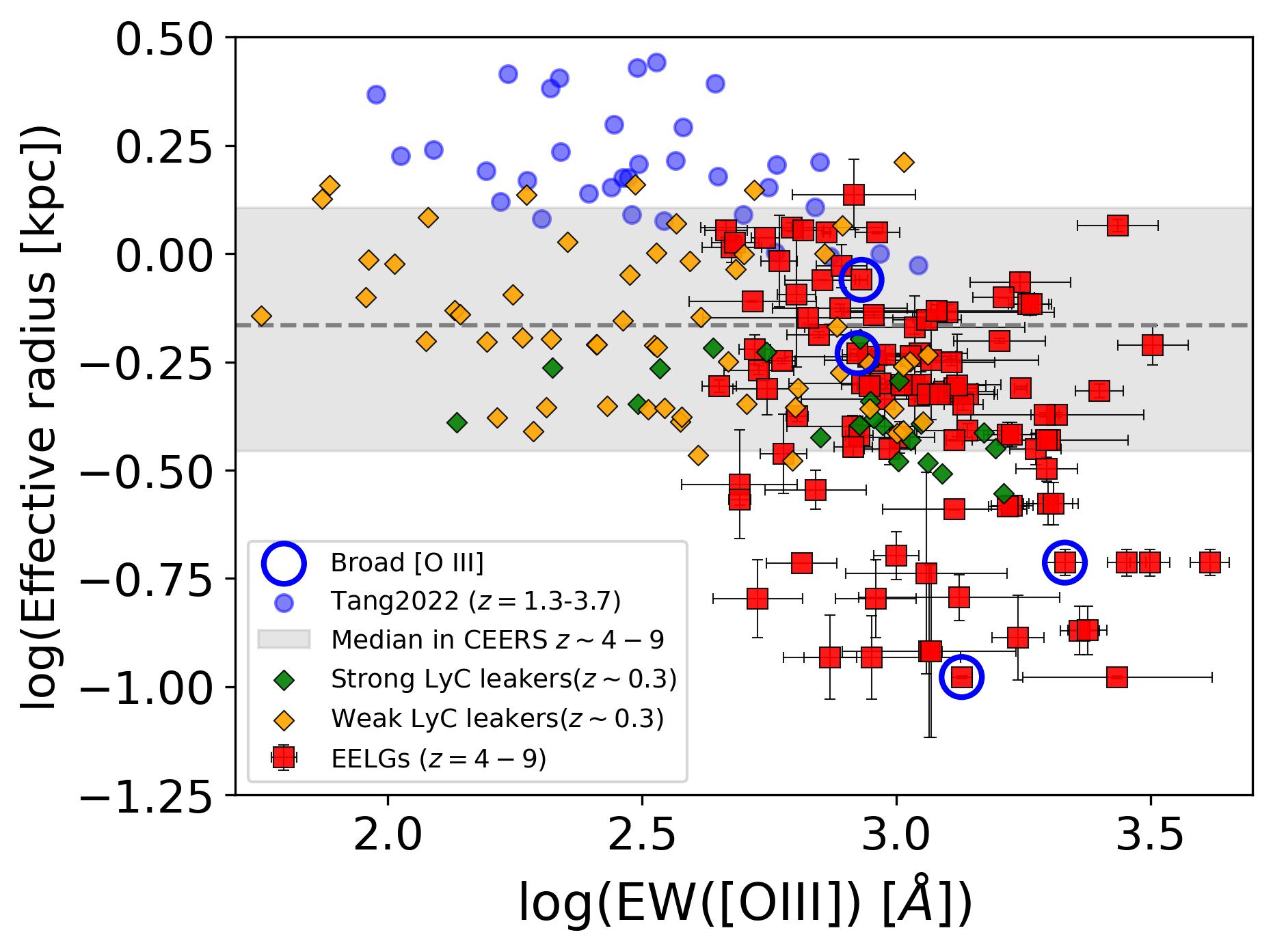}\\
    \includegraphics[width=0.5\linewidth]{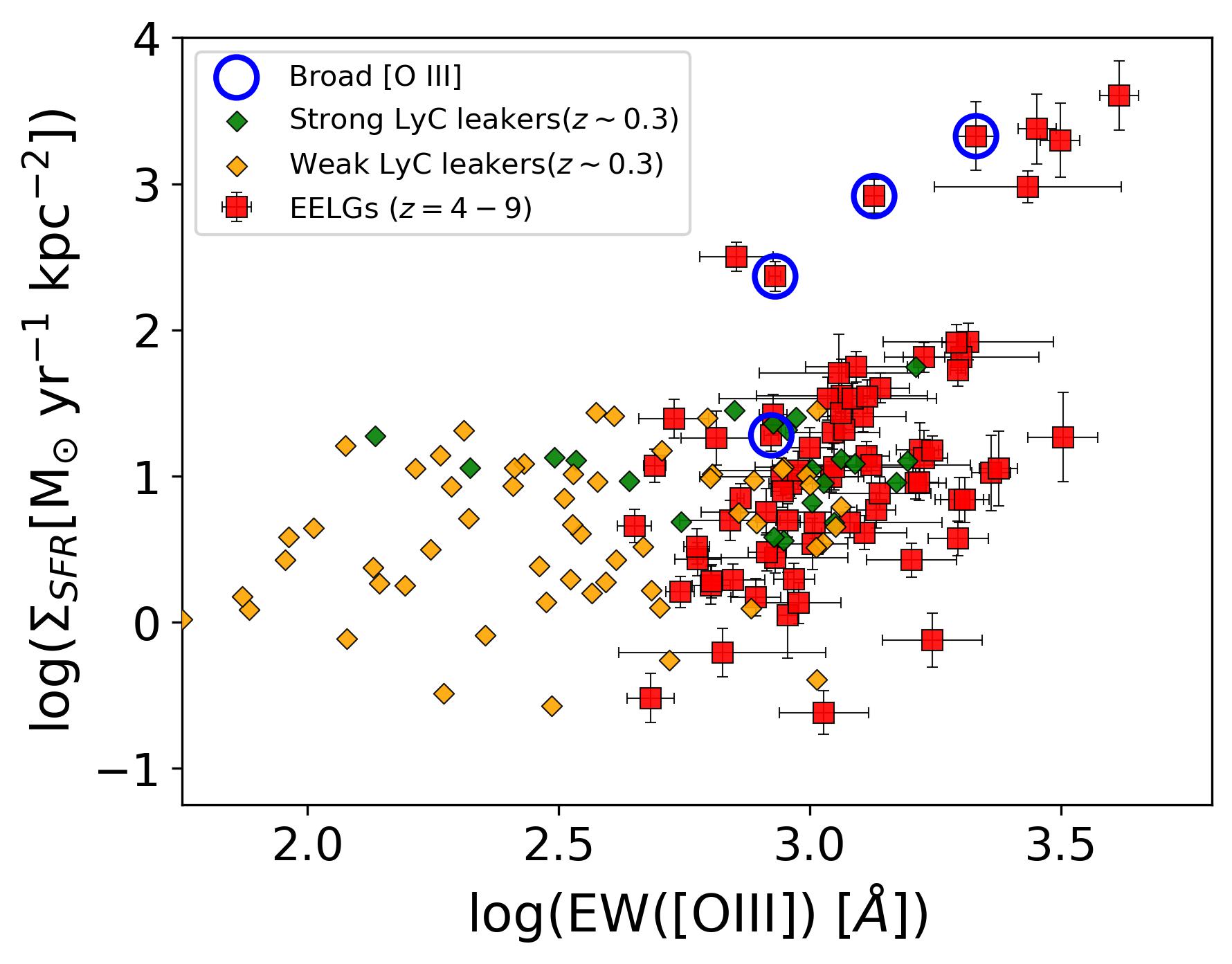}\,\includegraphics[width=0.5\linewidth]{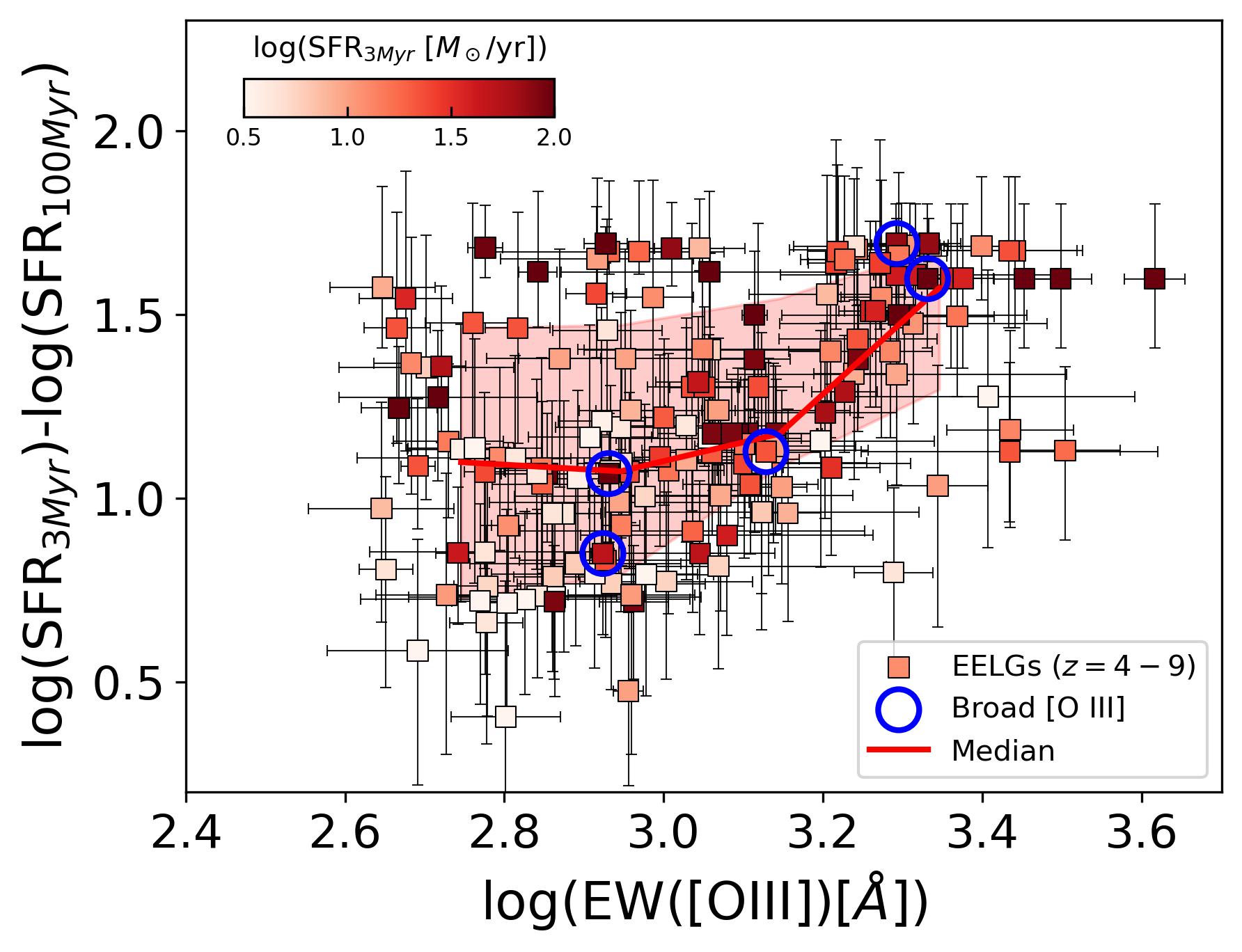}
    \caption{{Relation between EW(\oiii) and: \textit{Top left pane}l:} sSFR. The symbols are color-coded by E(B-V) from SED fitting. The triangle symbols are upper limits due to low \hbeta~S/N. The dashed line is the threshold for super-Eddington galaxies \citep{Ferrara2024}. {\textit{Top right panel}:}  r$_{\rm{opt}}$. The blue circles are EELGs at intermediate-$z$ \citep{Tang2022}. The horizontal gray dashed line is the median r$_{\rm{opt}}$ in CEERS in the same redshift range. The shaded region indicates the 16th–84th percentile range. {\textit{Bottom left panel}:} $\Sigma_{\rm SFR}$. {\textit{Bottom right panel}:} SFR$_{\rm 3Myr}$/SFR$_{\rm 100Myr}$, as a proxy of the burstiness. The galaxies in the sample are color-coded by the average SFR in the last 3Myr. The red line traces the median values of burstiness computed in bins of EW, while the shaded region is the 16th-and 84th percentiles. {\textit{In all panels}, green and orange} symbols are as in Fig. \ref{fig:MS}. {Open blue circles mark the position of EELGs identified in Sec. \ref{sec:blagn} with evidence of broad \oiii, suggesting the presence of stellar outflows.}
    }
    \label{fig:ew_ssfr}
\end{figure*}

In {the top left panel in} Fig. \ref{fig:ew_ssfr}, we show a strong correlation ($\rho=0.64$, $p\sim0$) between EW(\oiii) and sSFR, where the most extreme emitters show elevated sSFR. A similar trend is found with EW(\hbeta). While our sample is limited to high-EW galaxies, this increasing trend has also been reported for lower-EW galaxies at $0.1< z<0.94$ \citep{Amorin2015}. Our sample shows more extreme EWs and sSFRs than low-$z$ LyC leakers, more consistent with the subsample of strong leakers. The elevated sSFRs seem to be a potential driver of the high EWs observed in the sample.
In particular, the median sSFR in the sample is 43 Gyr$^{-1}$, with 16-84th percentile interval from 24 to 103 Gyr$^{-1}$, and most of the galaxies ($80\%$) show sSFR~>~25~Gyr$^{-1}$ with high EWs. This is the threshold for a galaxy to develop a radiation-driven outflow, ejecting both dust and
gas from the system, assuming {that the source is sufficiently compact} \citep[{effective radius $\lesssim200$~pc,}][]{Ferrara2024}. Our results imply a fraction of 19\% super-Eddington galaxies ({only assuming the threshold in sSFRs) in the total population of SF galaxies at $z\sim6$}, based on the parent sample defined for the photometric selection of EELGs, which contains 738 galaxies within $5.5 < z_{phot} < 6.5$.

We also analyze a possible relation between EW and galaxy compactness. To measure the rest-frame UV (r$_{\rm{UV}}$) and optical (r$_{\rm{opt}}$) effective radii of the galaxies in our sample, we use the Galfit catalog v2.0 from the CEERS collaboration (McGrath et al. \textit{in prep}.). Galfit \citep{Peng2002,Peng2010} was run for sources with F356W$<$28.5 mag using background-subtracted mosaics. To determine r$_{\rm{opt}}$, we considered the effective radius in the filter F277W, F356W, and F444W for galaxies at $z<6$, $6<z<8$, and $z>8$, respectively. For Galfit modeling, we restricted our analysis to galaxies with a good quality flag (so-called Flag = 0) in the corresponding filter, representing $66\%$ of the total sample. The r$_{\rm{opt}}$ are reported in Table \ref{lt_ewo3_err}. For this subsample, we also determined r$_{\rm{UV}}$ considering the effective radius in the filter F115W and F150W for galaxies at $z<6$ and $z>6$, respectively. We found a median r$_{\rm{UV}}$/r$_{\rm{opt}}$=0.96.

In {the top right panel in} Fig. \ref{fig:ew_ssfr} we show the relation between r$_{\rm{opt}}$ and EW(\oiii), which are moderately anti-correlated ($\rho=-0.37$, $p=1.2\times10^{-4}$), with galaxies with the most extreme EWs showing the most compact sizes. The median r$_{\rm{opt}}$ is 0.49~kpc with values from 104~pc to 3.70~kpc{, indicating that EELGs tend to be compact, which is a necessary condition for the development of radiation-driven outflows according to the AFM model. In particular, we note that 17\% of the EELGs have r$_{\rm{opt}}<$~200~pc.} Compared to extreme emitters at $z=1.3-3.7$ \citep[][]{Tang2022}, our galaxies are more extreme in terms of EWs and r$_{\rm{opt}}$, but they follow the trend observed at lower redshifts. We verified that the median r$_{\rm{opt}}$ of all sources in the Galfit catalog within the same redshift range and quality flag is 0.69~kpc, indicating that EELGs are, on average, more compact than galaxies at similar redshifts. {In addition, we note that all low-$z$ strong LyC leakers have sizes below the median size of high-$z$ galaxies, suggesting that compact sizes may not only drive the high EWs observed in the sample but also facilitate the escape of ionizing radiation.}

\begin{figure}[t!]
    \centering
    \includegraphics[width=\linewidth]{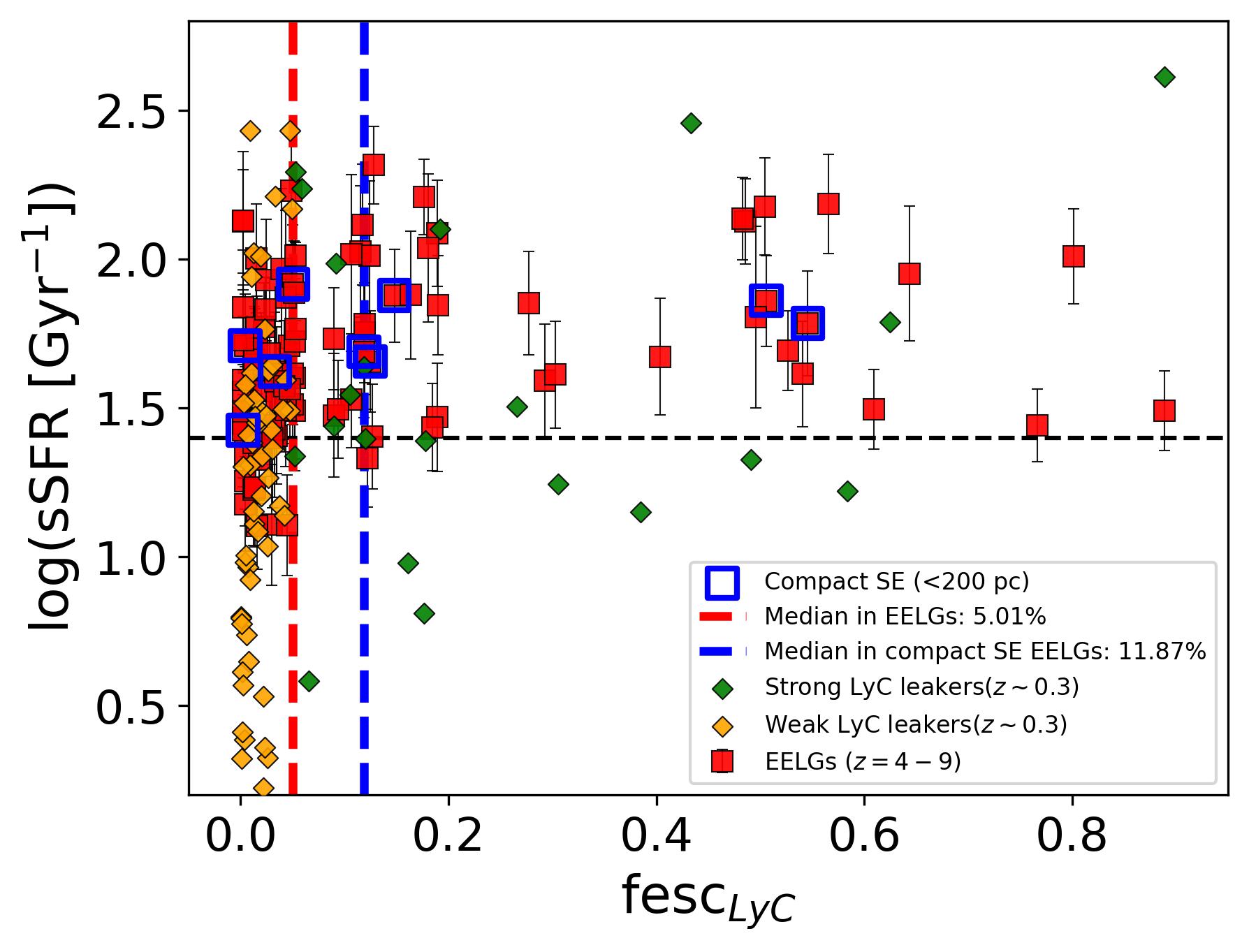}
    \caption{{Inferred \flyc~vs sSFR in the sample of EELGs (red squares). The horizontal black dashed line represents the 25 Gyr$^{-1}$ threshold for SE regime. The blue open squares mark the position of galaxies with compact ($< 200$~pc) SE activity. The dashed red and blue vertical lines represent the median \flyc~for the entire sample and compact SE, respectively. Green and orange symbols are as in Fig. \ref{fig:MS}.}}
    \label{fig:ssfr_fesc}
\end{figure}
\subsubsection{{LyC escape in EELGs with SE activity}}\label{sec:fesc_in_SE}
{We discuss our results in {Sec. \ref{sec:escapelyc} and Sec. \ref{sec:ssfr-size}} in the context of the AFM model. The fraction of SE galaxies
in the parent sample (19\%) agrees very well with the expected fraction of $22\%$ predicted by AFM at $z\sim6$ \citep{Ferrara2025}. The small difference may result from the approximate comparison between spectroscopic and photometric samples, or from some SE galaxies missed due to extremely low metallicities, which prevent them from being classified as EELGs \citep[e.g.,][]{Laseter2025}.

Galaxies in our sample also show relatively low and moderate E(B-V) ({top left panel in }Fig. \ref{fig:ew_ssfr} {and Table \ref{tab:sample-median}}) as expected by AFM, with the level of dust clearing depending on the evolutionary phase of the outflow, {leading to reduced attenuation at later stages. Consistent with this picture, the small subsample of five EELGs showing evidence of broad \oiii~emission (open blue circles in Fig. \ref{fig:ew_ssfr}), likely tracing stellar-driven outflows as discussed in Sec. \ref{sec:blagn}, exhibits sSFR values above the threshold for SE activity and compact sizes, suggesting a link between enhanced star formation, radiation-driven outflows, and the physical conditions that may facilitate LyC escape.}

Interestingly, {as shown in Fig. \ref{fig:ssfr_fesc}, for a given sSFR the inferred \flyc~values are diverse, with SE galaxies exhibiting both low and high \flyc. We also note that all EELGs with highly efficient \flyc$>$20\% show sSFR$>$25Gyr$^{-1}$, suggesting that SE star-formation may be a necessary, though not sufficient, condition for high LyC escape. In addition,} the median \flyc~of galaxies with sSFRs>25Gyr$^{-1}$ is {5.1}\% {(16-84th percentile interval: 1.3\% to 43.5\%)}, while for galaxies below is 1.3\% {(16-84th percentile interval: 0.44\% to 3.4\%)}, which suggests that elevated sSFRs enhance \flyc, in agreement with the AFM models. 

Additionally, an enhancement in \flyc~is observed in galaxies with sSFR $>25$ Gyr$^{-1}$ which also have $r_{\rm opt}<200$ pc ({open squares in Fig. \ref{fig:ssfr_fesc}}), {representing 15\% of EELGs}, with a median \flyc={11.8}\% {(16-84th percentile interval: 1.2\% to 40.5\%)}, while galaxies with sSFR $>25$ Gyr$^{-1}$ and $r_{\rm opt}>200$ pc, {representing 68\% of EELGs}, show a median \flyc={5.2}\% {(16-84th percentile interval: 1.4\%-48.4\%)}. {This apparent enhancement is consistent with results from low-$z$ LyC leakers, which suggest that compactness is linked to increased feedback efficiency when star formation is spatially concentrated \citep{Flury2022}. In such compact SF regions, feedback can efficiently deplete or disrupt the surrounding gas, enabling LyC photons to escape the local H II regions, propagate through the more diffuse ISM, and potentially leak into the IGM.}

\subsection{{SFR surface density}}\label{sec:sigma_sfr}
Given that these systems show compact effective radii, we expect they host high SFR surface densities $\left(\Sigma_{\rm{SFR}}=\dfrac{\rm{SFR}}{2\pi r_{UV}^2}\right)$. We derived this parameter for this sample, and found a median value of 10.7 \Msun~yr$^{-1}$ kpc$^{-2}$, with the 16-84th percentile interval between 1.9 and 56 \Msun~yr$^{-1}$ kpc$^{-2}$. This range is consistent with the high $\Sigma_{\rm{SFR}}$ measured in low-$z$ LyC leakers, as shown in {the bottom left panel in} Fig. \ref{fig:ew_ssfr}. {Also, the median value in the sample is consistent with the threshold of 10 \Msun~yr$^{-1}$ kpc$^{-2}$ above which nearly all strong LyC leakers appear in the LzLCS survey \citep{Flury2022}, suggesting a population of strong and weak LyC leakers in our EELG sample.} However, we note that 11\% of the sample show $\Sigma_{\rm{SFR}}>100$ \Msun~yr$^{-1}$ kpc$^{-2}$, {with a subsample of them showing evidence of broad \oiii~emission}. Assuming the Kennicutt-Schmidt law \citep{Kennicutt2012}, this implies a high total
baryonic mass of the gas (M$_{\rm gas}$) involved in star formation within the effective radius, with a median value of log(M$_{\rm gas}$/\Msun)=9.28 {(16-86th percentiles: 8.95 and 9.71)}. Given the low stellar masses of these systems, such large M$_{\rm gas}$ values correspond to high gas fractions $\left(f_{\rm{gas}}=\dfrac{\rm{M}_{\rm gas}}{\rm{M}_{\rm gas}+\rm{M}_{\star}}\right)$ ranging from 33\% to 98\%, with a median value of 87\%. These values are consistent with those estimated using similar methods for EELGs at $z=0.1-0.9$ \citep{Calabro2017}. In {the bottom left panel in} Fig. \ref{fig:ew_ssfr}, we show that $\Sigma_{\rm{SFR}}$ and EW(\oiii) are moderately correlated ($\rho=0.50$, $p=4\times10^{-6}$), which appears slightly stronger than the correlation with the effective radius. This suggests that the strength of the [O III] emission is more closely linked to the intensity of compact star formation than to the overall size of the galaxy.

Overall, elevated sSFRs and compact star-formation activity appear to drive high EWs. The high sSFRs and small effective radii, which place them in the super-Eddington regime, lead to an enhancement of \flyc~with moderate and low E(B–V) values.

\subsection{Burstiness in EELGs}\label{sec:burstiness}

{Bursty star formation is a key characteristic of EELGs. Analogs of high-$z$ EELGs in the local universe, such as compact SF galaxies with small linear radii of a few kpc in the Sloan Digital Sky Survey, exhibit highly bursty star-formation histories, resulting in rapid luminosity variations on Myr timescales \citep{Izotov2016burst}. Furthermore, \citet{Flury2025} shows that a two-stage star-formation burst, with an initial phase that generates feedback via stellar winds and supernovae and a second phase that produces LyC photons, is necessary for strong LyC escape by providing both mechanical and ionizing feedback capable of redistributing or removing LyC-obscuring gas and dust. These findings underscore the relevance of burstiness for EELGs, particularly concerning their potential to leak LyC photons. }

In this section, we analyze the SFH of EELGs based on the SED model. Given that SFHs indicate EELGs are experiencing a burst over the last 3~Myr, we investigate their burstiness by comparing this timescale with a longer period of 100~Myr. For this analysis, we consider the SFRs obtained from the non-parametric SED modeling. We estimated the average SFR in the last 3~Myr and in the last 100~Myr, and the uncertainties are estimated from the average of the 16th and 84th percentiles of the SFR distribution. We calculated the burstiness by the ratio SFR$_{\rm 3Myr}$/SFR$_{\rm 100Myr}$ which is shown in {the bottom right panel in} Fig. \ref{fig:ew_ssfr}.. We note that all EELGs have SFR$_{\rm 3Myr}>$SFR$_{\rm 100Myr}$ and exhibit a median log(SFR$_{\rm 3Myr}$)-log(SFR$_{\rm 100Myr}$)=1.16 ($\sigma=0.31$), which is an increase in SFR by a factor of $14.6$ over the last 3~Myr, indicating that they are young starbursts. This level of burstiness implies that, on average, 36\% of the stellar mass in EELGs is assembled within the last 3~Myr. Compared to EW(\oiii), we find a moderate correlation ($\rho=0.46$, $p\sim0$) such that galaxies with high EWs tend to show high SFR$_{\rm3Myr}$/SFR$_{\rm100Myr}$. Additionally, galaxies with high burstiness tend to show high SFR$_{\rm 3Myr}$ ($\rho=0.42$, $p=2.6\times10^{-9}$). This suggests that EW is regulated not only by the burstiness of the system but also by the most recent SFR. An enhancement in the burstiness of a system in galaxies with elevated EW(\halpha) has been found in samples of \halpha~emitters at $z\sim2$ and $z\sim4-6$ \citep{Navarro-Carrera2024}, and in low-mass galaxies $0.7 < z < 1.5$ \citep{Atek2022}. {The high burstiness observed in a subsample of EELGs with broad \oiii~emission and high EWs is consistent with a scenario in which these systems generate the feedback necessary to facilitate the escape of LyC radiation.}

\subsection{What are the primary drivers {of extreme EWs}?}\label{sec:partial_corr}
To isolate the intrinsic correlation between EW(\oiii) and each parameter explored in Sec. \ref{sec:ssfr-size}, \ref{sec:sigma_sfr} and \ref{sec:burstiness}, we performed a partial Spearman correlation analysis by regressing each variable against the others and correlating the resulting residuals. This eliminates the influence of the other parameters, enabling us to evaluate the direct correlations between each variable and EW(\oiii). To this analysis, we use the \textit{statsmodels} package. We find that EW(\oiii) shows a moderate positive correlation with both sSFR ($\rho=0.42$, $p=1.3\times10^{-4}$) and $\Sigma_{\rm{SFR}}$ ($\rho=0.41$, $p=2.2\times10^{-4}$) even after controlling for the other parameters, indicating that the EW(\oiii) is primarily driven by star formation activity. In contrast, no significant correlation is found with effective radius ($\rho=-0.09$, $p=0.43$) or burstiness ($\rho\sim0$, $p=0.93$), suggesting that structural parameters and specific degree of burstiness have little direct impact on EW(\oiii) in this sample. However, we note that our sample is biased toward extreme EWs, and the strength of the correlations may change when including galaxies with lower EWs. {A summary of the resulting partial corrrelations are reported in Table \ref{tab:correlations}.}

Overall, EELGs exhibit elevated sSFRs and $\Sigma_{\rm{SFR}}$. These properties suggest that intense star formation activity occurring within small spatial scales may be a key factor driving their extreme EWs. Additionally, high levels of burstiness, indicated by elevated  SFR$_{\rm 3Myr}$/SFR$_{\rm 100Myr}$ ratios, might further enhance the EWs, although our partial correlation analysis shows that sSFR and $\Sigma_{\rm{SFR}}$ are the primary drivers, while burstiness has no significant independent effect.

\section{Summary and conclusions}\label{sec:conclusions}
We analyzed a sample of 160 EELGs at $z\sim4-9$, including 127 observed with NIRSpec/PRISM and 81 with NIRSpec/G395M, drawn from the CAPERS, CEERS, and RUBIES surveys. These galaxies were initially identified as EELG candidates from NIRCam photometry \citep{Llerena2024EELG}. Our results can be summarized as follows: 
\begin{enumerate}
  \renewcommand{\labelenumi}{\roman{enumi})}
\item We found a high success rate of 89\% for the selection criteria proposed in \cite{Llerena2024EELG} to identify EELGs at $z\sim4-9$ with EW(\oiiihb)$>680$\r{A}. In the confirmed extreme EELGs, the median EW([OIII]+\hbeta)=$1616$\r{A} and the median EW(\halpha)=$763$\r{A}. Photometric redshifts are confirmed with a mean difference 
$z_{phot}-z_{spec} = 0.13$. 

\item EELGs show a median log(\Mstar/\Msun)=8.26 and are scattered above the main-sequence of SF galaxies at $z\sim6$ based on the Balmer-derived SFRs. They show a wide range of $\rm{E(B-V)}\sim0-0.46$ with no correlation with  EW(\oiii). They also show a wide range of $\beta$ slopes with a {median} $\beta=-2.0$. Only 7\% of EELGs have super-blue $\beta$ slopes $<-2.6$ that are bluer than the limit set by SF + nebular continuum, however, this fraction is not significant at the 1$\sigma$ level.

\item Emission-line diagnostics (OHNO, MEx diagrams) suggest that massive stars are the dominant ionizing source in EELGs. An AGN contribution cannot be fully excluded in 14\% of the sample, and evidence for BLAGN is found in a small fraction ($4$\%) of the sample. 

\item EELGs are efficient producers of ionizing photons with a {median} log(\xiion [Hz erg$^{-1}$]) ={25.37}, which is slightly higher than the mean value found for SF galaxies at similar redshift. However, the escape of LyC photons, indirectly inferred from Cox models, is not particularly efficient. Only {16}\% of EELGs show high escape fractions {$>$5\%} of both \lya~and LyC photons, while {half} ({49}\%) are likely not strong {LyC} leakers (\flyc<{5}\%). This is supported by the lack of \lya~emission in 82\% of the sample and the fact that the median inferred \flyc={5}\% is not particularly elevated. However, the escape of LyC photons is enhanced in EELGs with sSFR>25Gyr$^{-1}$ and r$_{\rm{opt}}$<200 pc with a median \flyc={11.8}\%. 

\item {These results imply that EELGs play a non-negligible role in the total ionizing emissivity required to sustain hydrogen reionization, with a median contribution of approximately $16-40$\%. We emphasize, however, that this estimate relies on extrapolations into the very-faint UV regime, where the median properties of SF galaxies and EELGs remain highly uncertain.}

\item EELGs exhibit very high sSFRs, with a median value of 43~Gyr$^{-1}$, along with very high $\Sigma_{\rm{SFR}}$, with a median value of 10.7 \Msun~yr$^{-1}$ kpc$^{-2}$ and compact r$_{\rm{opt}}$, with a median of 0.49~kpc. These properties suggest that intense star formation activity occurring within small spatial scales may be a key factor driving their extreme EWs. Additionally, high burstiness, traced by the ratio SFR$_{3Myr}$/SFR$_{100Myr}$, might enhance the observed EWs. However, sSFR and $\Sigma_{\rm{SFR}}$ are likely to be the primary drivers.
\end{enumerate}

Overall, EELGs are low-mass systems characterized by elevated and compact star formation with a recent burst. They are efficient producers of ionizing photons; however, while a subset of the population exhibits high escape fractions of \lya~and LyC photons, they are not uniformly strong leakers with \flyc\,>\,{5}\%. This may be attributed to the presence of dust within the ISM, as suggested by the relatively moderate (i.e., not extremely blue) UV continuum slopes observed in the sample since $\beta$ seems to be an important indicator of \flyc~\citep{Chisholm2022,Giovinazzo2025}. Notably, in compact super-Eddington galaxies, the escape of LyC photons is enhanced.

\begin{acknowledgements}   
{We thank the anonymous referee for the thorough review and constructive suggestions that significantly improved this paper. We also thank Sophia Flury for valuable feedback and insightful comments, which helped refine our interpretation of the results.}
     MLl acknowledges support from the INAF Large Grant 2022 “Extragalactic Surveys with JWST” (PI L. Pentericci), the PRIN 2022 MUR project 2022CB3PJ3 - First Light And Galaxy aSsembly (FLAGS) funded by the European Union – Next Generation EU, the INAF Mini-grant 2024 "Galaxies in the epoch of Reionization and their analogs at lower redshift" (PI M. Llerena), and the Large Grant RF 2023 F.O. 1.05.23.01.11 "The MOONS Extragalactic Survey". RA acknowledges support of Grant PID2023-147386NB-I00 funded by MICIU/AEI/10.13039/501100011033 and by ERDF/EU, and the Severo Ochoa award to the IAA-CSIC CEX2021-001131-S.
This work is based on observations made with the NASA/ESA/CSA \textit{James Webb Space Telescope}, obtained at the Space Telescope Science Institute, which is operated by the Association of Universities for Research in Astronomy, Incorporated, under NASA contract NAS5-03127. Support for program number GO-6368 was provided through a grant from the STScI under NASA contract NAS5-03127. The data were obtained from the Mikulski Archive for Space Telescopes (MAST) at the Space Telescope Science Institute. These observations are associated with program \#6368, and can be accessed via \href{https://archive.stsci.edu/doi/resolve/resolve.html?doi=10.17909/0q3p-sp24}{doi}. Some of the data products presented herein were retrieved from the Dawn JWST Archive (DJA). DJA is an initiative of the Cosmic Dawn Center (DAWN), which is funded by the Danish National Research Foundation under grant DNRF140. These observations are associated with programs \#1345, \#2750, and \#4233.

    This work has made extensive use of Python packages astropy \citep{astropy:2018}, numpy \citep{harris2020}, Matplotlib \citep{Hunter:2007} and LiMe \citep{Fernandez2024Lime}.
\end{acknowledgements}

%
%
\bibliographystyle{aa}
\bibliography{main}

\begin{appendix}

\section{Summary of properties in the sample}
In Table \ref{lt_ewo3_err}, we provide information about the galaxies in the sample. We report their identification ID from the CEERS photometric catalog and their spectroscopic ID (Spec ID) in each survey. We report their coordinates, the corresponding spectroscopic survey and disperser, spectroscopic redshifts, rest-frame EWs of bright emission lines (\hbeta, \oiii, and \halpha), E(B–V) from SED modeling, sSFRs, $\beta$ slopes, M$_{\rm UV}$, rest-frame optical effective radius, and the inferred \flyc~from the Cox models. The {NLAGN} candidates based on the OHNO diagram shown in Sec. \ref{sec:ohno-mex} are marked with a $^{\ddagger}$. The EELGs in the sample classified as BLAGN in Sec. \ref{sec:blagn} are marked with a $^{\dagger}$, {while the ones with evidence of a broad \oiii~component are marked with a $^{\mathsection}$}. 
\onecolumn
{\tiny
\renewcommand{\arraystretch}{0.97}
\setlength{\tabcolsep}{2pt}
\begin{longtable}{ccccccccccccccc}
\caption{Properties of the sample, in order of decreasing EW([OIII]).}\\[-2em]
\label{lt_ewo3_err}\\
\hline
ID &Spec ID&RA&DEC&Survey$^{*}$&$z_{spec}$& EW(\hbeta)&EW(\oiii)&EW(\halpha)&E(B-V)&log(sSFR)&$\beta$&M$_{UV}$&r$_{\rm{opt}}$&log(fesc$_{\rm{LyC}}$) \\
&&deg&deg& & & $\AA$&$\AA$&$\AA$&&Gyr$^{-1}$ &&&kpc\\
\hline
\endfirsthead
\caption{continued.}\\
\hline
ID&Spec ID &RA&DEC&Survey$^{*}$&$z_{spec}$&EW(\hbeta)&EW(\oiii)&EW(\halpha)&E(B-V)&log(sSFR)&$\beta$&M$_{UV}$ &r$_{\rm{opt}}$&log(fesc$_{\rm{LyC}}$)\\
&&deg&deg& & & $\AA$&$\AA$&$\AA$&&Gyr$^{-1}$ &&&kpc\\
\hline
\endhead
\hline
\endfoot
59920 &944720 & 214.88300 & 52.84042 & R-M & 7.823 & 501 $\pm$ 54 & 4131 $\pm$ 362 & -&0.12&2.09 $\pm$ 0.15&- &-&0.19 $\pm$ 0.01&-\\
44733 &956207 & 214.89393 & 52.87458 & R-M & 7.033 & 495 $\pm$ 171 & 3190 $\pm$ 509 & -&0.06&1.99 $\pm$ 0.22&- &-&0.61 $\pm$ 0.06&-\\
59920 &944720 & 214.88300 & 52.84042 & R-P & 7.829 & 248 $\pm$ 60 & 3150 $\pm$ 281 & -&0.12&1.78 $\pm$ 0.18&-2.12$^{+0.10}_{-0.10}$ &-20.55&0.19 $\pm$ 0.01&-0.26\\
59920 &1027 & 214.88300 & 52.84042 & C-P & 7.829 & 295 $\pm$ 42 & 2834 $\pm$ 249 & -&0.12&1.86 $\pm$ 0.15&- &-19.80&0.19 $\pm$ 0.01&-0.30\\
70972 &908453 & 215.12884 & 52.95519 & R-P & 7.778 & 424 $\pm$ 81 & 2761 $\pm$ 539 & -&0.07&2.13 $\pm$ 0.14&-1.84$^{+0.26}_{-0.26}$ &-19.25&-&-0.31\\
44935 &955977 & 214.89363 & 52.87401 & R-P & 7.034 & <419  & 2722 $\pm$ 499 & 2972 $\pm$ 878&0.08&2.07 $\pm$ 0.21&- &-19.36&1.16 $\pm$ 0.04&-\\
47504$^{\tiny\mathsection}$ &40556 & 214.92191 & 52.87619 & R-P & 5.651 & 401 $\pm$ 173 & 2718 $\pm$ 1165 & 1236 $\pm$ 92&0.16&1.71 $\pm$ 0.17&-1.68$^{+0.19}_{-0.19}$ &-18.93&0.11 $\pm$ 0.00&-2.35\\
70972 &908453 & 215.12884 & 52.95519 & R-M & 7.774 & 392 $\pm$ 78 & 2712 $\pm$ 531 & -&0.07&2.09 $\pm$ 0.14&- &-&-&-\\
9655 &85504 & 214.98882 & 52.95910 & CA-P & 7.489 & <794  & 2553 $\pm$ 1080 & -&0.04&<2.34 &-2.02$^{+1.01}_{-1.04}$ &-19.30&-&-\\
13353 &2658 & 214.87553 & 52.91494 & CA-P & 5.093 & 282 $\pm$ 66 & 2505 $\pm$ 272 & 1591 $\pm$ 255&0.05&2.18 $\pm$ 0.16&- &-19.11&0.48 $\pm$ 0.02&-0.30\\
75511 &25029 & 214.99131 & 52.89005 & R-M & 6.184 & - & 2375 $\pm$ 209 & 1569 $\pm$ 223&0.13&1.91 $\pm$ 0.16&- &-&0.13 $\pm$ 0.02&-\\
80207 &27069 & 215.00893 & 52.87681 & CA-P & 5.147 & 283 $\pm$ 35 & 2335 $\pm$ 249 & 817 $\pm$ 392&0.11&1.89 $\pm$ 0.17&-2.32$^{+0.23}_{-0.23}$ &-18.91&-&-1.29\\
75511 &25029 & 214.99131 & 52.89005 & R-P & 6.190 & 241 $\pm$ 38 & 2293 $\pm$ 204 & 1459 $\pm$ 204&0.13&1.88 $\pm$ 0.16&-1.77$^{+0.15}_{-0.16}$ &-19.47&0.13 $\pm$ 0.02&-0.83\\
98065 &920424 & 214.85299 & 52.77648 & R-M & 6.954 & 299 $\pm$ 91 & 2208 $\pm$ 320 & -&0.14&2.07 $\pm$ 0.29&- &-&-&-\\
78647$^{\tiny\mathsection}$ &919107 & 215.03719 & 52.90671 & R-P & 7.195 & 361 $\pm$ 49 & 2152 $\pm$ 198 & -&0.15&2.21 $\pm$ 0.13&-1.75$^{+0.50}_{-0.50}$ &-20.12&-&-0.75\\
59920 &1027 & 214.88300 & 52.84042 & C-M & 7.823 & 264 $\pm$ 27 & 2143 $\pm$ 190 & -&0.12&1.81 $\pm$ 0.15&- &-&0.19 $\pm$ 0.01&-\\
64206$^{\tiny\ddagger}$ &26883 & 215.10656 & 52.97582 & R-P & 6.193 & 268 $\pm$ 105 & 2070 $\pm$ 809 & 1321 $\pm$ 121&0.07&2.09 $\pm$ 0.18&-1.91$^{+0.12}_{-0.12}$ &-20.11&0.42 $\pm$ 0.00&-0.72\\
80216 &27083 & 215.00605 & 52.87480 & CA-P & 5.711 & 296 $\pm$ 115 & 2056 $\pm$ 794 & 1318 $\pm$ 169&0.11&2.01 $\pm$ 0.22&-1.68$^{+0.26}_{-0.26}$ &-18.39&-&-1.28\\
74822 &27121 & 215.02787 & 52.92017 & R-M & 5.239 & 228 $\pm$ 34 & 2035 $\pm$ 225 & 1537 $\pm$ 253&0.07&2.32 $\pm$ 0.13&- &-&0.26 $\pm$ 0.03&-\\
81061$^{\tiny\dagger\tiny\ddagger}$ &8488 & 215.03539 & 52.89067 & R-P & 8.688 & 253 $\pm$ 91 & 2005 $\pm$ 709 & -&0.16&2.01 $\pm$ 0.16&-1.92$^{+0.05}_{-0.05}$ &-21.88&0.37 $\pm$ 0.00&-0.10\\
74822 &27121 & 215.02787 & 52.92017 & R-P & 5.247 & 232 $\pm$ 34 & 1987 $\pm$ 220 & 1528 $\pm$ 255&0.07&2.31 $\pm$ 0.13&-2.06$^{+0.54}_{-0.55}$ &-18.11&0.26 $\pm$ 0.03&-0.89\\
53240 &49877 & 214.84929 & 52.84848 & R-M & 5.107 & 393 $\pm$ 69 & 1973 $\pm$ 273 & 1238 $\pm$ 186&0.15&1.98 $\pm$ 0.16&- &-&0.32 $\pm$ 0.02&-\\
81061$^{\tiny\dagger\tiny\ddagger}$ &8488 & 215.03539 & 52.89067 & R-M & 8.681 & 208 $\pm$ 10 & 1973 $\pm$ 70 & -&0.16&1.92 $\pm$ 0.16&- &-&0.37 $\pm$ 0.00&-\\
78647$^{\tiny\mathsection}$ &919107 & 215.03719 & 52.90671 & R-M & 7.193 & 469 $\pm$ 61 & 1961 $\pm$ 188 & -&0.15&2.32 $\pm$ 0.13&- &-&-&-\\
86830 &80916 & 214.89163 & 52.81594 & C-P & 5.676 & 442 $\pm$ 220 & 1960 $\pm$ 961 & 1791 $\pm$ 243&0.10&2.01 $\pm$ 0.25&-2.33$^{+0.23}_{-0.23}$ &-18.69&-&-0.91\\
64206$^{\tiny\ddagger}$ &26883 & 215.10656 & 52.97582 & R-M & 6.185 & 249 $\pm$ 19 & 1958 $\pm$ 126 & 1312 $\pm$ 123&0.07&2.08 $\pm$ 0.18&- &-&0.42 $\pm$ 0.00&-\\
19899 &55191 & 214.88128 & 52.88267 & R-M & 5.763 & 514 $\pm$ 102 & 1943 $\pm$ 220 & -&0.03&2.06 $\pm$ 0.22&- &-&-&-\\
54735 &47238 & 214.83164 & 52.82893 & R-P & 6.076 & 281 $\pm$ 39 & 1924 $\pm$ 232 & 1922 $\pm$ 363&0.10&2.04 $\pm$ 0.25&-1.70$^{+0.22}_{-0.22}$ &-19.16&-&-0.74\\
64730 &25470 & 215.08049 & 52.95477 & R-P & 5.064 & 341 $\pm$ 72 & 1876 $\pm$ 220 & 1173 $\pm$ 201&0.19&1.93 $\pm$ 0.20&- &-18.32&0.35 $\pm$ 0.03&-1.61\\
57054$^{\tiny\ddagger}$ &42959 & 214.86326 & 52.84053 & R-M & 5.677 & 217 $\pm$ 34 & 1868 $\pm$ 154 & 1339 $\pm$ 152&0.21&2.14 $\pm$ 0.23&- &-&-&-\\
55513 &45892 & 214.83370 & 52.82697 & R-M & 4.987 & 163 $\pm$ 36 & 1844 $\pm$ 165 & 1133 $\pm$ 115&0.17&1.79 $\pm$ 0.19&- &-&0.76 $\pm$ 0.04&-\\
55513 &45892 & 214.83370 & 52.82697 & R-P & 4.984 & 249 $\pm$ 41 & 1817 $\pm$ 164 & 1264 $\pm$ 132&0.17&1.84 $\pm$ 0.19&-1.77$^{+0.42}_{-0.41}$ &-19.10&0.76 $\pm$ 0.04&-2.60\\
87370$^{\tiny\ddagger}$ &1374 & 214.94390 & 52.85005 & C-M & 5.002 & 206 $\pm$ 9 & 1754 $\pm$ 50 & 997 $\pm$ 31&0.13&1.81 $\pm$ 0.15&- &-&0.49 $\pm$ 0.00&-\\
97883 &83779 & 214.82142 & 52.75484 & C-P & 4.315 & 222 $\pm$ 63 & 1752 $\pm$ 400 & 976 $\pm$ 77&0.14&1.65 $\pm$ 0.19&- &-18.59&0.86 $\pm$ 0.05&-1.86\\
74981 &26776 & 214.96108 & 52.87163 & R-P & 7.294 & 192 $\pm$ 70 & 1748 $\pm$ 340 & -&0.10&2.03 $\pm$ 0.22&-2.74$^{+0.99}_{-0.80}$ &-18.77&-&-0.94\\
7841 &81759 & 214.98543 & 52.96609 & CA-P & 5.039 & 185 $\pm$ 49 & 1735 $\pm$ 305 & 1413 $\pm$ 263&0.10&2.23 $\pm$ 0.18&-1.95$^{+0.54}_{-0.55}$ &-17.72&-&-1.31\\
100621 &81068 & 214.82051 & 52.73715 & C-P & 6.273 & 248 $\pm$ 42 & 1733 $\pm$ 202 & 1310 $\pm$ 221&0.12&1.92 $\pm$ 0.20&-1.63$^{+0.35}_{-0.35}$ &-18.39&0.13 $\pm$ 0.03&-1.30\\
87910 &23290 & 214.89472 & 52.81213 & CA-P & 4.889 & 249 $\pm$ 25 & 1687 $\pm$ 159 & 1002 $\pm$ 141&0.12&1.85 $\pm$ 0.17&-1.94$^{+0.04}_{-0.04}$ &-20.26&0.26 $\pm$ 0.00&-0.56\\
30373 &58008 & 214.77859 & 52.81455 & R-M & 6.542 & 250 $\pm$ 37 & 1687 $\pm$ 182 & 1332 $\pm$ 244&0.02&2.13 $\pm$ 0.17&- &-&0.38 $\pm$ 0.03&-\\
30373 &58008 & 214.77859 & 52.81455 & R-P & 6.542 & 231 $\pm$ 31 & 1655 $\pm$ 177 & 1506 $\pm$ 276&0.02&2.18 $\pm$ 0.17&-2.43$^{+0.19}_{-0.20}$ &-19.64&0.38 $\pm$ 0.03&-0.25\\
74016 &29785 & 215.00264 & 52.90762 & R-M & 6.195 & 162 $\pm$ 32 & 1653 $\pm$ 143 & 1187 $\pm$ 189&0.12&2.11 $\pm$ 0.20&- &-&0.26 $\pm$ 0.01&-\\
57054$^{\tiny\ddagger}$ &42959 & 214.86326 & 52.84053 & R-P & 5.684 & 150 $\pm$ 22 & 1646 $\pm$ 132 & 1295 $\pm$ 145&0.21&2.13 $\pm$ 0.23&-1.70$^{+0.49}_{-0.49}$ &-18.47&-&-2.60\\
16169 &5528 & 214.90667 & 52.92076 & CA-P & 4.558 & 228 $\pm$ 33 & 1624 $\pm$ 227 & 971 $\pm$ 203&0.12&1.75 $\pm$ 0.17&-2.23$^{+0.10}_{-0.11}$ &-20.26&0.79 $\pm$ 0.00&-0.92\\
54735 &47238 & 214.83164 & 52.82893 & R-M & 6.076 & 222 $\pm$ 31 & 1619 $\pm$ 195 & -&0.10&1.83 $\pm$ 0.25&- &-&-&-\\
17146 &79896 & 214.89259 & 52.90539 & CA-P & 5.125 & 314 $\pm$ 96 & 1604 $\pm$ 362 & 1100 $\pm$ 267&0.10&2.02 $\pm$ 0.27&- &-18.67&-&-0.97\\
16056$^{\tiny\ddagger}$ &2000 & 214.85965 & 52.88814 & C-P & 4.811 & 225 $\pm$ 48 & 1595 $\pm$ 330 & 727 $\pm$ 257&0.16&1.71 $\pm$ 0.16&-2.00$^{+0.11}_{-0.11}$ &-20.26&0.63 $\pm$ 0.01&-1.80\\
46552 &80244 & 214.90216 & 52.86976 & C-P & 7.004 & 286 $\pm$ 121 & 1574 $\pm$ 515 & 838 $\pm$ 460&0.02&1.80 $\pm$ 0.30&-1.57$^{+0.44}_{-0.45}$ &-19.26&-&-0.30\\
6224 &5927 & 215.01220 & 52.99299 & CA-P & 4.632 & 199 $\pm$ 48 & 1431 $\pm$ 137 & 1136 $\pm$ 161&0.11&1.51 $\pm$ 0.19&- &-18.93&-&-1.30\\
52156 &10399 & 214.85116 & 52.85567 & CA-P & 4.818 & 218 $\pm$ 18 & 1406 $\pm$ 102 & 1614 $\pm$ 636&0.19&1.66 $\pm$ 0.18&-1.53$^{+0.25}_{-0.24}$ &-18.34&-&-1.88\\
25074$^{\tiny\dagger}$ &397 & 214.83618 & 52.88268 & C-P & 6.013 & 181 $\pm$ 31 & 1383 $\pm$ 186 & 959 $\pm$ 358&0.08&1.69 $\pm$ 0.13&-1.99$^{+0.05}_{-0.05}$ &-21.24&0.47 $\pm$ 0.00&-0.28\\
84971 &19689 & 214.87164 & 52.81231 & CA-P & 5.123 & 244 $\pm$ 60 & 1380 $\pm$ 320 & 914 $\pm$ 93&0.07&1.56 $\pm$ 0.18&-1.86$^{+0.17}_{-0.18}$ &-18.93&0.39 $\pm$ 0.01&-1.32\\
27280 &439 & 214.82537 & 52.86307 & C-P & 7.181 & 266 $\pm$ 45 & 1373 $\pm$ 185 & -&0.03&1.93 $\pm$ 0.21&- &-&-&-\\
100152 &81063 & 214.79911 & 52.72512 & C-P & 6.089 & 182 $\pm$ 26 & 1355 $\pm$ 120 & 1165 $\pm$ 172&0.12&1.71 $\pm$ 0.22&-1.44$^{+0.25}_{-0.25}$ &-19.04&0.45 $\pm$ 0.01&-1.33\\
47504$^{\tiny\mathsection}$ &40556 & 214.92191 & 52.87619 & R-M & 5.642 & 196 $\pm$ 14 & 1344 $\pm$ 63 & -&0.16&1.64 $\pm$ 0.18&- &-&0.11 $\pm$ 0.00&-\\
6348 &79098 & 214.96007 & 52.95547 & CA-P & 5.089 & 188 $\pm$ 67 & 1328 $\pm$ 404 & 997 $\pm$ 390&0.03&1.64 $\pm$ 0.29&- &-&-&-\\
35306 &82043 & 214.71999 & 52.75025 & C-P & 4.329 & 193 $\pm$ 92 & 1328 $\pm$ 603 & 726 $\pm$ 47&0.06&1.33 $\pm$ 0.16&-2.50$^{+0.17}_{-0.17}$ &-18.96&0.16 $\pm$ 0.02&-0.91\\
76140 &927584 & 214.97837 & 52.87739 & R-M & 7.011 & 335 $\pm$ 73 & 1315 $\pm$ 172 & -&0.09&2.05 $\pm$ 0.19&- &-&0.50 $\pm$ 0.14&-\\
77845 &18741 & 214.98771 & 52.87574 & R-P & 5.690 & 167 $\pm$ 57 & 1302 $\pm$ 425 & 705 $\pm$ 314&0.08&1.53 $\pm$ 0.16&-1.90$^{+0.14}_{-0.14}$ &-19.84&0.26 $\pm$ 0.00&-0.97\\
81061$^{\tiny\dagger\tiny\ddagger}$ &1019 & 215.03539 & 52.89067 & C-M & 8.681 & 138 $\pm$ 14 & 1300 $\pm$ 47 & -&0.16&1.74 $\pm$ 0.16&- &-&0.37 $\pm$ 0.00&-\\
87370$^{\tiny\ddagger}$ &1374 & 214.94390 & 52.85005 & C-P & 5.002 & 184 $\pm$ 23 & 1299 $\pm$ 134 & 911 $\pm$ 34&0.13&1.77 $\pm$ 0.15&-2.13$^{+0.10}_{-0.10}$ &-20.53&0.49 $\pm$ 0.00&-1.65\\
15050$^{\tiny\dagger}$ &67001 & 214.91408 & 52.93303 & R-P & 4.518 & 191 $\pm$ 40 & 1284 $\pm$ 251 & 747 $\pm$ 336&0.09&1.33 $\pm$ 0.15&-2.18$^{+0.14}_{-0.13}$ &-19.52&0.56 $\pm$ 0.01&-1.74\\
25074$^{\tiny\dagger}$ &69412 & 214.83618 & 52.88268 & R-P & 6.005 & 174 $\pm$ 36 & 1276 $\pm$ 253 & 787 $\pm$ 382&0.08&1.50 $\pm$ 0.13&-2.21$^{+0.06}_{-0.06}$ &-21.35&0.47 $\pm$ 0.00&-0.22\\
38411 &41260 & 214.78801 & 52.78155 & R-M & 5.689 & - & 1264 $\pm$ 116 & 996 $\pm$ 119&0.05&1.54 $\pm$ 0.20&- &-&-&-\\
18197 &970128 & 214.88100 & 52.89121 & R-P & 6.491 & 212 $\pm$ 103 & 1261 $\pm$ 606 & 772 $\pm$ 81&0.05&1.67 $\pm$ 0.20&-1.81$^{+0.13}_{-0.13}$ &-20.26&0.73 $\pm$ 0.03&-0.39\\
2149 &3584 & 214.98875 & 52.99804 & C-P & 4.650 & 153 $\pm$ 39 & 1239 $\pm$ 287 & 823 $\pm$ 207&0.16&1.61 $\pm$ 0.17&-1.40$^{+0.10}_{-0.10}$ &-20.41&0.48 $\pm$ 0.01&-1.30\\
25074$^{\tiny\dagger}$ &397 & 214.83618 & 52.88268 & C-M & 6.002 & 167 $\pm$ 6 & 1217 $\pm$ 26 & 1014 $\pm$ 33&0.08&1.62 $\pm$ 0.13&- &-&0.47 $\pm$ 0.00&-\\
43111$^{\tiny\ddagger}$ &10716 & 214.95139 & 52.92510 & CA-P & 5.710 & 151 $\pm$ 65 & 1201 $\pm$ 507 & 765 $\pm$ 382&0.20&1.49 $\pm$ 0.17&-1.14$^{+0.42}_{-0.44}$ &-19.56&0.74 $\pm$ 0.02&-2.60\\
85072 &19851 & 214.86061 & 52.80368 & CA-P & 6.126 & 233 $\pm$ 90 & 1178 $\pm$ 451 & 719 $\pm$ 100&0.02&1.61 $\pm$ 0.18&-2.24$^{+0.10}_{-0.10}$ &-19.43&-&-0.52\\
71325$^{\tiny\dagger}$ &6411 & 215.10919 & 52.93978 & R-M & 4.882 & - & 1171 $\pm$ 57 & 1031 $\pm$ 73&0.05&1.43 $\pm$ 0.17&- &-&0.12 $\pm$ 0.06&-\\
43496 &11072 & 214.96605 & 52.93314 & CA-P & 6.145 & 202 $\pm$ 101 & 1171 $\pm$ 568 & 820 $\pm$ 212&0.06&1.39 $\pm$ 0.23&- &-&0.57 $\pm$ 0.18&-\\
71325$^{\tiny\dagger}$ &6411 & 215.10919 & 52.93978 & R-P & 4.888 & 236 $\pm$ 94 & 1159 $\pm$ 454 & 1390 $\pm$ 451&0.05&1.66 $\pm$ 0.17&-2.19$^{+0.16}_{-0.16}$ &-19.13&0.12 $\pm$ 0.06&-0.90\\
25074$^{\tiny\dagger}$ &69412 & 214.83618 & 52.88268 & R-M & 6.001 & 163 $\pm$ 9 & 1153 $\pm$ 26 & 799 $\pm$ 29&0.08&1.52 $\pm$ 0.13&- &-&0.47 $\pm$ 0.00&-\\
19193$^{\tiny\ddagger}$ &445 & 214.94161 & 52.92912 & C-P & 6.982 & 119 $\pm$ 19 & 1150 $\pm$ 178 & 694 $\pm$ 163&0.08&1.47 $\pm$ 0.21&-1.82$^{+0.13}_{-0.13}$ &-19.43&0.70 $\pm$ 0.00&-1.05\\
24622 &3040 & 214.78771 & 52.85042 & CA-P & 4.704 & 147 $\pm$ 57 & 1144 $\pm$ 418 & 591 $\pm$ 288&0.04&1.69 $\pm$ 0.22&-2.10$^{+0.17}_{-0.17}$ &-18.50&0.18 $\pm$ 0.10&-0.93\\
3391$^{\tiny\dagger}$ &69475 & 214.94919 & 52.96414 & R-P & 5.621 & 127 $\pm$ 57 & 1141 $\pm$ 489 & 1218 $\pm$ 67&0.36&2.13 $\pm$ 0.17&-2.02$^{+0.43}_{-0.43}$ &-18.71&-&-2.60\\
9482 &8945 & 214.99062 & 52.96119 & CA-P & 3.961 & 294 $\pm$ 81 & 1139 $\pm$ 304 & 809 $\pm$ 344&0.03&1.88 $\pm$ 0.21&-2.31$^{+0.17}_{-0.17}$ &-18.48&-&-0.79\\
98160 &80432 & 214.81206 & 52.74675 & C-P & 7.476 & 135 $\pm$ 23 & 1125 $\pm$ 113 & -&0.06&1.61 $\pm$ 0.18&-1.79$^{+0.15}_{-0.16}$ &-19.88&-&-0.27\\
84706 &19346 & 214.88680 & 52.82464 & CA-P & 6.193 & 163 $\pm$ 59 & 1118 $\pm$ 403 & 1093 $\pm$ 165&0.02&1.95 $\pm$ 0.23&-2.09$^{+0.10}_{-0.10}$ &-19.41&0.50 $\pm$ 0.03&-0.19\\
75478$^{\tiny\ddagger\tiny\mathsection}$ &24917 & 215.03208 & 52.91896 & R-P & 6.185 & 147 $\pm$ 34 & 1110 $\pm$ 240 & 671 $\pm$ 50&0.14&1.42 $\pm$ 0.15&-1.28$^{+0.13}_{-0.13}$ &-20.07&0.59 $\pm$ 0.01&-2.35\\
87192 &50773 & 214.95030 & 52.85571 & CA-P & 5.231 & 225 $\pm$ 38 & 1109 $\pm$ 145 & 1112 $\pm$ 217&0.01&2.14 $\pm$ 0.14&- &-19.62&-&-0.32\\
46330 &13794 & 214.95403 & 52.90786 & CA-P & 4.567 & 155 $\pm$ 22 & 1105 $\pm$ 136 & 743 $\pm$ 120&0.12&1.47 $\pm$ 0.18&-2.21$^{+0.07}_{-0.07}$ &-20.07&0.47 $\pm$ 0.01&-0.72\\
45212 &2315 & 214.91917 & 52.89350 & C-P & 5.766 & 154 $\pm$ 80 & 1087 $\pm$ 541 & 819 $\pm$ 65&0.07&1.44 $\pm$ 0.17&-1.95$^{+0.16}_{-0.16}$ &-19.72&0.68 $\pm$ 0.11&-1.53\\
76140 &927584 & 214.97837 & 52.87739 & R-P & 7.014 & 344 $\pm$ 65 & 1086 $\pm$ 140 & 1169 $\pm$ 242&0.09&1.84 $\pm$ 0.17&-2.54$^{+0.50}_{-0.48}$ &-20.30&0.50 $\pm$ 0.14&-0.72\\
45809 &80239 & 214.89605 & 52.86985 & C-P & 7.491 & <119  & 1068 $\pm$ 265 & -&0.14&<1.34 &- &-19.22&-&-\\
31826 &86480 & 214.79735 & 52.82069 & CA-P & 4.195 & 203 $\pm$ 49 & 1067 $\pm$ 218 & 733 $\pm$ 136&0.08&1.53 $\pm$ 0.25&- &-17.88&0.58 $\pm$ 0.03&-1.46\\
8666 &57270 & 214.98169 & 52.95906 & R-M & 5.081 & 114 $\pm$ 29 & 1023 $\pm$ 67 & 541 $\pm$ 52&0.18&1.49 $\pm$ 0.11&- &-&0.50 $\pm$ 0.01&-\\
6184 &974163 & 214.95008 & 52.94927 & R-P & 7.930 & 113 $\pm$ 30 & 1014 $\pm$ 162 & -&0.14&1.56 $\pm$ 0.26&- &-19.24&0.38 $\pm$ 0.03&-1.70\\
15637 &65746 & 214.91322 & 52.92958 & R-P & 4.506 & 136 $\pm$ 45 & 1007 $\pm$ 252 & 762 $\pm$ 75&0.04&1.40 $\pm$ 0.18&-2.70$^{+0.17}_{-0.17}$ &-19.35&-&-0.90\\
31338 &498 & 214.81305 & 52.83425 & C-P & 7.182 & 180 $\pm$ 37 & 1000 $\pm$ 102 & -&0.04&1.59 $\pm$ 0.19&-2.34$^{+0.19}_{-0.20}$ &-20.13&0.20 $\pm$ 0.03&-0.53\\
59817 &792 & 214.87177 & 52.83317 & C-M & 6.260 & 127 $\pm$ 18 & 989 $\pm$ 73 & 448 $\pm$ 56&0.20&1.34 $\pm$ 0.21&- &-&-&-\\
64730 &25470 & 215.08049 & 52.95477 & R-M & 5.065 & 139 $\pm$ 38 & 970 $\pm$ 112 & 556 $\pm$ 96&0.19&1.60 $\pm$ 0.20&- &-&0.35 $\pm$ 0.03&-\\
29539 &81694 & 214.83973 & 52.86289 & CA-P & 4.116 & 231 $\pm$ 47 & 950 $\pm$ 186 & 1360 $\pm$ 186&0.06&1.49 $\pm$ 0.19&-2.74$^{+0.32}_{-0.31}$ &-18.16&0.58 $\pm$ 0.03&-1.37\\
100312 &83856 & 214.80302 & 52.72654 & C-P & 4.561 & 266 $\pm$ 103 & 947 $\pm$ 357 & 965 $\pm$ 466&0.05&1.72 $\pm$ 0.21&-2.02$^{+0.26}_{-0.25}$ &-18.53&0.46 $\pm$ 0.10&-1.28\\
88854 &16429 & 214.94607 & 52.84163 & R-M & 5.241 & 139 $\pm$ 26 & 931 $\pm$ 87 & 602 $\pm$ 61&0.07&1.85 $\pm$ 0.18&- &-&0.50 $\pm$ 0.01&-\\
67762 &16915 & 215.07964 & 52.93826 & R-P & 5.060 & 122 $\pm$ 15 & 917 $\pm$ 92 & 633 $\pm$ 164&0.08&1.44 $\pm$ 0.12&-1.95$^{+0.05}_{-0.05}$ &-21.62&1.12 $\pm$ 0.01&-0.12\\
61419 &34809 & 214.89723 & 52.84385 & R-M & 9.001 & <276  & 911 $\pm$ 167 & -&0.11&<1.57 &- &-&0.16 $\pm$ 0.03&-\\
47379 &15045 & 214.93603 & 52.88679 & CA-P & 5.990 & 225 $\pm$ 31 & 909 $\pm$ 91 & 975 $\pm$ 170&0.04&1.78 $\pm$ 0.24&-1.91$^{+0.20}_{-0.20}$ &-19.02&-&-0.92\\
74993 &26352 & 215.02308 & 52.91531 & R-P & 6.179 & 115 $\pm$ 17 & 905 $\pm$ 50 & 630 $\pm$ 69&0.13&1.55 $\pm$ 0.18&-1.34$^{+0.33}_{-0.33}$ &-19.12&0.57 $\pm$ 0.01&-2.60\\
4176 &1953 & 214.99826 & 52.99475 & C-P & 4.615 & 148 $\pm$ 26 & 903 $\pm$ 40 & 613 $\pm$ 45&0.09&1.41 $\pm$ 0.16&- &-19.95&0.72 $\pm$ 0.01&-1.46\\
9010 &56600 & 214.97488 & 52.95267 & R-P & 5.076 & 212 $\pm$ 89 & 895 $\pm$ 359 & 835 $\pm$ 125&0.10&1.62 $\pm$ 0.20&-1.50$^{+0.46}_{-0.47}$ &-18.52&0.12 $\pm$ 0.03&-1.48\\
35645 &82052 & 214.76656 & 52.78227 & C-P & 5.160 & 114 $\pm$ 31 & 886 $\pm$ 96 & 614 $\pm$ 104&0.10&1.43 $\pm$ 0.23&-2.72$^{+0.39}_{-0.39}$ &-18.76&-&-1.49\\
21394 &355 & 214.80649 & 52.87883 & C-M & 6.101 & 106 $\pm$ 12 & 884 $\pm$ 48 & 664 $\pm$ 58&0.06&1.51 $\pm$ 0.19&- &-&0.49 $\pm$ 0.01&-\\
7985 &7498 & 214.95030 & 52.94046 & CA-P & 4.482 & 114 $\pm$ 34 & 883 $\pm$ 167 & 531 $\pm$ 62&0.12&1.40 $\pm$ 0.21&- &-17.80&-&-1.88\\
9290 &3587 & 215.01064 & 52.97618 & C-P & 3.919 & <110  & 879 $\pm$ 330 & 590 $\pm$ 228&0.02&1.27 $\pm$ 0.15&-2.83$^{+0.27}_{-0.27}$ &-19.49&0.58 $\pm$ 0.02&-\\
4745 &4399 & 214.96883 & 52.97107 & CA-P & 4.728 & 233 $\pm$ 64 & 860 $\pm$ 233 & -&0.06&1.50 $\pm$ 0.16&-2.32$^{+0.10}_{-0.10}$ &-19.17&-&-1.03\\
24654 &3070 & 214.80249 & 52.86077 & CA-P & 4.211 & 263 $\pm$ 104 & 857 $\pm$ 264 & 317 $\pm$ 47&0.09&1.11 $\pm$ 0.20&-2.23$^{+0.35}_{-0.35}$ &-18.18&-&-1.52\\
88854 &16429 & 214.94607 & 52.84163 & R-P & 5.247 & 119 $\pm$ 41 & 855 $\pm$ 283 & 842 $\pm$ 86&0.07&2.00 $\pm$ 0.18&-1.98$^{+0.22}_{-0.22}$ &-19.34&0.50 $\pm$ 0.01&-1.81\\
66855$^{\tiny\ddagger\tiny\mathsection}$ &19489 & 215.12261 & 52.97321 & R-M & 6.393 & 92 $\pm$ 5 & 854 $\pm$ 23 & -&0.07&1.36 $\pm$ 0.13&- &-&0.87 $\pm$ 0.00&-\\
46308 &91392 & 214.89723 & 52.86775 & CA-P & 5.176 & 167 $\pm$ 32 & 850 $\pm$ 134 & 826 $\pm$ 305&0.06&1.87 $\pm$ 0.29&-1.96$^{+0.53}_{-0.52}$ &-18.13&-&-1.36\\
96682$^{\tiny\ddagger}$ &22500 & 214.86884 & 52.79748 & CA-P & 5.214 & 124 $\pm$ 9 & 846 $\pm$ 53 & 613 $\pm$ 187&0.17&1.83 $\pm$ 0.11&-1.79$^{+0.06}_{-0.06}$ &-20.26&0.38 $\pm$ 0.01&-1.63\\
74181$^{\tiny\ddagger}$ &19394 & 214.95844 & 52.87509 & CA-P & 6.176 & 121 $\pm$ 31 & 846 $\pm$ 211 & 674 $\pm$ 291&0.21&1.73 $\pm$ 0.18&-1.25$^{+0.13}_{-0.13}$ &-19.45&-&-2.60\\
75478$^{\tiny\ddagger\tiny\mathsection}$ &24917 & 215.03208 & 52.91896 & R-M & 6.181 & 119 $\pm$ 6 & 839 $\pm$ 28 & 643 $\pm$ 37&0.14&1.40 $\pm$ 0.15&- &-&0.59 $\pm$ 0.01&-\\
57786 &97175 & 214.86708 & 52.83907 & CA-P & 5.203 & 135 $\pm$ 32 & 836 $\pm$ 105 & 823 $\pm$ 135&0.07&1.68 $\pm$ 0.25&- &-17.91&-&-1.55\\
87160 &22404 & 214.94191 & 52.84989 & CA-P & 5.222 & 172 $\pm$ 49 & 825 $\pm$ 230 & 1205 $\pm$ 295&0.02&2.11 $\pm$ 0.20&-2.17$^{+0.17}_{-0.17}$ &-19.00&1.37 $\pm$ 0.26&-0.93\\
59112 &39176 & 214.85326 & 52.82296 & R-M & 6.112 & 102 $\pm$ 14 & 823 $\pm$ 71 & 470 $\pm$ 67&0.03&1.69 $\pm$ 0.20&- &-&0.36 $\pm$ 0.02&-\\
87607 &22902 & 214.89620 & 52.81507 & CA-P & 4.379 & 117 $\pm$ 38 & 819 $\pm$ 244 & 695 $\pm$ 236&0.08&1.38 $\pm$ 0.18&-1.89$^{+0.17}_{-0.17}$ &-18.29&0.40 $\pm$ 0.02&-1.90\\
80182 &27050 & 215.00818 & 52.87645 & CA-P & 5.150 & 173 $\pm$ 42 & 808 $\pm$ 143 & 782 $\pm$ 214&0.06&1.57 $\pm$ 0.22&-2.91$^{+0.35}_{-0.34}$ &-18.23&-&-1.37\\
24169 &71096 & 214.83231 & 52.88411 & R-P & 3.886 & 187 $\pm$ 33 & 780 $\pm$ 89 & 863 $\pm$ 71&0.03&1.49 $\pm$ 0.20&-1.81$^{+0.40}_{-0.40}$ &-18.49&0.94 $\pm$ 0.11&-1.28\\
76343 &22230 & 214.98981 & 52.88463 & CA-P & 5.091 & 140 $\pm$ 44 & 775 $\pm$ 236 & 679 $\pm$ 301&0.01&1.44 $\pm$ 0.15&-2.41$^{+0.12}_{-0.12}$ &-19.36&0.75 $\pm$ 0.02&-0.73\\
87541 &21865 & 214.94387 & 52.84935 & R-P & 5.234 & 127 $\pm$ 19 & 750 $\pm$ 79 & 763 $\pm$ 117&0.03&1.60 $\pm$ 0.27&-2.06$^{+0.29}_{-0.29}$ &-18.62&-&-1.28\\
9010 &56600 & 214.97488 & 52.95267 & R-M & 5.061 & <184  & 740 $\pm$ 89 & 588 $\pm$ 94&0.10&1.47 $\pm$ 0.20&- &-&0.12 $\pm$ 0.03&-\\
59403 &16014 & 214.85234 & 52.82106 & CA-P & 4.273 & 148 $\pm$ 66 & 730 $\pm$ 308 & 509 $\pm$ 53&0.05&1.11 $\pm$ 0.17&-2.34$^{+0.22}_{-0.22}$ &-18.71&-&-1.35\\
67762 &16915 & 215.07964 & 52.93826 & R-M & 5.058 & 85 $\pm$ 4 & 729 $\pm$ 12 & 500 $\pm$ 10&0.08&1.34 $\pm$ 0.12&- &-&1.12 $\pm$ 0.01&-\\
48882 &38182 & 214.90763 & 52.85764 & R-P & 5.168 & <87  & 726 $\pm$ 82 & 516 $\pm$ 91&0.11&1.18 $\pm$ 0.18&-1.57$^{+0.53}_{-0.53}$ &-18.74&-&-\\
87541 &21865 & 214.94387 & 52.84935 & R-M & 5.219 & 130 $\pm$ 31 & 725 $\pm$ 77 & 575 $\pm$ 89&0.03&1.47 $\pm$ 0.27&- &-&-&-\\
66855$^{\tiny\ddagger\tiny\mathsection}$ &19489 & 215.12261 & 52.97321 & R-P & 6.400 & 81 $\pm$ 15 & 714 $\pm$ 120 & 543 $\pm$ 177&0.07&1.49 $\pm$ 0.13&-2.26$^{+0.04}_{-0.04}$ &-22.12&0.87 $\pm$ 0.00&-0.05\\
43895 &11439 & 214.93516 & 52.90881 & CA-P & 4.802 & 73 $\pm$ 13 & 704 $\pm$ 102 & 657 $\pm$ 39&0.12&1.25 $\pm$ 0.15&-1.94$^{+0.19}_{-0.18}$ &-18.79&0.65 $\pm$ 0.01&-2.35\\
90671 &80083 & 214.96128 & 52.84236 & C-P & 8.638 & 147 $\pm$ 35 & 703 $\pm$ 110 & -&0.16&1.77 $\pm$ 0.22&- &-19.28&-&-1.81\\
3391$^{\tiny\dagger}$ &69475 & 214.94919 & 52.96414 & R-M & 5.620 & 70 $\pm$ 18 & 695 $\pm$ 39 & -&0.36&1.62 $\pm$ 0.21&- &-&-&-\\
63949 &19999 & 215.07849 & 52.95742 & CA-P & 4.757 & 117 $\pm$ 29 & 694 $\pm$ 159 & 571 $\pm$ 174&0.03&1.10 $\pm$ 0.14&-2.26$^{+0.15}_{-0.15}$ &-18.84&0.28 $\pm$ 0.03&-1.81\\
53321 &11407 & 214.88357 & 52.87248 & CA-P & 5.091 & 99 $\pm$ 30 & 693 $\pm$ 72 & 437 $\pm$ 61&0.12&1.23 $\pm$ 0.19&-1.67$^{+0.27}_{-0.27}$ &-18.11&-&-1.89\\
82853 &30177 & 215.01304 & 52.86613 & CA-P & 4.403 & 134 $\pm$ 68 & 670 $\pm$ 319 & 988 $\pm$ 459&0.02&1.45 $\pm$ 0.18&-2.49$^{+0.24}_{-0.24}$ &-18.47&0.71 $\pm$ 0.04&-2.60\\
39375 &39348 & 214.79241 & 52.77950 & R-P & 5.226 & 128 $\pm$ 24 & 656 $\pm$ 62 & 936 $\pm$ 119&0.10&1.97 $\pm$ 0.27&-1.35$^{+0.30}_{-0.29}$ &-19.54&1.13 $\pm$ 0.06&-1.40\\
88515 &24097 & 214.95649 & 52.85255 & CA-P & 4.659 & 91 $\pm$ 17 & 651 $\pm$ 104 & 423 $\pm$ 194&0.05&1.43 $\pm$ 0.19&-2.31$^{+0.16}_{-0.16}$ &-18.22&0.19 $\pm$ 0.00&-2.60\\
43544 &50293 & 214.89893 & 52.88539 & R-M & 4.810 & - & 638 $\pm$ 31 & 423 $\pm$ 29&0.14&1.15 $\pm$ 0.18&- &-&0.42 $\pm$ 0.01&-\\
86889 &22085 & 214.94708 & 52.85497 & CA-P & 4.549 & 123 $\pm$ 26 & 636 $\pm$ 55 & 692 $\pm$ 77&0.03&1.38 $\pm$ 0.16&- &-&0.80 $\pm$ 0.31&-\\
61253 &80710 & 214.88498 & 52.83604 & C-P & 6.547 & <99  & 633 $\pm$ 100 & 637 $\pm$ 167&0.12&1.12 $\pm$ 0.19&- &-&-&-\\
64762 &25262 & 215.12482 & 52.98572 & R-P & 4.472 & 97 $\pm$ 44 & 623 $\pm$ 257 & 534 $\pm$ 49&0.11&1.34 $\pm$ 0.18&-2.00$^{+0.28}_{-0.29}$ &-19.03&1.15 $\pm$ 0.03&-2.36\\
57710 &14815 & 214.81160 & 52.79985 & CA-P & 6.099 & 97 $\pm$ 14 & 601 $\pm$ 58 & 534 $\pm$ 90&0.06&1.40 $\pm$ 0.23&- &-18.77&-&-1.51\\
55139 &93206 & 214.81126 & 52.81250 & CA-P & 5.651 & 120 $\pm$ 16 & 599 $\pm$ 63 & 540 $\pm$ 86&0.11&1.23 $\pm$ 0.20&-2.33$^{+0.43}_{-0.42}$ &-18.57&0.34 $\pm$ 0.07&-1.86\\
62230 &33287 & 214.85444 & 52.80952 & R-M & 5.243 & 72 $\pm$ 10 & 597 $\pm$ 30 & 448 $\pm$ 34&0.16&1.47 $\pm$ 0.12&- &-&-&-\\
82417 &36016 & 214.99383 & 52.85461 & CA-P & 7.280 & <105  & 597 $\pm$ 199 & -&0.16&<0.92 &- &-&-&-\\
74993 &26352 & 215.02308 & 52.91531 & R-M & 6.180 & 72 $\pm$ 11 & 596 $\pm$ 35 & 417 $\pm$ 60&0.13&1.37 $\pm$ 0.19&- &-&0.57 $\pm$ 0.01&-\\
68788 &13262 & 215.08658 & 52.93647 & R-M & 4.808 & - & 588 $\pm$ 48 & 655 $\pm$ 72&0.04&1.42 $\pm$ 0.22&- &-&0.96 $\pm$ 0.24&-\\
31339 &499 & 214.81301 & 52.83417 & C-M & 7.171 & <138  & 579 $\pm$ 112 & -&0.05&<1.67 &- &-&-&-\\
28223 &973612 & 214.83698 & 52.86700 & R-M & 7.163 & 109 $\pm$ 20 & 576 $\pm$ 79 & -&0.07&1.65 $\pm$ 0.18&- &-&-&-\\
100210 &40066 & 214.86248 & 52.76995 & CA-P & 5.215 & <93  & 556 $\pm$ 85 & 842 $\pm$ 173&0.02&1.77 $\pm$ 0.27&-2.24$^{+0.50}_{-0.48}$ &-18.52&0.49 $\pm$ 0.07&-\\
6306$^{\tiny\ddagger}$ &6009 & 215.01330 & 52.99324 & CA-P & 5.077 & 69 $\pm$ 5 & 552 $\pm$ 36 & 369 $\pm$ 99&0.12&1.18 $\pm$ 0.16&-1.69$^{+0.10}_{-0.11}$ &-19.95&1.09 $\pm$ 0.00&-2.35\\
96184 &24306 & 214.85451 & 52.79026 & R-P & 4.574 & 81 $\pm$ 14 & 537 $\pm$ 85 & 711 $\pm$ 266&0.05&1.73 $\pm$ 0.17&-2.31$^{+0.08}_{-0.08}$ &-19.93&0.54 $\pm$ 0.01&-1.05\\
61419 &34809 & 214.89723 & 52.84385 & R-P & 9.004 & 225 $\pm$ 66 & 534 $\pm$ 109 & -&0.11&1.68 $\pm$ 0.26&- &-&0.16 $\pm$ 0.03&-\\
16205 &977908 & 214.86303 & 52.88943 & R-M & 6.576 & - & 526 $\pm$ 37 & 10 $\pm$ 4&0.11&1.40 $\pm$ 0.19&- &-&0.60 $\pm$ 0.05&-\\
2166 &1912 & 215.01084 & 53.01333 & C-P & 5.104 & 80 $\pm$ 26 & 521 $\pm$ 150 & 556 $\pm$ 28&0.24&1.47 $\pm$ 0.16&-2.06$^{+0.22}_{-0.22}$ &-20.99&0.77 $\pm$ 0.00&-2.60\\
86030 &80374 & 214.89807 & 52.82489 & C-P & 7.180 & 102 $\pm$ 40 & 504 $\pm$ 127 & -&0.03&1.76 $\pm$ 0.29&-2.04$^{+0.53}_{-0.53}$ &-19.32&-&-1.28\\
47521 &42630 & 214.91832 & 52.87936 & R-M & 5.665 & 88 $\pm$ 21 & 492 $\pm$ 24 & 453 $\pm$ 50&0.03&0.96 $\pm$ 0.13&- &-&0.27 $\pm$ 0.01&-\\
55144 &93221 & 214.86470 & 52.85050 & CA-P & 4.349 & 129 $\pm$ 45 & 491 $\pm$ 128 & 1872 $\pm$ 311&0.10&1.46 $\pm$ 0.18&-2.51$^{+0.38}_{-0.39}$ &-18.01&0.29 $\pm$ 0.08&-2.35\\
40979 &36473 & 214.79873 & 52.77641 & R-M & 5.239 & 195 $\pm$ 32 & 482 $\pm$ 52 & 631 $\pm$ 85&0.10&1.83 $\pm$ 0.33&- &-&1.06 $\pm$ 0.04&-\\
69753 &10366 & 215.14136 & 52.97019 & R-M & 5.245 & <11879  & 474 $\pm$ 64 & 290 $\pm$ 53&0.29&1.42 $\pm$ 0.24&- &-&1.03 $\pm$ 0.08&-\\
8674 &2355 & 215.00849 & 52.97797 & C-P & 6.124 & 148 $\pm$ 46 & 465 $\pm$ 51 & 481 $\pm$ 61&0.47&1.59 $\pm$ 0.21&- &-20.08&-&-2.60\\
39375 &39348 & 214.79241 & 52.77950 & R-M & 5.224 & 106 $\pm$ 23 & 463 $\pm$ 44 & 676 $\pm$ 88&0.10&1.83 $\pm$ 0.27&- &-&1.13 $\pm$ 0.06&-\\
9623 &9080 & 214.98958 & 52.95979 & CA-P & 5.162 & 65 $\pm$ 33 & 450 $\pm$ 204 & 329 $\pm$ 41&0.04&1.07 $\pm$ 0.18&-2.53$^{+0.22}_{-0.22}$ &-18.89&-&-2.60\\
44441 &48045 & 214.96870 & 52.92965 & R-P & 8.722 & 81 $\pm$ 17 & 449 $\pm$ 56 & -&0.05&1.60 $\pm$ 0.23&-2.18$^{+0.14}_{-0.14}$ &-20.37&0.87 $\pm$ 0.14&-0.72\\
37321 &14257 & 214.80521 & 52.79907 & CA-P & 4.833 & 191 $\pm$ 25 & 448 $\pm$ 35 & 577 $\pm$ 278&0.07&1.59 $\pm$ 0.20&-1.92$^{+0.14}_{-0.14}$ &-19.34&0.49 $\pm$ 0.02&-1.48\\
19984 &54857 & 214.87256 & 52.87595 & R-M & 5.669 & 135 $\pm$ 35 & 444 $\pm$ 67 & 751 $\pm$ 123&0.07&1.83 $\pm$ 0.24&- &-&-&-\\
74818 &931914 & 214.98850 & 52.89181 & R-P & 7.103 & <128  & 442 $\pm$ 93 & -&0.11&<1.31 &-1.21$^{+0.53}_{-0.55}$ &-18.90&-&-\\
43544 &11130 & 214.89893 & 52.88539 & CA-P & 4.820 & 35 $\pm$ 9 & 437 $\pm$ 41 & 466 $\pm$ 37&0.14&1.20 $\pm$ 0.18&-1.24$^{+0.22}_{-0.21}$ &-18.26&0.42 $\pm$ 0.01&-2.60\\
49117 &36695 & 214.98711 & 52.91274 & R-P & 4.556 & 58 $\pm$ 8 & 428 $\pm$ 38 & 405 $\pm$ 73&0.16&1.27 $\pm$ 0.13&-2.00$^{+0.07}_{-0.07}$ &-20.92&1.20 $\pm$ 0.01&-2.36\\
75646 &24484 & 215.02100 & 52.91035 & R-M & 5.022 & 121 $\pm$ 35 & 420 $\pm$ 52 & 311 $\pm$ 67&0.08&1.32 $\pm$ 0.25&- &-&-&-\\
89417 &14715 & 214.94097 & 52.83481 & R-M & 5.193 & 54 $\pm$ 14 & 414 $\pm$ 33 & 303 $\pm$ 33&0.07&0.89 $\pm$ 0.17&- &-&0.52 $\pm$ 0.03&-\\
27536 &63849 & 214.82339 & 52.86037 & R-P & 5.754 & <99  & 394 $\pm$ 54 & 262 $\pm$ 71&0.02&1.19 $\pm$ 0.23&- &-19.30&0.98 $\pm$ 0.10&-2.36\\
44441 &48045 & 214.96870 & 52.92965 & R-M & 8.720 & <49  & 387 $\pm$ 50 & -&0.05&<1.30 &- &-&0.87 $\pm$ 0.14&-\\
44441 &11964 & 214.96870 & 52.92965 & CA-P & 8.726 & 77 $\pm$ 12 & 376 $\pm$ 45 & -&0.05&1.58 $\pm$ 0.23&- &-&0.87 $\pm$ 0.14&-\\
29581 &59668 & 214.83063 & 52.85577 & R-M & 4.817 & - & 347 $\pm$ 148 & 1638 $\pm$ 117&0.14&2.05 $\pm$ 0.13&- &-&0.24 $\pm$ 0.01&-\\
12215 &1430 & 214.86850 & 52.91620 & CA-P & 4.606 & <53  & 311 $\pm$ 131 & 659 $\pm$ 56&0.12&1.48 $\pm$ 0.17&-1.73$^{+0.66}_{-0.68}$ &-19.69&0.22 $\pm$ 0.02&-\\
27594 &63761 & 214.81885 & 52.85695 & R-M & 5.099 & <112  & 307 $\pm$ 44 & 247 $\pm$ 38&0.09&0.89 $\pm$ 0.17&- &-&0.47 $\pm$ 0.02&-\\
78953 &14864 & 214.99914 & 52.87693 & R-M & 5.136 & 71 $\pm$ 14 & 301 $\pm$ 25 & 320 $\pm$ 43&0.03&0.94 $\pm$ 0.15&- &-&0.72 $\pm$ 0.03&-\\
45913 &44556 & 214.97770 & 52.92708 & R-P & 6.153 & <51  & 274 $\pm$ 132 & 340 $\pm$ 81&0.03&1.17 $\pm$ 0.23&-2.51$^{+0.40}_{-0.40}$ &-19.30&0.97 $\pm$ 0.03&-\\
48882 &38182 & 214.90763 & 52.85764 & R-M & 5.164 & 99 $\pm$ 27 & 269 $\pm$ 35 & 288 $\pm$ 55&0.11&0.93 $\pm$ 0.18&- &-&-&-\\
46186 &82300 & 214.90056 & 52.87161 & C-P & 4.719 & 76 $\pm$ 25 & 258 $\pm$ 69 & 298 $\pm$ 129&0.06&0.87 $\pm$ 0.14&- &-19.16&0.59 $\pm$ 0.04&-1.89\\
81601 &37936 & 215.00228 & 52.86474 & CA-P & 7.273 & <64  & 251 $\pm$ 51 & -&0.09&<0.90 &- &-17.94&-&-\\
49011 &37135 & 214.93817 & 52.87862 & R-M & 5.786 & <78  & 137 $\pm$ 23 & 258 $\pm$ 58&0.08&0.97 $\pm$ 0.21&- &-&-&-\\
27536 &63849 & 214.82339 & 52.86037 & R-M & 5.754 & - & 131 $\pm$ 23 & -&0.02&<0.97 &- &-&0.98 $\pm$ 0.10&-\\
70867$^{\tiny\dagger}$ &1236 & 215.14529 & 52.96729 & C-P & 4.496 & <21  & 119 $\pm$ 19 & 130 $\pm$ 26&0.15&0.72 $\pm$ 0.15&-1.98$^{+0.20}_{-0.20}$ &-19.33&0.86 $\pm$ 0.00&-2.60\\
18197 &970128 & 214.88100 & 52.89121 & R-M & 6.487 & 10 $\pm$ 3 & 75 $\pm$ 26 & 590 $\pm$ 64&0.05&1.55 $\pm$ 0.20&- &-&0.73 $\pm$ 0.03&-\\
88902 &45430 & 214.90575 & 52.81278 & CA-P & 4.489 & - & - & 1083 $\pm$ 170&0.02&1.74 $\pm$ 0.22&- &-&0.16 $\pm$ 0.05&-\\
48859 &80671 & 214.92161 & 52.86771 & C-P & 5.735 & - & - & 1657 $\pm$ 306&0.10&1.87 $\pm$ 0.21&- &-19.50&-&-\\
16620 &63183 & 214.85692 & 52.88300 & R-P & 4.433 & - & - & 992 $\pm$ 206&0.01&0.89 $\pm$ 0.10&- &-&0.10 $\pm$ 0.02&-\\
16205 &977908 & 214.86303 & 52.88943 & R-P & 6.580 & - & - & 133 $\pm$ 47&0.11&2.21 $\pm$ 0.18&-1.53$^{+0.27}_{-0.27}$ &-20.36&0.60 $\pm$ 0.05&-\\
70867$^{\tiny\dagger}$ &1236 & 215.14529 & 52.96729 & C-M & 4.485 & - & - & 121 $\pm$ 11&0.15&0.60 $\pm$ 0.15&- &-&0.86 $\pm$ 0.00&-\\
5040 &3585 & 215.02321 & 53.00797 & C-M & 3.868 & - & - & 704 $\pm$ 60&0.12&1.70 $\pm$ 0.13&- &-&0.13 $\pm$ 0.01&-\\
78973 &1305 & 215.04087 & 52.90623 & C-M & 4.282 & - & - & 18102 $\pm$ 1808&0.06&2.74 $\pm$ 0.15&- &-&0.50 $\pm$ 0.03&-\\
64762 &25262 & 215.12482 & 52.98572 & R-M & 4.468 & - & - & 443 $\pm$ 38&0.11&1.26 $\pm$ 0.18&- &-&1.15 $\pm$ 0.03&-\\
49117 &36695 & 214.98711 & 52.91274 & R-M & 4.557 & - & - & 283 $\pm$ 7&0.16&1.16 $\pm$ 0.13&- &-&1.20 $\pm$ 0.01&-\\
75857$^{\tiny\dagger}$ &24489 & 214.97813 & 52.87952 & R-M & 4.553 & - & - & 744 $\pm$ 14&0.16&1.62 $\pm$ 0.13&- &-&0.71 $\pm$ 0.01&-\\
78973 &14545 & 215.04087 & 52.90623 & R-M & 4.281 & - & - & 332 $\pm$ 24&0.06&1.00 $\pm$ 0.15&- &-&0.50 $\pm$ 0.03&-\\
77845 &18741 & 214.98771 & 52.87574 & R-M & 5.674 & - & - & 753 $\pm$ 52&0.08&1.49 $\pm$ 0.16&- &-&0.26 $\pm$ 0.00&-\\
52164 &52613 & 214.79375 & 52.81475 & R-M & 4.510 & - & - & 1523 $\pm$ 75&0.01&2.07 $\pm$ 0.18&- &-&-&-\\
55189 &46423 & 214.82801 & 52.82411 & R-M & 4.690 & - & - & 305 $\pm$ 33&0.17&0.49 $\pm$ 0.18&- &-&3.70 &-\\
96184 &24306 & 214.85451 & 52.79026 & R-M & 4.563 & - & - & 492 $\pm$ 23&0.05&1.59 $\pm$ 0.17&- &-&0.54 $\pm$ 0.01&-\\
972 &74288 & 214.92394 & 52.95791 & R-M & 4.462 & - & - & 2404 $\pm$ 180&0.05&1.62 $\pm$ 0.14&- &-&0.56 $\pm$ 0.05&-\\
6079 &63238 & 214.95853 & 52.95591 & R-M & 4.574 & - & - & 550 $\pm$ 35&0.09&1.36 $\pm$ 0.16&- &-&1.15 $\pm$ 0.20&-\\
6454 &62125 & 214.97495 & 52.96537 & R-M & 4.615 & - & - & 723 $\pm$ 88&0.12&1.68 $\pm$ 0.23&- &-&-&-\\
15050$^{\tiny\dagger}$ &67001 & 214.91408 & 52.93303 & R-M & 4.508 & - & - & 943 $\pm$ 34&0.09&1.42 $\pm$ 0.15&- &-&0.56 $\pm$ 0.01&-\\
15637 &65746 & 214.91322 & 52.92958 & R-M & 4.511 & - & - & 726 $\pm$ 65&0.04&1.38 $\pm$ 0.17&- &-&-&-\\
13730 &70345 & 214.85527 & 52.89870 & R-M & 4.353 & - & - & 454 $\pm$ 27&0.18&1.24 $\pm$ 0.17&- &-&0.94 $\pm$ 0.01&-\\
24169 &71096 & 214.83231 & 52.88411 & R-M & 3.867 & - & - & 513 $\pm$ 56&0.03&1.26 $\pm$ 0.20&- &-&0.94 $\pm$ 0.11&-\\
28312 &62302 & 214.82923 & 52.86109 & R-M & 4.707 & - & - & 640 $\pm$ 40&0.17&1.15 $\pm$ 0.13&- &-&-&-\\
47168 &41367 & 214.91793 & 52.87556 & R-M & 4.726 & - & - & 1483 $\pm$ 40&0.10&1.92 $\pm$ 0.15&- &-&0.32 $\pm$ 0.00&-\\
56599 &44510 & 214.87269 & 52.85145 & R-M & 4.283 & - & - & 855 $\pm$ 75&0.07&1.51 $\pm$ 0.19&- &-&-&-\\
\end{longtable}
\begin{flushleft}
{\footnotesize
\textbf{Notes:}~$^{*}$ Survey and disperser abbreviations: CA-P: CAPERS-PRISM, C-P: CEERS-PRISM, R-P: RUBIES-PRISM, C-M: CEERS-MR, R-M: RUBIES-MR.   
$^{\dagger}$ Classified as BLAGN. $^{\ddagger}$ Possible AGN from OHNO diagram. $^{\mathsection}$ EELG with evidence of broad [O III] 
}
\end{flushleft}}
\FloatBarrier 
\twocolumn

\section{Gaussian fitting of \oiiihb~lines}\label{sec:App_o3}
{Due to the wavelength-dependent spectral resolution of NIRSpec/PRISM, the \oiiihb~lines may appear blended or partially blended depending on the redshift of the source. In contrast, the higher spectral resolution of NIRSpec/MR (R$\sim$1000) allows the three lines to be well separated. Fig. \ref{fig:z_fitting} shows two examples of the Gaussian modeling of \oiiihb~lines using LiMe described in Section~\ref{sec:spec_z} for galaxies observed with NIRSpec/PRISM. }

{In the top panel, the \oiiihb~lines of galaxy ID 9482 are blended. This galaxy is the lowest-redshift source ($z_{spec}=3.96$) in the CAPERS subsample, and the \oiiihb~lines fall at observed wavelengths corresponding to R$\sim$70. In the bottom panel, the spectrum of galaxy ID 44441, the highest-redshift source ($z_{\rm spec}=8.73$), shows the unblended \oiiihb~lines at observed wavelengths corresponding to R$\sim$270.  These two cases correspond to the more extreme cases in terms of spectral resolution for these lines.}

{As described in Section~\ref{sec:spec_z}, \oiiihb~lines are modeled as blended Gaussians, meaning that each line is represented by an individual Gaussian fit simultaneously. For the \oiiihb\ complex, we assume the theoretical \oiii$\lambda5007$/\oiii$\lambda4959$ flux ratio of 2.94 and adopt the same velocity width for all lines. As shown in Fig.~\ref{fig:z_fitting}, this approach allows us to recover the individual line components even in the lowest-resolution case.}
\begin{figure}[t!]
    \centering
    \includegraphics[width=\linewidth]{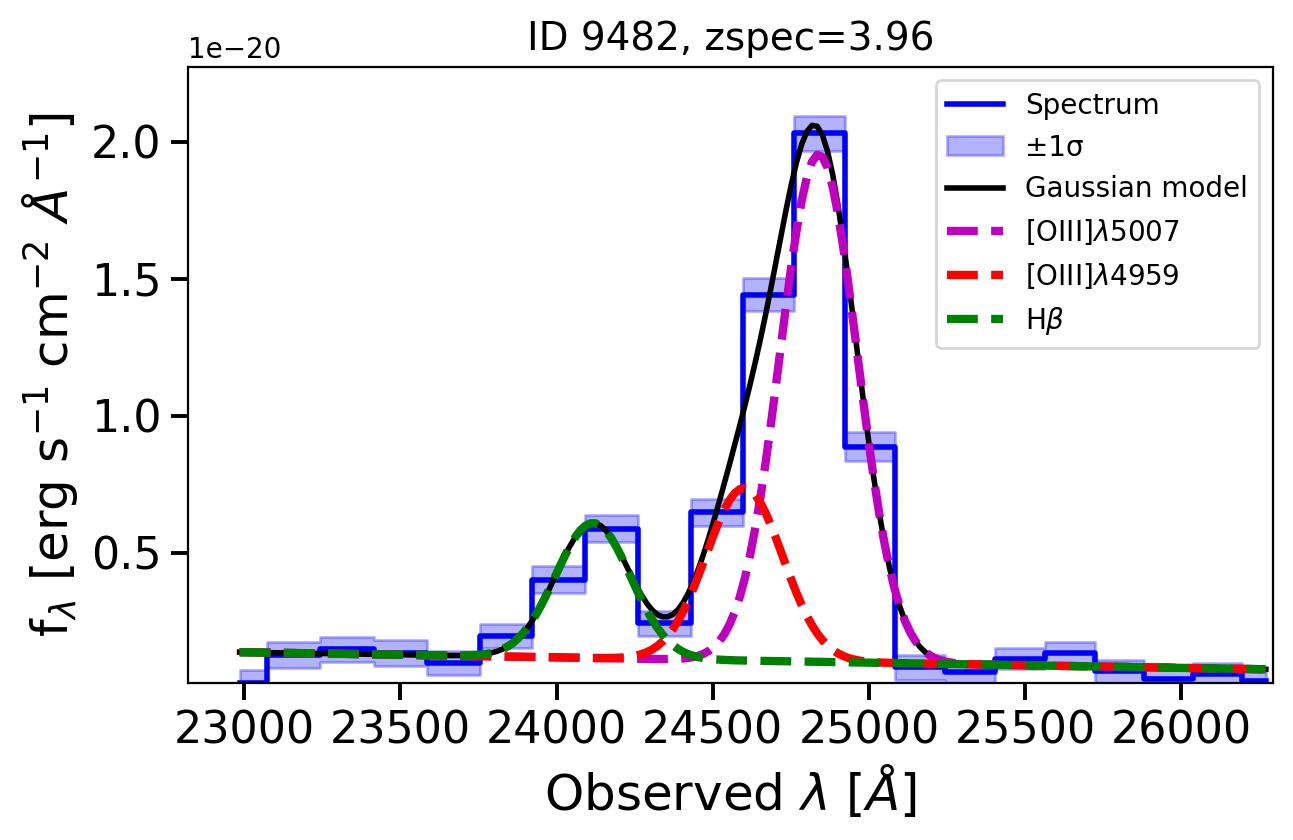}\\
    \includegraphics[width=\linewidth]{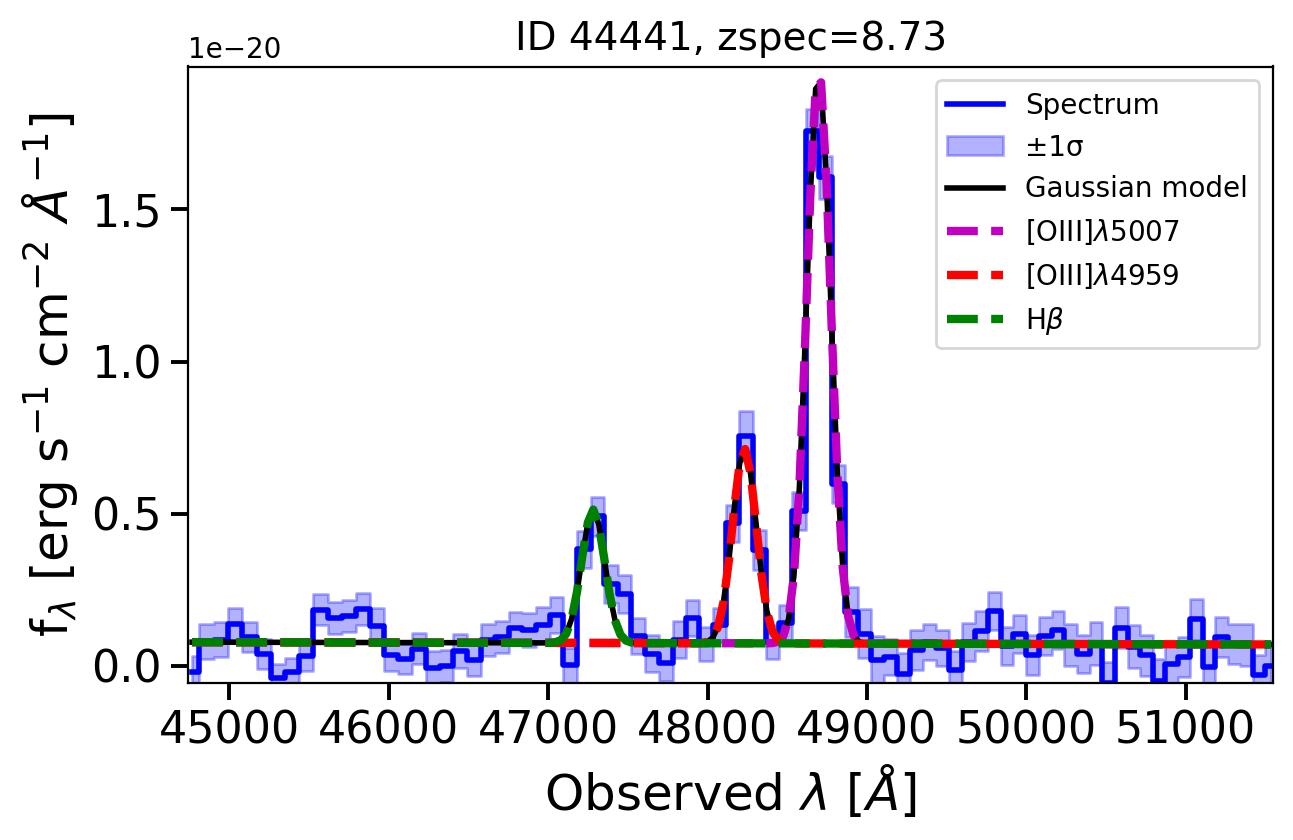}
    \caption{{Showcases of Gaussian modeling of the \oiiihb\ emission-line complex. \textit{Top panel}: The lowest-redshift source in the CAPERS subsample, illustrating the case where the \oiiihb~lines are observed with a resolution R$\sim70$. \textit{Bottom panel}: The highest-redshift source in the CAPERS subsample, representing the opposite extreme where the \oiiihb~emission is observed with a resolution R$\sim270$. \textit{In both panels}, the observed NIRSpec/PRISM spectrum is shown as a blue solid line, with the shaded region indicating the 1$\sigma$ uncertainty. The solid black curve represents the total best-fitting model to the \oiiihb\ complex, while the magenta, red, and green dashed lines represent the individual Gaussian components of \oiii$\lambda5007$, \oiii$\lambda4959$, and \hbeta, respectively.}}
    \label{fig:z_fitting}
\end{figure}

\section{Median spectrum of EELGs}\label{sec:stack}

To show the typical spectral features of EELGs, we constructed a median stacked spectrum of all 127 galaxies in our sample with PRISM observations. The stacking was performed using a non-weighted approach. Each individual spectrum was first shifted to the rest-frame using its spectroscopic redshift and then interpolated onto a common wavelength grid between 1000 and 7000~\r{A} and considering the median redshift of the sample, $z=5.2$. The spectra were subsequently normalized to the median flux within the 2700-2800~\r{A} range. At each wavelength, the final stacked flux density was computed as the median of the individual flux densities, applying a 3-$\sigma$ clipping to exclude outliers.

The 1-$\sigma$ uncertainty on the stacked spectrum was estimated via bootstrap resampling: for each wavelength, the median flux was recalculated for 1000 random resamplings of the individual spectra, and the standard deviation of these medians was taken as the error. The resulting stacked spectrum is presented in Fig. \ref{fig:stack}, where we indicate the positions of prominent emission lines, many of which are analyzed in detail in this paper. This approach allows us to highlight the typical spectral features of the EELG population. 

We perform a similar approach with the 81 galaxies in the sample with MR spectra. In this case, we considered a common wavelength grid between 3600 and 7000~\r{A} and considering the median redshift of the sample, $z=5.2$. The individual spectra were normalized to the median flux within the 5100-6400~\r{A} range. The resulting spectrum is shown in Fig. \ref{fig:stack_mr}.

\begin{figure*}[t!]
    \centering
    \includegraphics[width=0.7\linewidth]{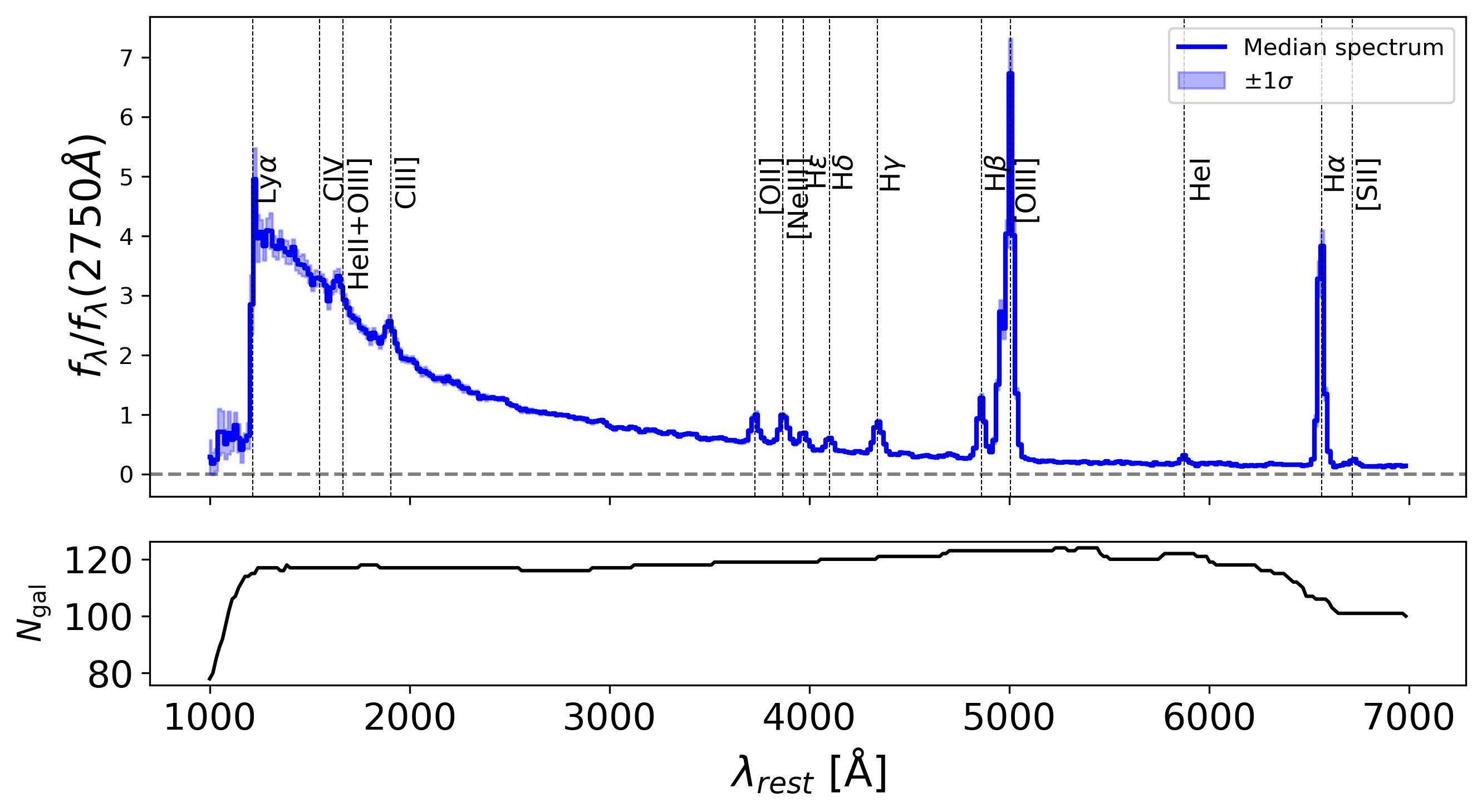}
    \caption{Median spectrum of the EELGs in the sample with PRISM observations. The shaded region is the 1$\sigma$ uncertainty. The vertical lines mark the position of typical emission lines. The bottom panel shows the number of galaxies considered for the median spectrum.}
    \label{fig:stack}
\end{figure*}

\begin{figure*}[t!]
    \centering
    \includegraphics[width=0.7\linewidth]{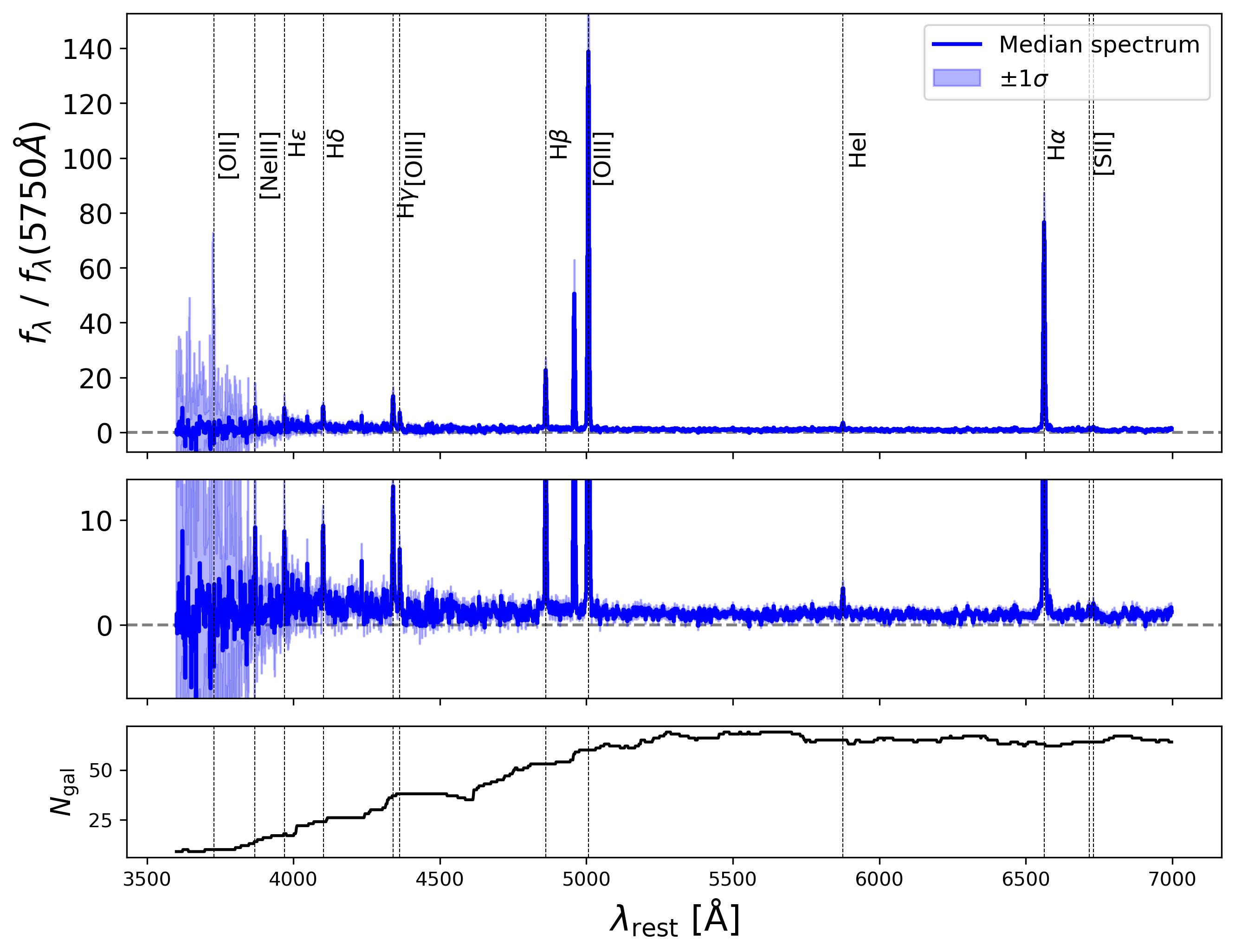}
    \caption{Median spectrum of the EELGs in the sample with MR observations. The shaded region is the 1$\sigma$ uncertainty. The vertical lines mark the position of typical emission lines. The middle panel presents a zoomed-in view of the flux density for clarity. The bottom panel shows the number of galaxies considered for the median spectrum.}
    \label{fig:stack_mr}
\end{figure*}

\section{{Fluxes and EWs of emission lines}}
\begin{figure}[t!]
    \centering
    \includegraphics[width=\linewidth]{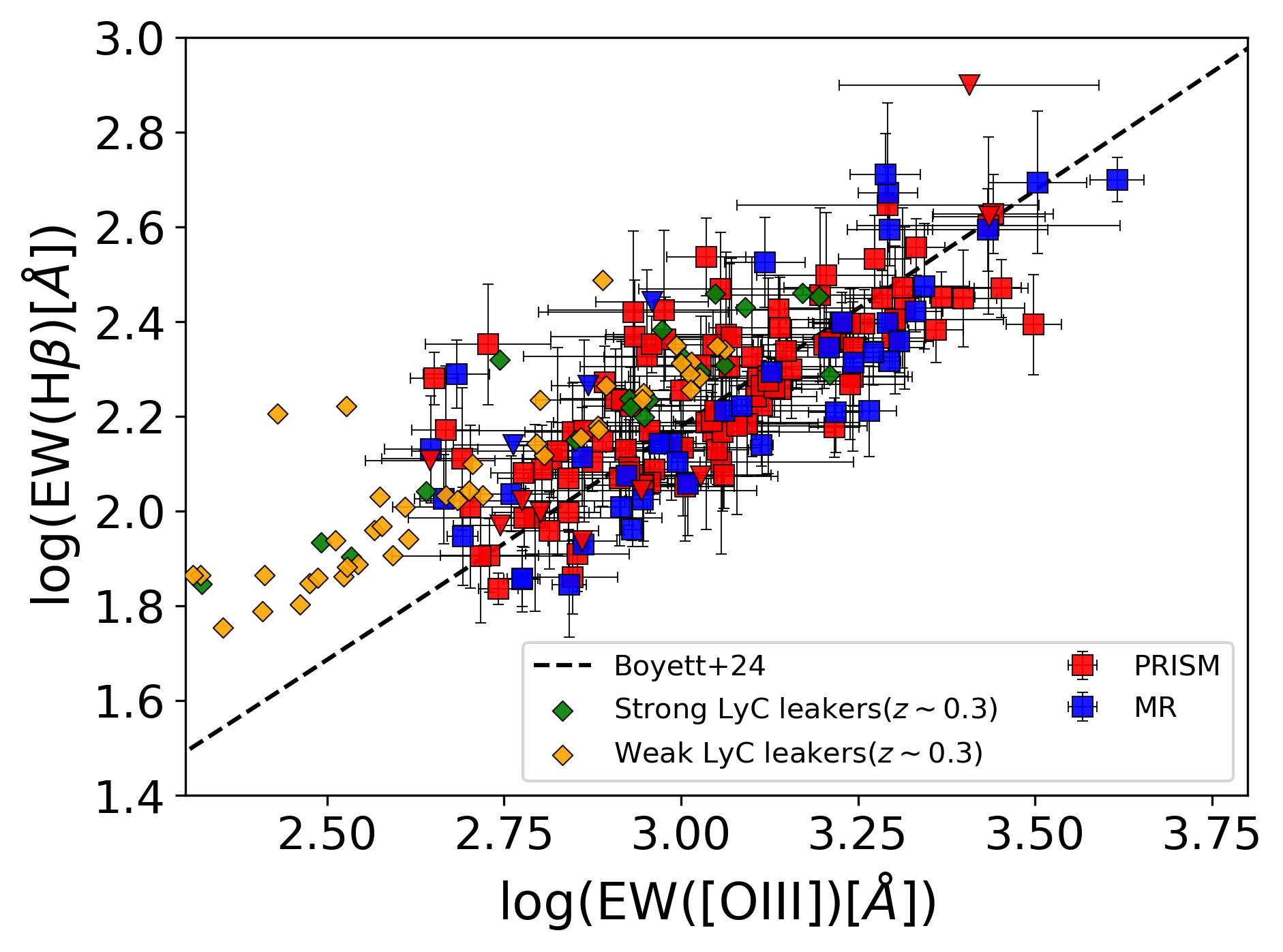}
    \caption{Comparison between EWs in the sample {with PRISM (red squares) and MR (blue squares) spectra}. The triangle symbols are upper limits due to low S/N of \hbeta. The black dashed line is the relation from \cite{Boyett2024} for galaxies at $3<z<9.5$. {Green and orange} symbols are as in Fig. \ref{fig:MS}.}
    \label{fig:ew_balmer_o3}
\end{figure}
In Fig. \ref{fig:ew_balmer_o3} we show the comparison between EW(\oiii) and EW(\hbeta). We find that galaxies exhibiting high EW(\oiii) also tend to show correspondingly high EWs in the Balmer lines. This strong correlation ($\rho=0.78$, $p\sim0$) suggests a common source of ionization for both \oiii~and the Balmer emission, most likely originating from intense star-forming activity. Our findings are consistent with recent results from the JADES survey, which examined EELGs over $3 < z < 9.5$ and reported similar EW behavior \citep{Boyett2024}. Furthermore, we note that a similar EW(\oiii)–EW(\hbeta) relationship has been observed in LyC leakers at $z \sim 0.3$, where the emission line strengths are comparable to those in our EELG sample.

In Fig. \ref{fig:fluxes_prism_mr} we show a comparison between the observed fluxes (top panel) and EWs (bottom panel) in the subsample of galaxies that have PRISM and MR spectra. Regarding the observed fluxes, we find a very strong correlation ($\rho={0.94}$, $p\sim0$) between the fluxes measured in the PRISM and the MR spectra for \oiii, \halpha~and \hbeta. We notice, however, that the fluxes measured in the PRISM spectra tend to be slightly higher than those measured in the MR spectra, with a median difference of {0.06}~dex ($\sigma=0.14$). Comparing the EWs of the same lines, we also find a very strong correlation ($\rho={0.95}$, $p\sim0$) with a median difference of {0.06}~dex ($\sigma=0.17$) between the EWs measured from the PRISM and the MR spectra. This larger offset and scatter reflect the limitations in determining the faint continuum level in these galaxies.

\begin{figure}[t!]
    \centering
    \includegraphics[width=\linewidth]{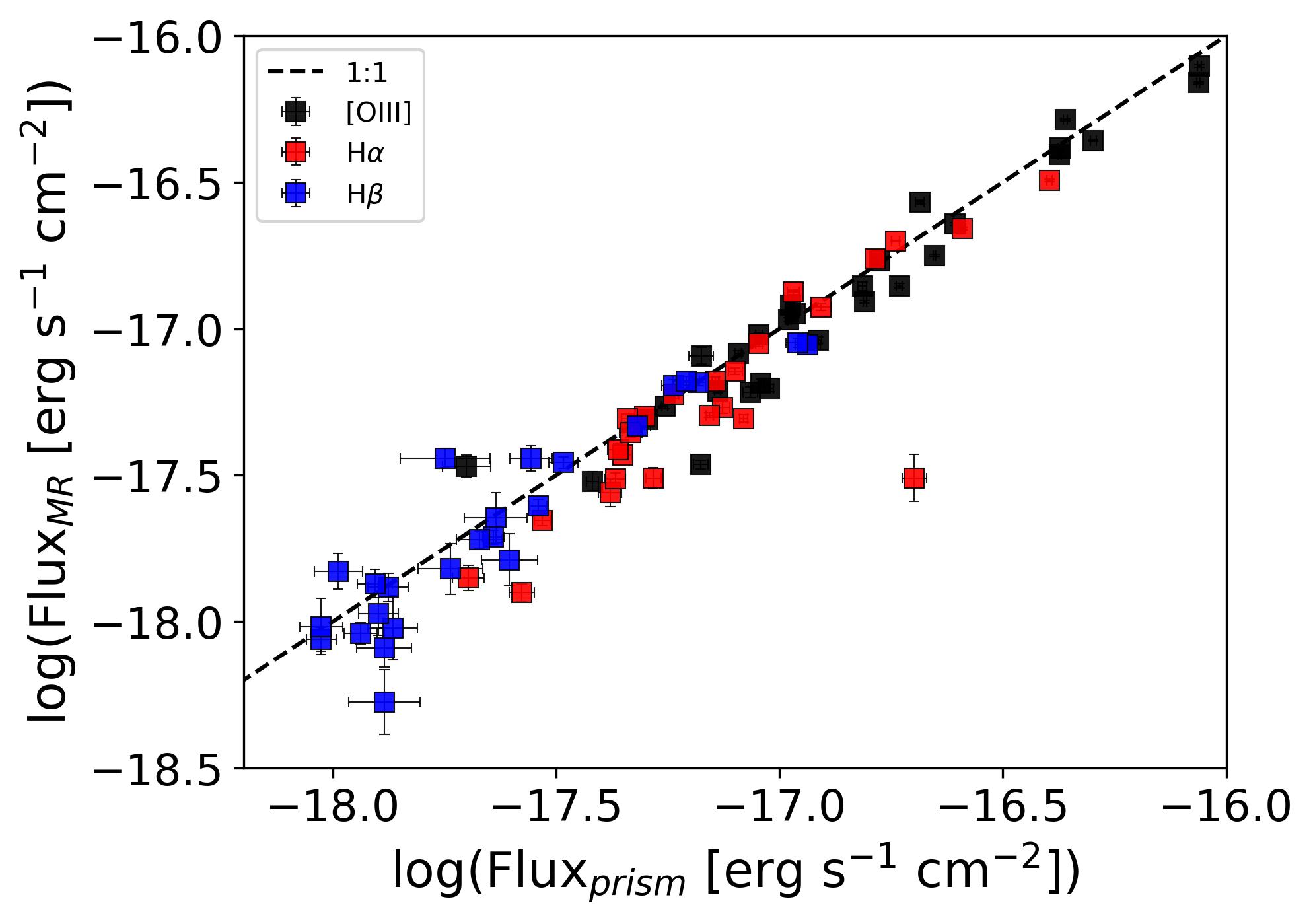}\\
    \includegraphics[width=\linewidth]{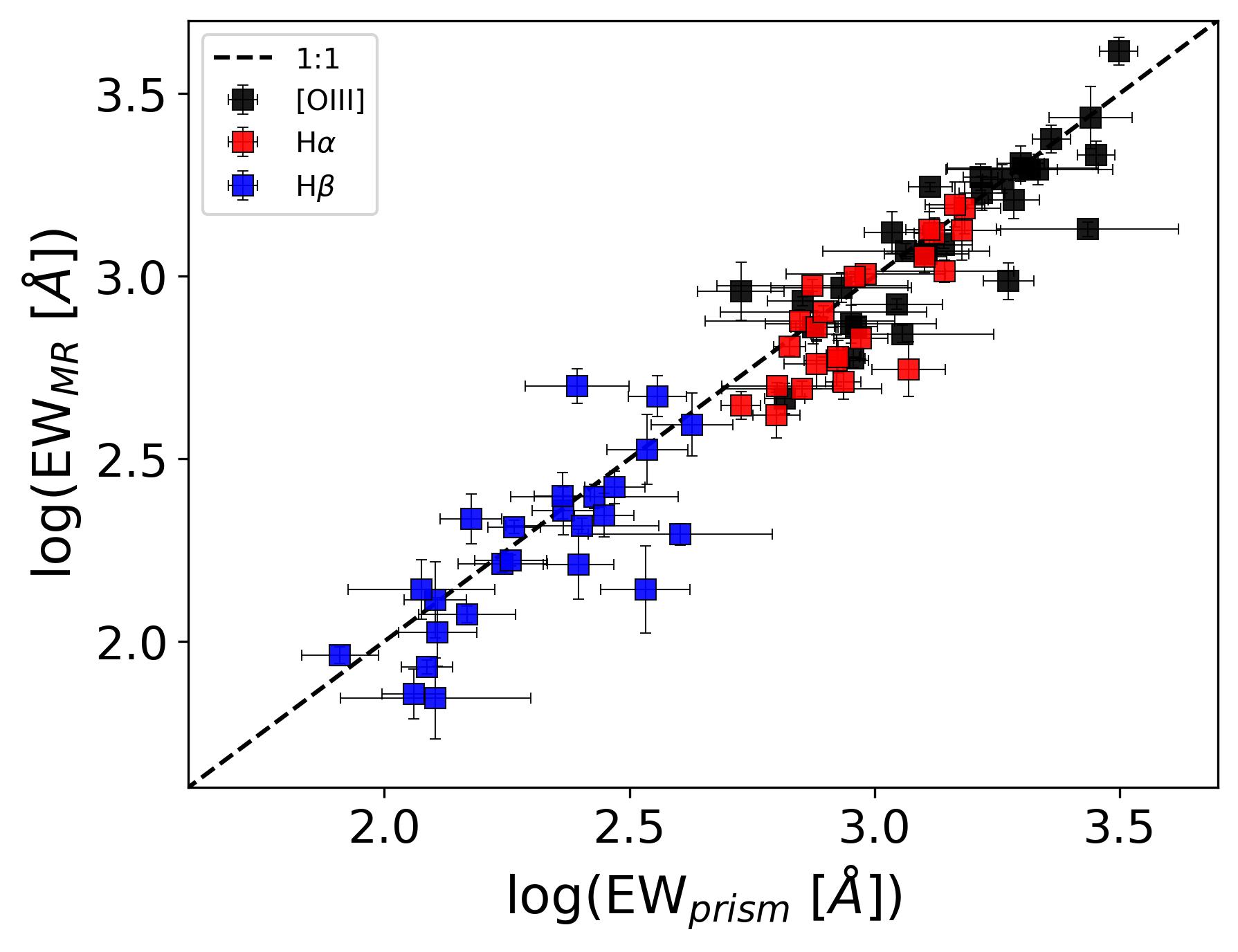}\\
    \caption{Comparison of fluxes (top panel) and EWs (bottom panel) of emission lines from PRISM and MR. In black, \oiii, in red \halpha, and in blue \hbeta. The dashed black line is the 1:1 relation.}
    \label{fig:fluxes_prism_mr}
\end{figure}

\section{SFRs from Balmer lines}\label{sec:sfr10Myr}
In Fig. \ref{fig:sfr_balmer_sed}, we show the comparison between the SFR obtained from SED fitting (average in the last 10 Myr) and from Balmer lines. We find a good agreement ($\rho=0.85$, $p\sim0$) between both parameters with a median log(SFR$_{\rm10Myr}$)-log(SFR$_{\rm Balmer}$)=${-0.16}$~dex ($\sigma=0.24$~dex). This indicates that Balmer lines are tracers of the average SFR in the last 10Myr in our sample. 

\begin{figure}[t!]
    \centering
    \includegraphics[width=\linewidth]{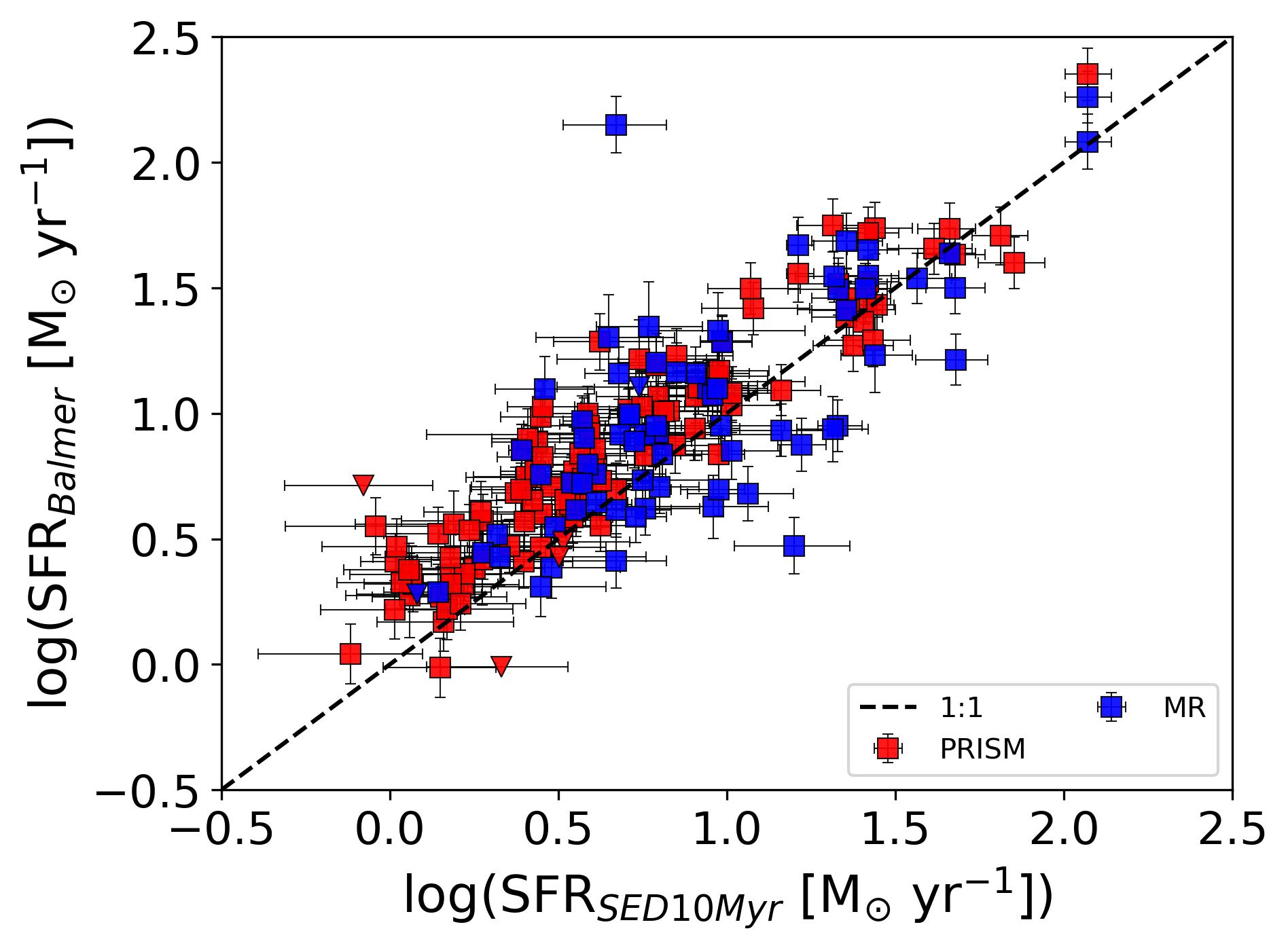}
    \caption{Comparison between SFRs from SED fitting (averaged over the last 10 Myr) and Balmer lines {in the sample with PRISM (red squares) and MR (blue squares) spectra}. Triangles indicate upper limits due to low \hbeta\ S/N. The dashed black line shows the 1:1 relation.}
    \label{fig:sfr_balmer_sed}
\end{figure}
\section{$\beta$ from photometry vs spectra}

In Fig. \ref{fig:delta_beta}, we show the difference in the UV $\beta$ slopes using the methods described in Sec. \ref{sec:beta-muv} where we estimate $\beta$ using the spectra or the photometry. These differences have been found in other works \citep[e.g.,][]{Morales2025} when using only spectra or photometry to estimate the $\beta$ slopes. We note that no wavelength-dependent correction was applied to the spectral flux calibration in this study. This choice is motivated by our finding that, on average, the correction factor remains consistent across all available filters, indicating that a single scaling provides a satisfactory match between photometric and spectroscopic measurements. Nevertheless, we cannot fully rule out a subtle wavelength dependence of the correction, which could contribute to some of the observed discrepancies between $\beta$ estimated using spectra or photometry.
\begin{figure}[t!]
    \centering
    \includegraphics[width=\linewidth]{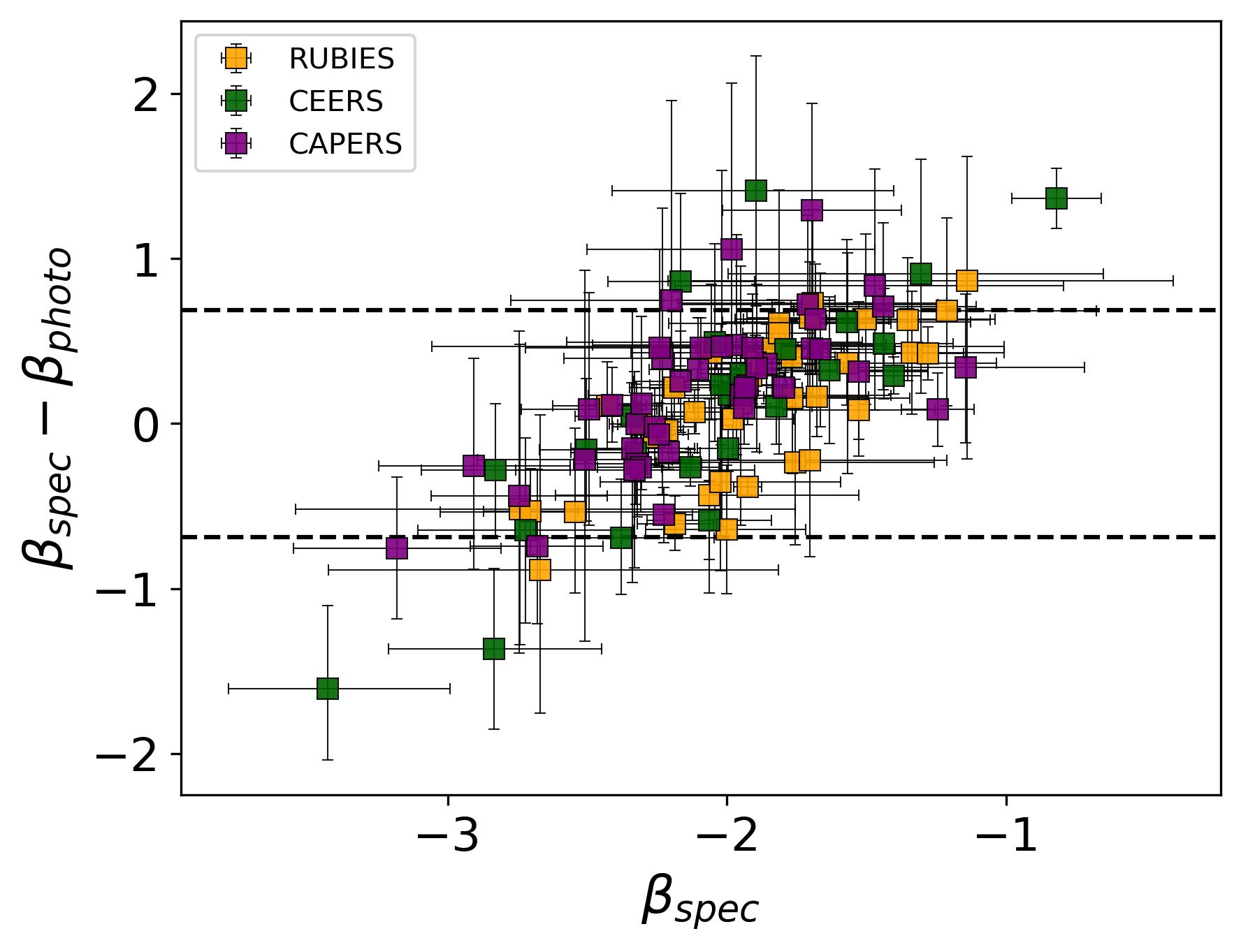}
    \caption{Difference in the UV $\beta$ slopes estimated from spectra and photometry {in the sample with PRISM spectra from the RUBIES (in orange), CEERS (in green) and CAPERS (in purple) surveys}. The black dashed lines represent the median difference $\pm$ 1$\sigma$ of the observed scatter.}
    \label{fig:delta_beta}
\end{figure}

\end{appendix}
\end{document}